\documentclass[reprint,aip,jcp,onecolumn,longbibliography,nofootinbib]{revtex4-2}
\usepackage{graphicx}
\usepackage{dcolumn}
\usepackage{subfigure}
\usepackage{bm}
\usepackage{amssymb}
\usepackage{amsmath}
\usepackage{gensymb}
\usepackage{mathtools}
\usepackage{cases}
\usepackage{xcolor}
\usepackage{soul}
\usepackage[latin1]{inputenc}
\usepackage{comment}
\newcommand{\be}{\begin{equation}}
\newcommand{\ee}{\end{equation}}

\DeclareMathOperator{\sech}{sech}
\DeclareMathOperator{\csch}{csch}
\DeclarePairedDelimiter\floor{\lfloor}{\rfloor}

\begin{document}
\title{Single file dynamics of tethered random walkers}
\author{Santos Bravo Yuste}
\email{santos@unex.es}
\affiliation{
Departamento de F\'{\i}sica and Instituto de Computaci\'on Cient\'{\i}fica Avanzada (ICCAEX), Universidad de Extremadura, E-06006 Badajoz, Spain}
\author{A. Baumgaertner}
\email{artur.baumgaertner@uni-due.de}
\affiliation{Faculty of Physics, University of Duisburg-Essen, 47048 Duisburg, Germany}
\author{E. Abad}
\email{eabad@unex.es (Corresponding author)}
\affiliation{Departamento de F\'{\i}sica Aplicada and Instituto de Computaci\'on Cient\'{\i}fica Avanzada (ICCAEX), Centro Universitario de M\'erida, Universidad de Extremadura, E-06800 M\'erida, Spain}
\date{\today}
\begin{abstract}
We consider the single file dynamics of $N$ identical random walkers moving with diffusivity $D$ in one dimension (walkers bounce off each other when attempting to overtake). Additionally, we require that the separation between neighboring walkers does not exceed a threshold value $\Delta$ and therefore call them ``tethered walkers'' (they behave as if bounded by strings that fully tighten when reaching the maximum length $\Delta$). For a finite $\Delta$, we study the diffusional relaxation to the equilibrium state and characterize the latter [the long-time relaxation is exponential with a characteristic time that scales as $(N\Delta)^2/D$]. In particular, our approximate approach for the $N$-particle probability distribution yields the one-particle distribution function of the central and edge particles [the first two positional moments are given as power expansions in $\Delta/\sqrt{4Dt}$]. For $N=2$, we find an exact solution (both in the continuum and on-lattice case) and use it to test our approximations for one-particle distributions, positional moments, and correlations. For finite $\Delta$ and arbitrary $N$, edge particles move with an effective long-time diffusivity $D/N$, in sharp contrast with the $1/\ln(N)$-behavior observed when $\Delta=\infty$. Finally, we compute the probability distribution of the equilibrium system length and associated entropy. We find that the force required to change this length by a given amount is linear in this quantity; the (entropic) spring constant is $6k_BT/(N\Delta^2)$. In this respect, the system behaves as an ideal polymer. The main analytical results are confirmed using Monte Carlo simulations.
\end{abstract}
%\pacs{05.40.Fb, 02.50.-r}
\maketitle

\section{Introduction}

Diffusive transport in confined geometries continues to be a hot topic in many different fields, notably biology. For example, narrow channels are characterized by a width of the order of only a few particle diameters; hence, the simple paradigm of non-interacting diffusive particles clearly does not apply in this case, as excluded-volume effects and the influence of the confinement on other distance-dependent interactions must be taken into account.

Effectively one-dimensional natural and human-made systems in which diffusing particles are prevented from overtaking one another are abundant \cite{Ryabov2016, Benichou2018}.  This so-called single-file diffusion (SFD) preserves the initial particle ordering; an immediate consequence is that diffusive mixing is strongly hindered, implying that a strong memory of the initial condition persists for long times. Another important consequence is that the transport properties become strongly dependent on the particle density, as the mean free path is greatly influenced by changes in this quantity. As pointed out by B\'enichou \emph{et al.} \cite{Benichou2018}, these strong density effects can bring about drastic changes not only in transport coefficients, but also in the time dependence of key statistical quantities.

Unfortunately, we cannot account for all the important theoretical and experimental contributions to the problem of SFD. Therefore, we restrict ourselves to a non-exhaustive list of a few of them.  From an experimental point of view, SFD is relevant for transport in porous media \cite{Gupta1995, Meersmann2000}, hard rod systems \cite{Jepsen1965, Levitt1973},  traffic jams \cite{Chowdhury2000}, protein diffusion in DNA \cite{Graneli2006, Li2009}, dynamics of trailing ants \cite{John2009}, sliding of ribosomes in messenger RNA \cite{Bressloff2013}, polymer translocation \cite{Muthukumar2011, Perkins1994}, diffusion in zeolites \cite{Hahn1995, Kukla1996, Keffer1999, Kaerger2012}, nanotubes \cite{Cheng2007}, and confined colloidal systems \cite{Wei2000, Zia2010, Kollmann2003}. Among the theoretical aspects are the calculation of propagators \cite{Rodenbeck1998, Aslangul1998, Aslangul1999, Kollmann2003, Lizana2008, Lizana2009}, the effect of different types of bias (owing to autonomous motion or external forces) \cite{Illien2013}, density profiles \cite{Poncet2021}, large deviation functions \cite{Krapivsky2014, Krapivsky2015, Lapolla2018, Mallmin2021}, a broad class of correlation functions \cite{Grabsch2022, Grabsch2024}, and  first-passage properties \cite{Sanders2012, Lapolla2022}. For more details see, for example, Ryabov's monograph \cite{Ryabov2016}, the recent review by B\'enichou \emph{et al.} \cite{Benichou2018}, and the references therein.

The problem of diffusing particles subject to hard-core interactions in one dimension was considered
by Harris in a pioneering study \cite{Harris1965}. This author found that the positional probability density function (pdf) of a tracer particle remains Gaussian, but its mean square displacement (MSD) displays a subdiffusive behavior of the form $\langle x^2  \rangle  \propto t^{1/2}$, as opposed to the normal diffusive behavior observed for non-interacting particles. Kollmann demonstrated in Ref.~\onlinecite{Kollmann2003} that this behavior occurs for any type of interaction between particles provided that mutual passage is excluded. The independence from the details of the interaction potential between particles was exploited by Lizana \emph{et al.} \cite{Lizana2010}  to develop a harmonization method. In this method, the true interparticle potential (which includes a hard-core part) is replaced with a harmonic potential. This substitution facilitates the analytical treatment of SFD systems.

Beyond the observed onset of anomalous diffusion \cite{Metzler2000, Metzler2004}, we have already mentioned above that another hallmark of single file systems is their non-Markovian character due to correlations between particles, which results in strong memory effects \cite{Metzler2014a, Metzler2014b}. In this context, the importance of finite-size effects is remarkable. As previously mentioned, for the infinite system considered in Ref.~\onlinecite{Harris1965} the behavior of the tracer particle in the long-time asymptotic regime is subdiffusive; in contrast, for a finite system, subdiffusion is only transient, and a crossover to normal diffusion takes place as $t\to\infty$. While the thermodynamic limit is very popular in the existing literature on both continuum and on-lattice SFD, the case of a finite number of particles $N$ has special relevance for biological systems, as narrow channels typically simultaneously harbor a very limited number of particles \cite{Levitt1986, Bauer2006, Abad2009}. In this context, an interesting, recently addressed problem with practical implications for many real systems concerns the computation of occupation times of a finite set of particles performing SFD \cite{Benichou2015}.

In the present work, we address the problem of SFD in the case where the interacting potential is non-analytic and $N$ is finite.  Before outlining the contents of each section, let us briefly discuss some works that are of special relevance for our purposes. The intimate connection between reflected Brownian motion \cite{Skorokhod1961, Grebenkov2007a, Grebenkov2007b, Grebenkov2019} and single file transport has been exploited by a number of authors to study the properties of many systems of interest. R\"odenbeck \emph{et al.} devised a method based on the reflection principle to calculate the exact propagators for a general class of SFD systems \cite{Rodenbeck1998}.  They took advantage of the fact that the derivative normal to the plane defined by equating the coordinates of any two particles must vanish (zero-flux condition), and that the corresponding solution can be constructed by the method of images.  At approximately the same time, Aslangul \cite{Aslangul1998} used a factorization ansatz based on the product of Gaussian propagators and step functions to obtain an exact solution for the $N$-particle propagator that satisfies the zero-flux condition in the aforementioned planes. Using the propagator, Aslangul was able to compute the drift and diffusion properties arising from many-body interactions. In a subsequent work \cite{Aslangul1999}, he compared the off-lattice solution with the lattice solution and found that the signature of the discrete spatial support persisted for a long time in the solution via long-lived subdominant contributions.

\begin{figure}[ht]
\begin{center}
\includegraphics[width=0.5\textwidth,angle=0]{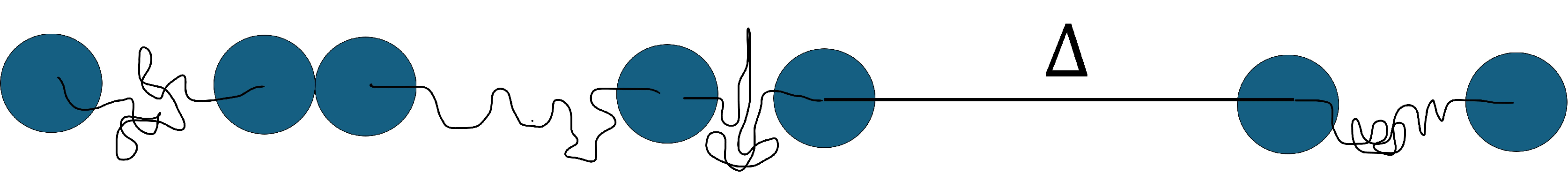}
\end{center}
\caption{\label{tethered} Representation of a string-bead system with maximum interparticle separation $\Delta$.}
\end{figure}

Here, we go one step further and investigate the effect of imposing a second constraint on diffusing particles subject to hard-core interactions, namely, that the separation distance between each particle and either of its nearest-neighbors never exceeds a fixed value $\Delta$. Hereafter we refer to such particles as ``tethered walkers''. To visualize our setting, the walkers can be thought of as being linked to one another by strings that progressively tighten as they move away from each other but have no influence on motion as long as the allowed maximum separation distance $\Delta$ (length of the string) is not exceeded (see Fig.~\ref{tethered}). In this case, the string exerts an infinite instantaneous attractive force that prevents further growth of the separation distance; in other words, each walker carries with her a symmetric infinite potential well (a hard-wall potential) of width $\Delta$ which limits the motion of both nearest neighbors. In this sense, loosely speaking our model can be viewed as a ``string-bead'' model rather than a spring-bead one.

Systems of tethered walkers may be physically realized in a variety of one-dimensional or quasi-onedimensional settings, such as stretched DNA molecules diffusing in nanochannels, where connectivity and confinement impose local constraints on the diffusive motion (see, e.g., Ref.~\onlinecite{Dai2016} and references therein). Another relevant scenario concerns SFD in nanopores,
where electrostatic or steric interactions limit interparticle spacing, as is the case in channels exhibiting multi-ionic conduction \cite{Hille2001}. Colloidal particles diffusing in microchannels also exhibit hard-core interactions \cite{Lutz2004}, which may coexist with short-range attractive interactions, such as those induced by polymers in solution \cite{Dzubiella2003}. In all of the aforementioned systems, the combination of hard-core repulsion and a mechanism that prevents neighboring particles from exceeding a maximal distance may mimic the effective tethering introduced in our model.

Let us briefly anticipate how our ``string-bead system'' behaves. Unless otherwise specified, we will consider for simplicity a fully packed initial condition in which all the walkers start at the origin (to implement this initial condition, we will additionally assume without real loss of generality that the walkers are point-like, implying that the radii of the particles depicted in Fig. \ref{tethered} become vanishingly small). As the walkers diffuse, they first separate from each other without crossing until the action of the $\Delta$-constraint starts to become noticeable. This trade-off between the single-file constraint and the $\Delta$-constraint results in relaxation towards an equilibrium state that is attained after a time of the order $ (N\Delta)^2/D$, where $D$ is the coefficient of diffusion of isolated and freely moving particles. In the following, we aim to study particle diffusion as a mechanism of relaxation to the equilibrium steady state. For a large $N$, we will characterize the latter from the point of view of statistical thermodynamics and also assess the extent to which this system resembles a linear chain in a viscous medium, that is, a polymer chain in a solvent \cite{Doi1986,Grosberg2002}.

In the above context, we stress the difficulty of dealing with the non-analytic interaction potential prescribed by the $\Delta$-constraint, as opposed to other types of interactions such as the harmonic potential that connects neighboring beads in the celebrated Rouse model (see e.g. Ref.~\onlinecite{Delfau2012}). There is a significant difference between the latter and the interaction potential associated with our tethered walkers.  From the point of view of the calculations, an exact solution in the latter case only seems possible for a two-particle system ($N=2$). Thus, even though we will focus on the approximate treatment for arbitrary $N$, the starting point of our analysis will be the $N=2$ case, which provides important generic insights about the relaxation behavior, and also allows one to assess the goodness of our approximation by comparison with the exact solution.

 In the absence of the $\Delta$-constraint, that is, for $\Delta=\infty$, one expects significant differences between the behavior of the edge particles and those in the bulk even in the long time regime. Aslangul's analysis for this case \cite{Aslangul1998} shows that at long times the central particle has an effective diffusivity equal to $D/N$, whereas the effective diffusivity of the edge particle decreases only as $1/\ln(N)$. In contrast, we find that this difference in behavior disappears as soon as $\Delta$ becomes finite, since all the particles then diffuse with the same effective diffusivity $D/N$ at sufficiently long times. We then complement our study of diffusional relaxation by calculating the probability distribution of the equilibrium system length and associated entropy. The force required to change the length by a given amount is linear in this quantity, the (entropic) spring constant being equal to $6k_BT/(N\Delta^2)$. In this respect, the system behaves like a polymer modeled by an ideal chain. Finally, we confirm our main results using Monte Carlo (MC) simulations.

The remainder of this paper is organized as follows.
In Sec.~\ref{ExactSol}, an exact solution for $N=2$ is provided and discussed in detail. Sec.~\ref{sec2} describes the approximate factorization ansatz used for the analytical computation of the positional pdf and other quantities of interest in the general case of $N$-walkers. Sec.~\ref{sec3} focuses on the approximate solution obtained in this manner for the specific case $N=2$. An in-depth comparison with the exact solution is also carried out. Next,  Sec.~\ref{sec5} deals with the $N=3$ case. Sec.~\ref{sec6} is devoted to the approximate solution for arbitrary $N$ and its relaxation to the equilibrium state. The latter is comprehensively studied in Sec.~\ref{sec8}. The main conclusions and perspectives are summarized in Sec.~\ref{sec9}. Appendix A gives the extension of the exact solution for $N=2$ when both particles are initially separated by a finite distance. Finally, appendices B and C are devoted to the approximate calculation of higher odd-order moments in the $N=2$ case as well as to the on-lattice solution for this case.

\section{$\bm{N=2}$ case: Exact solution}
\label{ExactSol}

The two-particle case is amenable to an exact solution and, therefore, worth a separate treatment. In Ref.~\onlinecite{Aslangul1999}, an exact solution (both on- and off-lattice) was given for the case $\Delta=\infty$, that is, when the separation between the walkers can become arbitrarily large (the solution was later extended to the case of walkers with opposite bias by Potts \emph{et al.} \cite{Potts2011}). In the following, we first recall the known exact result for $\Delta=\infty$, and then address the case with finite $\Delta$. We will mainly focus on the continuum limit (off-lattice) solution, but also provide some on-lattice results at a later stage.

Consider a pair of point-like walkers (hereafter termed particle 1 and 2) diffusing on the real line with common diffusivity $D$, their respective positions being $x_1$ and $x_2$. Unless otherwise specified, we assume that both walkers start at the same initial position, that is, $x_1^{(0)}=x_2^{(0)}=0$. The walkers perform SFD, i.e., they bounce off each other when they attempt to overtake. Thus, $x_2\ge x_1$ at all times. We now introduce an additional constraint by assuming that the walkers are ``tethered'', i.e., they cannot separate from each other by more than a distance $\Delta$, implying that $|x_1-x_2|\le \Delta$ at all times. As soon as one of the walkers attempts to violate this condition, the $\Delta$-constraints acts a tether that impedes the walker to move further away from the other one.
As we will see, for $\Delta<\infty$ and $t\to\infty$ the system tends to an equilibrium state characterized by a purely diffusive motion of the center-of-mass (c.o.m.) and constant values of the moments of the particles separation distance.

It is possible to obtain an exact solution for the joint positional pdf $p(x_1,x_2,t;\Delta)$ by mapping the problem to the problem of diffusion of a single particle in the plane with tilted reflected boundaries. To illustrate how the method works, we first consider the well-known $\Delta=\infty$ case.

\subsection{Case of infinite range ($\bm{\Delta=\infty}$)}
The evolution of both particles can be interpreted as a two-dimensional diffusion of a single fictitious particle with coordinates $(x_1,x_2)$, constrained to move in the half-plane above the reflecting line $x_2 = x_1$. Since both particles are assumed to start at the origin, the initial condition of the fictitious particle is, of course, $(x_1^{(0)},x_2^{(0)})\equiv (0,0)$.

To determine the pdf $p_\infty(x_1,x_2,t)\equiv p(x_1,x_2,t;\Delta=\infty$) for the two-dimensional coordinate $(x_1,x_2)$ of the fictitious particle at time $t$, we take as a starting point the analogous quantity $p_{\infty}^\text{hor}(x_1, x_2, t)$ when the reflecting line coincides with the horizontal axis given by the condition $x_2=0$. This latter pdf is well-known:
\begin{equation}
\label{horiboundsol}
p_{\infty}^\text{hor}(x_1,x_2,t)=\frac{e^{-(x_1^2+x_2^2)/4Dt}}{2\pi Dt}\,\Theta(x_2),
\end{equation}
where $\Theta(\cdot)$ stands for the Heaviside step function. Note that the above solution relies upon the observation that the boundary condition only affects the vertical degree of freedom. Because the $x_1$- and $x_2$-evolutions are uncoupled, the solution \eqref{horiboundsol} can be obtained as the product of the free one-dimensional Gaussian $e^{-x_1^2/4Dt}/\sqrt{4\pi Dt}$ for the horizontal coordinate $x_1$ times the one-dimensional solution $\Theta(x_2)\,e^{-x_2^2/4Dt}/\sqrt{\pi Dt}$ corresponding to diffusion on the halfline $x_2>0$ with a reflecting endpoint at $x_2=0$.

The solution \eqref{horiboundsol} satisfies the zero normal flux condition $\left.\partial_{x_2} p(x_1,x_2,t)\right|_{x_2=0}$=0. In contrast, the solution we seek must satisfy a similar boundary condition on the tilted reflecting line, i.e., $\left.{\bm n}_\perp \cdot \nabla p(x_1,x_2,t)\right|_{x_2=x_1}=0$, where ${\bm n}_\perp \equiv (1/\sqrt{2},-1/\sqrt{2})$ denotes the unit vector perpendicular to the reflecting line. To obtain the desired solution, we must now rotate the solution \eqref{horiboundsol} by 45\degree about an axis perpendicular to $(x_1,x_2)$ that goes through the origin. This rotation is performed via the unitary transformation $x_1\to (x_2+x_1)/\sqrt{2}$, $x_2\to (x_2-x_1)/\sqrt{2}$, which yields
\begin{equation}
\label{rotsol}
p_\infty(x_1,x_2,t)=\frac{e^{-(x_1^2+x_2^2)/4Dt}}{2\pi Dt}\,\Theta(x_2-x_1).
\end{equation}
This is precisely the solution obtained by Aslangul with his product ansatz for the $N=2$ case \cite{Aslangul1998}, from which we conclude that his procedure is \emph{exact} in this case.

The exact computation of the corresponding integer order moments
\begin{equation}
\langle x_2^m \rangle\equiv \int_{-\infty}^\infty dx_1 \int_{-\infty}^\infty dx_2 \, x_2^m p_\infty (x_1,x_2,t)= (-1)^m \langle x_1^m \rangle
\end{equation}
is now straightforward. We obtain
\begin{equation}
\langle x_2^m \rangle = \left\{\begin{array}{cc}
                       \pi^{-1/2} \Gamma(\tfrac{m+1}{2}) (4Dt)^{m/2},& \mbox {for $m$ even}, \\
                       2^{1/2}\pi^{-1} {}_2F_1 (\tfrac{1}{2},\tfrac{1-m}{2}, \tfrac{3}{2},\tfrac{1}{2}) \Gamma(\tfrac{m}{2}+1) (4Dt)^{m/2}, & \mbox {for $m$ odd,}
                     \end{array}\right.
\end{equation}
where ${}_2F_1(\cdot)$ denotes the confluent hypergeometric function. The expression for even $m$ can be further simplified using Legendre's duplication formula $\Gamma(z) \Gamma(z+1/2)=2^{1-2z} \sqrt{\pi} \, \Gamma(2z)$.  In terms of the rescaled spatial variable, we find
\begin{equation}
\langle u_2^m \rangle =\frac{\langle x_2^m \rangle}{(4Dt)^{m/2}}= \left\{\begin{array}{cc}
                       2^{-m} m!/(m/2)!,& \mbox {for $m$ even}, \\
                       2^{1/2}\pi^{-1} {}_2F_1 (\tfrac{1}{2},\tfrac{1-m}{2}, \tfrac{3}{2},\tfrac{1}{2}) \Gamma(\tfrac{m}{2}+1), & \mbox {for $m$ odd.}
                     \end{array}\right.
\end{equation}
These results confirm Aslangul's findings for the particular cases $m=1$ and $m=2$, namely, $\langle x_2 \rangle=(2 D t/\pi)^{1/2}$ and $\langle x_2^2 \rangle= 2Dt$.

 \subsection{Case of finite $\Delta$}

 Once again, the dynamics can be viewed as the two-dimensional diffusion of a single fictitious particle confined by two parallel reflecting lines at $x_2=x_1$  and $x_2=x_1+\Delta$ and starting at $(0,0)$.  This is the motion inside a tilted strip of thickness $\Delta/\sqrt{2}$. The probability $p(x_1,x_2,t;\Delta)$ of finding the single diffusing particle at position $(x_1,x_2)$ at time $t$ on this tilted strip is obviously related to the probability $p^\text{hor}(x_1,x_2,t;\Delta)$  for the horizontal strip limited by two parallel reflecting lines at $x_2=0$ and $x_2=\Delta/\sqrt{2}$. The non-tilted solution $p^\text{hor}(x_1,x_2,t;\Delta)$ can be e.g. straightforwardly obtained from the solution for diffusion confined by two fully reflecting parallel \emph{planes} oriented along the $x_1$-axis (see Eq.~(18) on p. 374 of Ref.~\onlinecite{CarslawJaeger1959}). Integrating out the transversal coordinate $x_3$ in this solution yields
\begin{equation}
\label{doubleboundsol}
p^\text{hor}(x_1,x_2,t;\Delta)=  \frac{\sqrt{2}}{\Delta}\left\{1+2\sum_{n=1}^\infty e^{-2n^2\pi^2 Dt/\Delta^2}
\cos\left(\frac{n\pi\sqrt{2}}{\Delta}x_2\right) \right\}R(x_2, {\scriptstyle \frac{\Delta}{\sqrt{2}}}) \times \frac{e^{-x_1^2/4Dt}}{\sqrt{4\pi Dt}}.
\end{equation}
Here, $R(x,y)$ denotes the rectangular function of width $y$, defined as follows:
\begin{equation}
 R(x,y)=
\begin{cases}
1 & \mbox{for } 0<x<y, \\
0 & \mbox{elsewhere}.
\end{cases}
\end{equation}
 The right-hand side (rhs) of Eq.~\eqref{doubleboundsol} is again the product of two solutions, namely, a free Gaussian corresponding to the horizontal degree of freedom and another solution corresponding to diffusion on a finite interval with reflecting endpoints at $x_2=0$ and $x_2=\Delta/\sqrt{2}$. Using the method of images, Eq.~\eqref{doubleboundsol} can be conveniently rewritten as follows [cf. Eq.~(19) on p. 374 in Ref.~\onlinecite{CarslawJaeger1959}]:
\begin{equation}
p^\text{hor}(x_1,x_2,t;\Delta)=\left\{\sum_{n=-\infty}^\infty e^{(-n^2\Delta^2+\sqrt{2} n x_2 \Delta)/2Dt}\right\}R(x_2,
{\scriptstyle \frac{\Delta}{\sqrt{2}}}) \times
\frac{e^{-(x_1^2+x_2^2)/4Dt}}{2\pi Dt}.
\end{equation}
To find the exact solution of our problem, we now perform the anticlockwise rotation of $p_\text{hor}$ by 45\degree, implying that the resulting solution must fulfil the boundary conditions
$\left.{\bm n}_\perp \cdot \nabla p(x_1,x_2,t)\right|_{x_2=x_1}=0$ and $\left.
{\bm n}_\perp \cdot \nabla p(x_1,x_2,t)\right|_{x_2=x_1+\Delta}=0$. This yields
\begin{equation}
\label{exactpdf}
p(x_1,x_2,t;\Delta)= \Upsilon(x_2-x_1,\Delta,t) \,\Theta(\Delta-(x_2-x_1)) \, p_\infty(x_1,x_2,t),
\end{equation}
where we have introduced the quantity
\begin{equation}
\Upsilon(x_2-x_1,\Delta,t)\equiv
\sum_{n=-\infty}^\infty \exp\left(\frac{-n^2\Delta^2+ n (x_2-x_1)\Delta}{2Dt}\right)
\end{equation}
and used the decomposition $R(x_2-x_1, \Delta)=\Theta(x_2-x_1) \Theta(\Delta-(x_2-x_1))$ of the rectangular function as a product of two Heaviside functions. The prefactor $\Upsilon(x_2-x_1,\Delta,t) \,\Theta(x_1+\Delta-x_2)$ describes the effect of the finite reach of motion owing to the $\Delta$-constraint. This
prefactor can also be expressed as follows:
\begin{equation}
\Upsilon(x_2-x_1,\Delta,t)=\frac{(2\pi Dt)^{1/2}e^{(x_2-x_1)^2/8Dt}}{\Delta}
\,\vartheta_3\left(-\frac{\pi (x_2-x_1)}{2\Delta}, e^{-2\pi^2 Dt/\Delta^2}\right) ,
\end{equation}
where $\vartheta_3(\cdot, \cdot)$ is an elliptic function.

Results for the reduced pdfs
\begin{align}
\label{exactreduced1}
p_1^{(1)}(x_1,t,\Delta)=& \int_{-\infty}^\infty dx_2 \, p(x_1,x_2,t;\Delta),\\
\label{exactreduced2}
p_2^{(1)}(x_1,t,\Delta)=& \int_{-\infty}^\infty dx_1 \, p(x_1,x_2,t;\Delta),
\end{align}
can now be obtained by inserting \eqref{exactpdf} into the above expressions and performing the corresponding integration. One finally obtains
\begin{align}
p_1^{(1)}(x_1,t,\Delta)=&\left\{\sum_{n=-\infty}^\infty e^{-n^2\Delta^2/(4Dt)} e^{-n x_1 \Delta/(2Dt)}\right.\times \nonumber \\
&
\label{exactp1x}
\times\left.\left[
\text{Erf}\left(\frac{x_1+\Delta-n\Delta}{\sqrt{4Dt}}\right)
-\text{Erf}\left(\frac{x_1-n\Delta}{\sqrt{4Dt}}\right)
\right]\right\}
\frac{e^{-x_1^2/4Dt}}{\sqrt{4\pi Dt}}
\end{align}
for the left particle, where $\text{Erf}(z)=(2/\sqrt{\pi})\int_0^z du\, e^{-u^2}$ is the error function. The corresponding expression for the right particle is  $p_1^{(2)}(x_2,t,\Delta)=-p_1^{(1)}(x_2,t,-\Delta)$.

Finally, we note that it is straightforward to obtain an exact expression for $p(x_1,x_2,t;\Delta)$ in the case of arbitrary non-zero initial particle separation (see Appendix A).

\subsection{Exact moments}
\label{subs-exact-moments}

We now discuss the behavior of the first-order
and second-order positional moments, as well as of the particle correlations.

\subsubsection{Interparticle distance and first-order positional moment}

An important parameter for measuring the relaxation of our system towards the equilibrium steady state is the distance between the two particles $L=x_2-x_1$. The $n$-th moment of this quantity is defined by the following integral:
\begin{equation}
\langle L^n \rangle=\int_{-\infty}^\infty dx_1 \int_{-\infty}^\infty  dx_2 \, (x_2-x_1)^n \, p(x_1,x_2,t)
\label{nth-moment-L}
\end{equation}
where $p(x_1,x_2,t;\Delta)$ is given by Eq.~\eqref{exactpdf}.  An easy way to simplify this integral is to switch to c.o.m. and relative coordinates. To this end, we define $X\equiv (x_1+x_2)/2$, yielding $x_1=X-(L/2)$ and $x_2=X+(L/2)$. The corresponding Jacobian is $J(X,L)\equiv \partial (x_1,x_2)/\partial (X,L)=1$. Using this transformation in \eqref{nth-moment-L}, we see that the integral decouples into two independent integrals,
\begin{equation}
\label{comreldec}
\langle L^n \rangle=\int_{-\infty}^\infty
dX \, P(X,t)\int_0^\Delta dL \, L^n P(L,t;\Delta).
\end{equation}
Here,
\begin{equation}
P(X,t)\equiv \frac{e^{-X^2/(2Dt)}}{\sqrt{2\pi D t}}, \qquad -\infty < X<\infty
\end{equation}
and
\begin{equation}
P(L,t)\equiv\frac{1}{\Delta}\vartheta_3({\scriptstyle -\frac{\pi L}{2\Delta}}, \zeta), \qquad 0 \le L \le \Delta
\end{equation}
are, respectively, the pdfs for the c.o.m. coordinate and the interparticle distance (for compactness, in the latter we have introduced the definition $\zeta=\zeta(t,\Delta) \equiv \exp{[-2\pi^2 Dt/\Delta^2]}$). In \eqref{comreldec}, the integral over $X$ obviously reduces to unity because of the normalization property of the pdf, and one is then left with
\begin{equation}
\langle L^n \rangle=\frac{1}{\Delta} \int_0^\Delta dL\, L^n
\vartheta_3({\scriptstyle -\frac{\pi L}{2\Delta}}, \zeta).
\end{equation}
We may now use the series representation
\begin{equation}
\vartheta_3(u, q)=1+2\sum_{m=1}^\infty q^{m^2} \cos{\left(2m u\right)}
\end{equation}
of the elliptic function to find
\begin{equation}
\langle L^n \rangle= \frac{\Delta^n}{n+1}+\frac{2}{\Delta}
\sum_{m=1}^\infty \zeta^{m^2}
\int_0^\Delta dL\, L^n  \cos{\left(\frac{\pi mL}{\Delta}\right)}.
\end{equation}
The integrals in the sum of the rhs can be expressed in terms of hypergeometric functions. However, more transparent expressions are obtained for the first low-order moments. In particular, for the first moment $n=1$, we obtain
\begin{equation}
\langle L \rangle=2\langle x_2 \rangle=\frac{\Delta}{2}-\frac{4\Delta}{\pi^2}
\sum_{m=1}^\infty \frac{\zeta^{(2m-1)^2}}{ (2m-1)^2},
\end{equation}
or, in terms of the approach to the mean particle separation $\langle L\rangle_{eq}=\Delta/2$ in the equilibrium state (attained in the $t\to\infty$ limit),
\begin{equation}
\label{rhoseries}
\rho\equiv \frac{\langle L \rangle}{\langle L\rangle_{eq}}-1=-\frac{8}{\pi^2}
\sum_{m=1}^\infty \frac{\zeta^{(2m-1)^2}}{ (2m-1)^2}.
\end{equation}
Writing out the first two terms explicitly, one gets
\begin{equation}
\rho(\zeta)=-\frac{8}{\pi^2}\,\zeta-\frac{8}{9\pi^2}\,\zeta^9+{\cal O}(\zeta^{25}).
\label{rhoseriestrunc}
\end{equation}
For the second-order moment $\langle L^2 \rangle$ of the separation distance, we now make use of the fact that
\begin{equation}
\int_0^\Delta dx\, x^2  \cos{\left(\frac{\pi mx}{\Delta}\right)}=
\frac{2\Delta^3 (-1)^m}{\pi^2 m^2}
\end{equation}
to find
\begin{equation}
\label{Second-L-moment}
\frac{\langle L^2\rangle}{\Delta^2} =\frac{1}{3}+\frac{4}{\pi^2}\sum_{m=1}^\infty (-1)^m \frac{\zeta^{m^2}}{m^2}.
\end{equation}
Note that, for $t=0$, we recover the correct initial value $\langle L^2\rangle=0$.

Finally, we express $\langle x_2 \rangle=\langle L \rangle/2$ in terms of time as
\begin{equation}
\label{exactFM}
\frac{\langle x_2 \rangle}{\Delta}=-\frac{\langle x_1 \rangle}{\Delta}=
\frac{1}{4}-\frac{2}{\pi^2}
\sum_{m=1}^\infty \frac{e^{-2\pi^2(2m-1)^2 Dt/\Delta^2}}{ (2m-1)^2},
\end{equation}
that is, one has exponential relaxation to the limiting value in the long-time regime.

To extract the short-time behavior from the rhs of Eq.~\eqref{exactFM}, we take its Laplace transform, which yields
\begin{equation}
\frac{\langle \widetilde{x}_2(s) \rangle}{\Delta} \equiv \frac{1}{\Delta}\int_0^\infty dt\,e^{-st}\langle x_2(t)\rangle
=\frac{1}{4 s}-\frac{\Delta^2}{\pi^4 D}
\sum_{m=1}^\infty \frac{1}{ (2m-1)^2 ( (2m-1)^2+(\Delta^2 s)/(2\pi^2D))}.
\end{equation}
We can now take advantage of the result (Ref.~\onlinecite{Prudnikov1998}, p. 689)
 \begin{equation}
\sum_{m=1}^\infty \frac{1}{ (2m-1)^2 ( (2m-1)^2+\mu^2)}=\frac{\pi^2}{8\mu^2}-\frac{\pi}{4\mu^3}\tanh{\frac{\pi \mu}{2}}
\end{equation}
to obtain
\begin{equation}
\langle \widetilde{x}_2(s) \rangle=\sqrt{\frac{D}{2s^3}}\tanh{\sqrt{\frac{\Delta^2 s}{8D}}}=\sqrt{\frac{D}{2s^3}}\left\{1-2 e^{-\sqrt{\frac{\Delta^2 s}{2D}}}+2 e^{-\sqrt{\frac{2\Delta^2 s}{D}}}+{\cal O}\left(e^{-\sqrt{\frac{9\Delta^2 s}{2D}}}\right)\right\}
.
\end{equation}
To retrieve the short-time behavior, we use the inverse Laplace transform
 \begin{equation}
{\cal L}^{-1}_s \left\{ \frac{e^{-C s^{1/2}}}{s^{3/2}} \right\}=\frac{2\sqrt{t}}{\pi} e^{-C^2/4t}
+C \,\text{Erf}\left(\frac{C}{2\sqrt{t}}\right)-C, \qquad C>0.
\end{equation}
Finally, invoking the well-known asymptotic expansion of the error function  for large arguments \cite{Abramowitz1972},
\begin{equation}
\label{exp-erf}
\text{Erf}(z)=1-\frac{e^{-z^2}}{{\sqrt{\pi} z}}\left[ \sum_{k=0}^{N-1} (-1)^k \frac{(2k-1)!!}{(2z^2)^k}+{\cal O}(|z|^{2-2N})\right], \quad z\to\infty,
\end{equation}
one eventually finds
\begin{equation}
\label{correctfom}
\langle x_2 \rangle =\frac{\langle L \rangle}{2}\approx\sqrt{\frac{2}{\pi}Dt}\,\left(1-\frac{8Dt}{\Delta^2}\,e^{-\Delta^2/8Dt}\right).
\end{equation}

\subsubsection{Second-order positional moment}
\label{sopm}

The second-order moment follows immediately from the above decomposition in c.o.m. and relative coordinates. One has
\begin{equation}
\langle x_2^2 \rangle=\langle x_1^2 \rangle=\int_{-\infty}^\infty dx_1 \int_{-\infty}^\infty  dx_2 \, x_2^2 \, p(x_1,x_2,t;\Delta).
\end{equation}
Taking into account that $p(x_1,x_2,t;\Delta) dx_1 dx_2=p(X,L,t;\Delta) dX dL=p(X,t)p(L,t;\Delta) dX dL$, one has
\begin{equation}
\langle x_2^2 \rangle=\int_{-\infty}^\infty \int_0^\Delta dX dL \, \left(X+\frac{L}{2} \right)^2 \, p(X,t) p(L,t;\Delta).
\end{equation}
Thus,
\begin{equation}
\langle x_2^2 \rangle=\langle X^2 \rangle+\langle X L \rangle+
\frac{\langle L^2 \rangle}{4}=\langle X^2 \rangle+\frac{\langle L^2 \rangle}{4},
\end{equation}
where we have used $\langle X L \rangle=\langle X\rangle \langle L \rangle$ and the fact that $\langle X \rangle =0$ to derive the last equality. The c.o.m. performs a purely diffusive motion with diffusivity $D/2$; therefore, $\langle X^2 \rangle=Dt$, implying that
\begin{equation}
\label{x22t}
\langle x_2^2 \rangle=Dt+\frac{\langle L^2 \rangle}{4}
=Dt+\frac{\Delta^2}{12}+\frac{\Delta^2}{\pi^2}
\sum_{m=1}^\infty \frac{(-1)^m}{m^2} \, e^{-2\pi^2 m^2 Dt/\Delta^2},
\end{equation}
where we have used the expression for $\langle L^2 \rangle$ given by Eq.~\eqref{Second-L-moment}.
As one can see, the  variance $\text{Var}(x_2)\equiv \langle x_2^2\rangle-\langle x_2\rangle^2=\text{Var}(x_1)$ (also termed ``centered MSD'' hereafter) is directly related to the variance $\text{Var}(L)\equiv \langle L^2\rangle-\langle L\rangle^2$ of the interparticle distance:
\begin{equation}
\text{Var}(x_2)=Dt+\frac{\text{Var}(L)}{4}.
\end{equation}
To extract the short-time behavior, we need to assess how $\langle L^2 \rangle$ behaves in this limit. According to Eq.~\eqref{Second-L-moment} the Laplace transform of $\langle L^2\rangle$ is
\begin{equation}
\langle \widetilde{L^2} (s)\rangle\equiv\int_0^\infty dt\,e^{-st}\langle L^2(t)\rangle=\frac{\Delta^2}{3 s}+\frac{4\Delta^2}{\pi^2}\sum_{m=1}^\infty \frac{ (-1)^m}{m^2 \left(\frac{2 \pi^2 D m^2}{\Delta^2}+s\right)}.
\end{equation}
We now use the result (see Ref.~\onlinecite{Prudnikov1998}, p. 686)
\begin{equation}
\sum_{m=1}^\infty  \frac{(-1)^m}{m^2 \left(m^2+\mu^2\right)}=
\frac{1}{2\mu^4}-\frac{\pi^2}{12\mu^2}-\frac{\pi \csch(\pi\mu)}{2\mu^3}.
\end{equation}
From the resulting large-$s$ behavior for $\langle \widetilde{L^2} (s)\rangle$, one then has
\begin{equation}
\langle \widetilde{L^2} (s)\rangle=\frac{4D}{s^2}-\sqrt{\frac{32\Delta^2 D}{s^3}}
\left\{e^{-\sqrt{\frac{\Delta^2u}{2D}}}+{\cal O}\left( e^{-\sqrt{\frac{9\Delta^2s}{2D}}}\right)\right\}.
\end{equation}
Inverting term by term, one sees that the above large-$s$ behavior in Laplace space corresponds to the following short-time behavior:
\begin{equation}
\langle {L^2} \rangle\approx 4Dt\left(1-\sqrt{\frac{128Dt}{\pi \Delta^2}}e^{-\Delta^2/(8Dt)}\right).
\end{equation}
Thus, at early times,
\begin{equation}
\label{correctvar}
\text{Var}(x_2) \approx 2\left(1-\frac{1}{\pi}\right)Dt-\sqrt{\frac{128(Dt)^3}{\pi \Delta^2}}e^{-\Delta^2/(8Dt)}.
\end{equation}
The first term corresponds to the result obtained by Aslangul \cite{Aslangul1998} for $\Delta=\infty$. In the opposite limit of long times, it is easy to see that $ \text{Var}(x_2)\to Dt=2(D/2)t$, i.e., the diffusivity of each particle is decreased by a factor of two with respect to its value $D$ for free particles.

\subsubsection{Correlator}
\label{ssscorr}

We are now in a position to obtain an exact expression for the two-particle correlation function $C(t)\equiv \langle x_1 x_2 \rangle-\langle x_1 \rangle \langle x_2 \rangle$. By substituting $x_1=X-(L/2)$ and $x_2=X+(L/2)$ in the corresponding expression, we obtain
\begin{equation}
C(t)=\langle X^2 \rangle -\frac{\text{Var}(L)}{4}=Dt-\frac{\text{Var}(L)}{4}=2Dt-\text{Var}(x_2).
\end{equation}
Thus, we see that the correlator, MSD, and variance of the relative coordinate $L$ are intimately connected to one another. In the short-time limit, we have
\begin{equation}
C(t)\approx \frac{2}{\pi}Dt+\sqrt{\frac{128(Dt)^3}{\pi \Delta^2}}e^{-\Delta^2/(8Dt)}.
\end{equation}
In contrast, one has $C(t)\approx Dt$ at long times.

\section{$\bm{N}$-particle pdf}
\label{sec2}

We now generalize the $N=2$ setting by considering a collection of $N\ge 2$ point walkers with a common diffusivity $D$ on the real line, their respective positions being $x_1,x_2,\ldots,x_N$. Denoting the initial coordinate of the $j$-th walker by $x_j^{(0)}$, we henceforth assume $x_1^{(0)}\le x_2^{(0)}, \ldots, x_{N-1}^{(0)}\le x_N^{(0)}$. These walkers perform SFD, which implies that they bounce off each other when they attempt to overtake. Thus, the initial ordering is preserved, that is, $x_1\le x_2, \ldots, x_{N-1}\le x_N$ at all times, and the system is similar in this respect to that studied in Ref.~\onlinecite{Aslangul1998}. In this reference, the study is restricted to the fully packed initial condition $x_1^{(0)}=x_2^{(0)}=\ldots = x_N^{(0)}=0$. Here, we will also mainly focus on this case, but our formulation holds for a general initial condition. As in the $N=2$ case, we now include as a novelty an additional constraint, which consists in preventing the walkers from traveling too far from each other. Thus, the $j$-th walker is not allowed to move further than a distance $\Delta$ from either neighbor, that is, one must always have $\left|x_j-x_{j\pm1}\right|\le \Delta$. As soon as any walker attempts to move further away from its neighbor, the $\Delta$-constraint acts as a tether that prevents her from doing so. As in the $N=2$ case, for $N>2$ the SFD of such tethered walkers is drastically different from walkers subject only to the non-crossing restriction. Again, as $t\to\infty$ the system tends to an equilibrium state with a purely diffusive c.o.m. and fixed moments of the interparticle distances. In contrast, when $\Delta=\infty$, these distances remain time-dependent for arbitrarily long times \cite{Aslangul1998}.

Inspired by the analytical approach described in Ref.~\onlinecite{Aslangul1998}, in what follows we assume that the $N$-particle positional pdf of the tethered walker system can be approximated by the product
\begin{equation}
\label{factansatz}
p_x(x_1,x_2,\ldots,x_N,t;\Delta)=A^{(N)} \prod_{i=1}^N G(x_i,t|x_i^{(0)}) \prod_{i=1}^{N-1} R(x_{i+1}-x_i,\Delta),
\end{equation}
where $A^{(N)}$ is a normalization constant,  $G(x,t|x^{(0)})$ is the free-particle Green function (Gaussian distribution)
\begin{equation}
\label{Gauss}
G(x,t|x^{(0)})=\frac{e^{-(x-x^{(0)})^2/4Dt}}{\sqrt{4\pi Dt}}.
\end{equation}
Equation \eqref{Gauss} accounts for the normal diffusion of particles with diffusion coefficient $D$  when they move freely.  The rectangular function ensures that the particles do not cross each other and that the nearest neighbors are separated by no more than a distance $\Delta$.

Before proceeding further, we emphasize once again that the ansatz \eqref{factansatz} is only an approximation. We will explore its limitations at the appropriate stage using the exact solution available for $N=2$ as well as numerical simulations for arbitrary $N$. On the other hand, it is intuitively clear that the cutoffs introduced by the rectangular functions are too simple to fully capture the effect of the perturbation induced by the constraints imposed on the walkers' motion. In particular, such constraints result in the loss of gaussianity characteristic of free diffusion. Such a loss becomes most evident at times long enough to have a large number of collisions between walkers; at such times, the $\Delta$ constraint play a significant role by abruptly and repeatedly interrupting the walkers' rattling motion. The lack of gaussianity is by no means surprising; the solution of a single Brownian particle diffusing inside a finite interval with fixed reflecting endpoints is already quite different from a single Gaussian, which is an acceptable approximation only at times short enough to ensure that the interaction with the endpoints remains negligibly small. On the other hand, at the single particle level, our problem can be viewed as a much more complicated version of the aforementioned interval problem in which both endpoints (given by the positions of the diffusing nearest-neighbors) move stochastically.

We can understand in an alternative way why the solution given by Eq.~\eqref{factansatz} is only an approximation. To do this, we will consider this approximation for $N=2$ in the limit in which the diffusion coefficient of one of the particles (say particle 1) goes to zero, that is, we will consider that particle 1 does not move (even though the diffusivities of both particles were assumed to be equal, we relax this condition here solely for the purpose of the subsequent argument). Let us assume particle 1 is fixed at $x_1=0$. Then, our approximation for the pdf of the position of particle 2, $p_2^{(1)}(x_2,t)$, which starts at $x_2=0$ at $t=0$, is (up to a normalization constant) $G(x_2,t|0) R(x_2,\Delta)$. This solution satisfies the diffusion equation and is zero for $x_2 \leq 0$ and for $x_2 \geq \Delta$, as it should be. It also satisfies the boundary condition that the probability flux of particle 2 is zero at $x_2=0$, a condition that is due to the fact that particle 1 is impenetrable (it does not absorb particle 2, but reflects it). The approximate solution
$p_2^{(1)}(x_2,t)$
satisfies this condition since $dG/dx=0$ for $x_2=0$ (given that $G(x_2,t|0)$ is maximum at $x_2=0$). On the other hand, the probability flux must also be zero at $x_2=\Delta$ for the same reasons that it must be at $x_2=0$. However, our proposed solution $p_2^{(1)}(x_2,t)$ does not satisfy this condition, since $dG/dx \neq 0$ at $x_2=\Delta$. In summary, the solution we propose is not exact because, although it satisfies the basic condition of being zero outside the interval prohibited by the restriction of tethered impenetrable particles, it does not satisfy the condition of zero probability flux at the extreme values of this prohibited interval (we could achieve this condition by using the procedure of the method of images: the zero flux condition at $\Delta$ is achieved by adding the term $G(x_2,t|2\Delta)$ to the solution; a solution that in turn would have to be corrected so that it satisfies the zero flux condition at $x_2=0$, and so on). A similar argument applies to the case $N>2$, in which a larger number of zero-flux boundary conditions need to be satisfied \cite{Rodenbeck1998}.

The calculations become significantly less cumbersome if one switches to dimensionless variables $u=x/\sqrt{4Dt}$ and $\delta\equiv\delta(t)=\Delta/(4Dt)^{1/2}$, which denote the scaled coordinate and the reach of the interaction, respectively. However, the coordinate scaling must ensure probability conservation, that is,
\begin{equation}
p_x(x_1,\ldots,x_N,t;\Delta)\, dx_1\ldots dx_N= p(u_1,\ldots,u_N;\delta)\, du_1\ldots du_N,
\end{equation}
which implies
\begin{equation}
 p(u_1,u_2,\ldots,u_N;\delta)=(4Dt)^{N/2} p_x(x_1,\ldots,x_N;\Delta=\delta\,\sqrt{4Dt}),
\end{equation}
with $x_i=u_i\,\sqrt{4Dt}$, or, equivalently,
 \begin{equation}
\label{pulobal}
 p(u_1,u_2,\ldots,u_N; \delta)=\frac{A^{(N)}}{\pi^{N/2}} \prod_{i=1}^N e^{-u_i^2} \prod_{i=1}^{N-1} R(u_{i+1}-u_i,\delta).
\end{equation}

At the single-particle level, one can define the reduced one-particle pdf $p_n^{(1)}(x_n,t,\Delta)$
via a multiple integral:
\begin{align}
p_n^{(1)}(x_n,t,\Delta)&=\int_{-\infty}^\infty \ldots\int_{-\infty}^\infty dx_1\ldots dx_{n-1} dx_{n+1}\ldots dx_N \, p(x_1,x_2,\ldots,x_N,t;\Delta) \nonumber \\
&=(4Dt)^{1/2}\int_{-\infty}^\infty  \ldots\int_{-\infty}^\infty du_1\ldots du_{n-1} du_{n+1}\ldots  du_N\, p(u_1,u_2,\ldots,u_N;\delta)
\nonumber \\
&= (4Dt)^{1/2} p_n^{(1)}(u_n,\delta),
\label{pnpn}
\end{align}
where, for brevity, we have dropped the subindex ``$x$'' from $p_x(x_1,x_2,\ldots,x_N,t;\Delta)$ [the nature of the pdf can be inferred from the symbols appearing in its argument].
In what follows, unless stated otherwise [see Eqs.~\eqref{p1x} and \eqref{p2x}], we will further simplify the notation
by also omitting the explicit reference to $\Delta$ in this quantity as well as in the reduced pdfs $p_n^{(1)}$. The one-particle moments defined in terms of the original and scaled coordinates are related to one another as follows:
 \begin{equation}
\langle x_n^m\rangle(t)=\int_{-\infty}^\infty dx_n x_n^m \, p_n^{(1)}(x_1,t)
=(4Dt)^{m/2}\int_{-\infty}^\infty du_n \, u_n^m p_n^{(1)}(u_1,\delta)
=(4Dt)^{m/2}\langle u_n^m\rangle(\delta)
\label{mxmu}
\end{equation}
In particular, the centered MSD $\text{Var}(x_n) \equiv \langle x_n^2\rangle-\langle x_n\rangle^2$ of the $n$-th walker position can now be expressed in terms of rescaled moments:
\begin{equation}
\text{Var}(x_n)= \text{Var}(u_n) 4D t= 4D_\text{eff}(t) \; t,
\end{equation}
where
\begin{equation}
\label{Deff}
D_\text{eff}(t)=D\; \text{Var}(u_n)
\end{equation}
The quantity $\text{Var}(u_n)\equiv \langle u_n^2\rangle-\langle u_n\rangle^2$ (and therefore also $D_\text{eff}$) depends on time, indicating the existence of anomalous diffusion. In fact, the term ``transient anomalous diffusion'' is pertinent here, because at very long times the diffusion becomes normal (albeit with an $N$-dependent effective diffusivity). We shall discuss this behavior in more detail at the appropriate stage.

\section{$\bm{N=2}$ case: Approximate results}
\label{sec3}

In this section, we obtain approximate results for the continuum case based on factorization ansatz \eqref{factansatz}, which will then be compared with the exact solution obtained in Sec. \ref{ExactSol}.

\subsection{Positional pdf}
\label{sec:pdfN2approx}

We start by noting that the exact two-particle pdf $p(x_1,x_2,t)$ given by Eq.~\eqref{exactpdf} can be rewritten in terms of the approximate pdf $p_\text{appr}(x_1,x_2,t)$  [which, in full notation, corresponds to $p_x(x_1,x_2,t;\Delta)$ as given by Eq.~\eqref{factansatz}] as follows:
\begin{equation}
p(x_1,x_2,t)=\Upsilon(x_2-x_1,\Delta,t) p_\text{appr}(x_1,x_2,t).
\end{equation}
Thus, the factor $\Upsilon(x_2-x_1,\Delta,t)$ is a measure of the quality of our proposal \eqref{factansatz}.

\begin{figure}[t!]
\begin{center}
 \includegraphics[width=0.46\textwidth,angle=0]{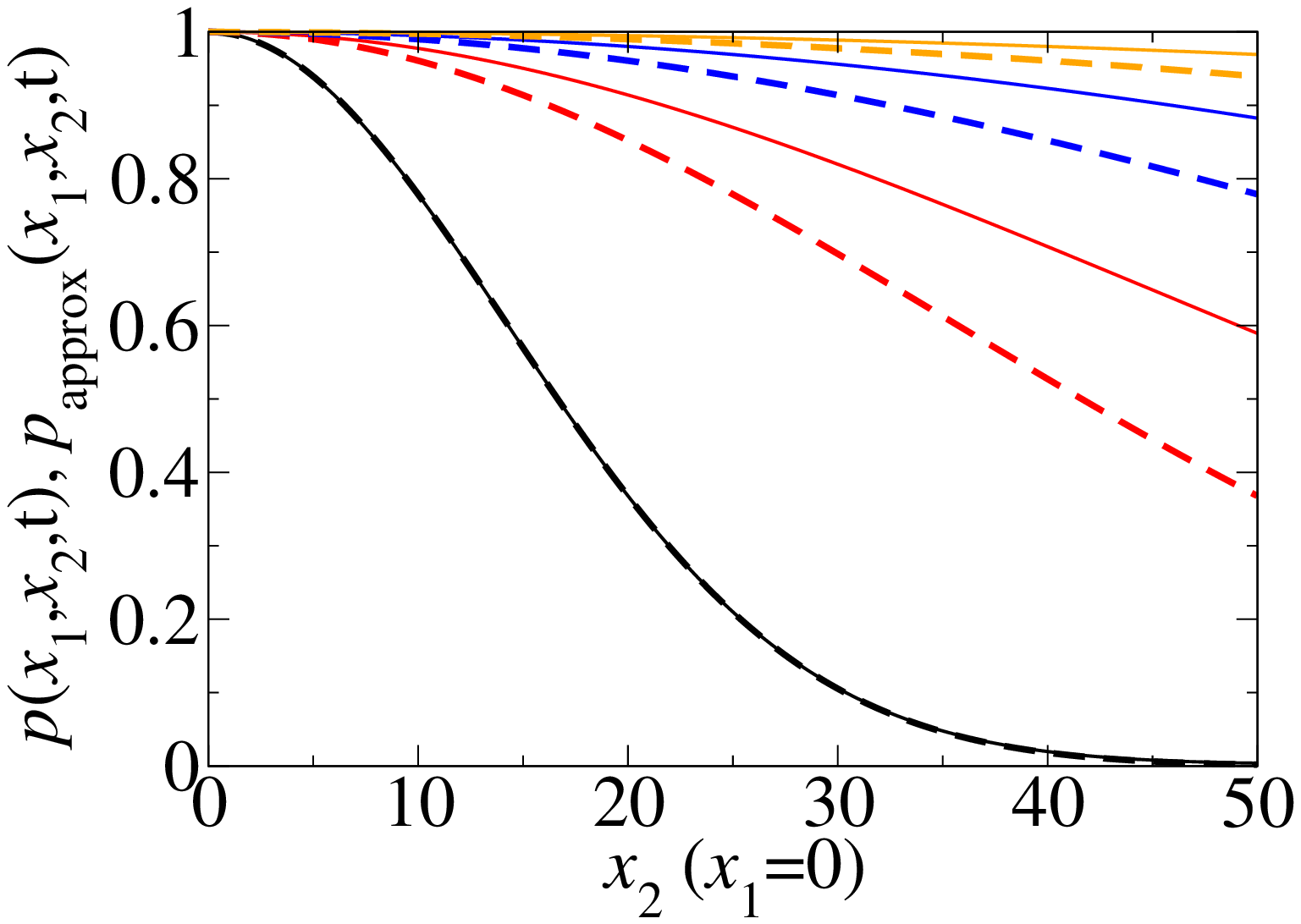}
\end{center}
\caption{\label{upsilonfigure} This figure illustrates the role of $\Upsilon$ by comparing $p(x_1,x_2,t)$ and $p_\text{appr}(x_1,x_2,t)$.  Solid curves correspond to $p(x_1,x_2,t)$, whereas dashed lines correspond to $p_\text{appr}(x_1,x_2,t)$. We have taken, from top to bottom, $t=10000,2500,625$, and $100$. For short times ($t=100$), $p(x_1,x_2,t)$ and $p_\text{appr}(x_1,x_2,t)$ are seen to be almost identical. We have chosen $D=1$ and $\Delta=50$.}
\end{figure}

Figure \ref{upsilonfigure} shows a comparison between the exact solution
$p(x_1,x_2,t)$ and $p_\text{appr}(x_1,x_2,t)$ as a function of $x_2-x_1$ for a fixed value of $\Delta$ and different times. Our approximation shows an extremely good agreement with the exact solution at short times; at intermediate times it clearly underestimates the real pdf, but the agreement improves again at long times and becomes acceptable provided that the value of $x_2-x_1$ is not too large.

According to Eq.~\eqref{pulobal} the scaled two-particle pdf is
 \begin{equation}
 p(u_1,u_2)\equiv p(u_1,u_2; \delta)=\frac{A^{(2)}}
 {\pi}\, e^{-u_1^2} e^{-u_2^2} R(u_{2}-u_1,\delta).
\end{equation}
The reduced one-particle pdfs are defined by Eq.~\eqref{pnpn}, which in the present case yields
\begin{align}
p_1^{(1)}(u_1,\delta)&= \frac{A^{(2)}}
 {\pi} \, e^{-u_1^2}\, \int_{u_1}^{u_1+\delta} du_2 \,e^{-u_2^2},\\
p_2^{(1)}(u_2,\delta)&= \frac{A^{(2)}}
 {\pi} \, e^{-u_2^2}\, \int_{u_2-\delta}^{u_2} du_1 \,e^{-u_1^2}
\end{align}
(recall that in our notation subscripts 1 and 2 refer to the left and right particles, respectively). The integrals above (as well as the normalization constant
$A^{(2)}$) can be expressed in terms of error functions. One finds
\begin{equation}
\label{p1N2u}
p_1^{(1)}(u_1,\delta)= \frac{\text{Erf}\left(u_1+\delta \right)
-\text{Erf}\left(u_1\right)}{\sqrt{\pi}\,\text{Erf}(\delta/\sqrt{2})}
\, e^{-u_1^2}
\end{equation}
and $p_2^{(1)}(u_2,\delta)=p_1^{(1)}(u_2,-\delta)$.
We have used the result \cite{Abramowitz1972}
 \begin{equation}
\int_{-\infty}^\infty du\, e^{-u^2} \text{Erf}(u+C) = \sqrt{\pi} \; \text{Erf}(C/\sqrt{2}).
\label{IntforNorm}
\end{equation}
In terms of the original coordinate, the above expressions give (in full notation)
\begin{equation}
\label{p1x}
p_1^{(1)}(x_1,t,\Delta)= \frac{1}{\text{Erf}(\Delta/\sqrt{8Dt})}
\left[
\text{Erf}\left( \frac{x_1+\Delta}{\sqrt{4Dt}}\right)
-\text{Erf}\left( \frac{x_1}{\sqrt{4Dt}}\right)
\right]
\frac{e^{-x_1^2/4Dt}}{\sqrt{4\pi Dt}}
\end{equation}
for the left particle and
\begin{equation}
\label{p2x}
p_2^{(1)}(x_2,t,\Delta)=p_1^{(1)}(x_2,t,-\Delta)
\end{equation}
for the right particle.

It is interesting to note that, leaving aside the normalization constant, our approximation \eqref{p1x} is obtained from the exact expression \eqref{exactp1x} by retaining only the first term ($n=0$) of the series. The terms with $n\neq 0$ are corrective decay modes whose contribution becomes most significant in the intermediate-time regime.

\begin{figure}[ht]
\begin{center}
\includegraphics[width=0.46\textwidth,angle=0]{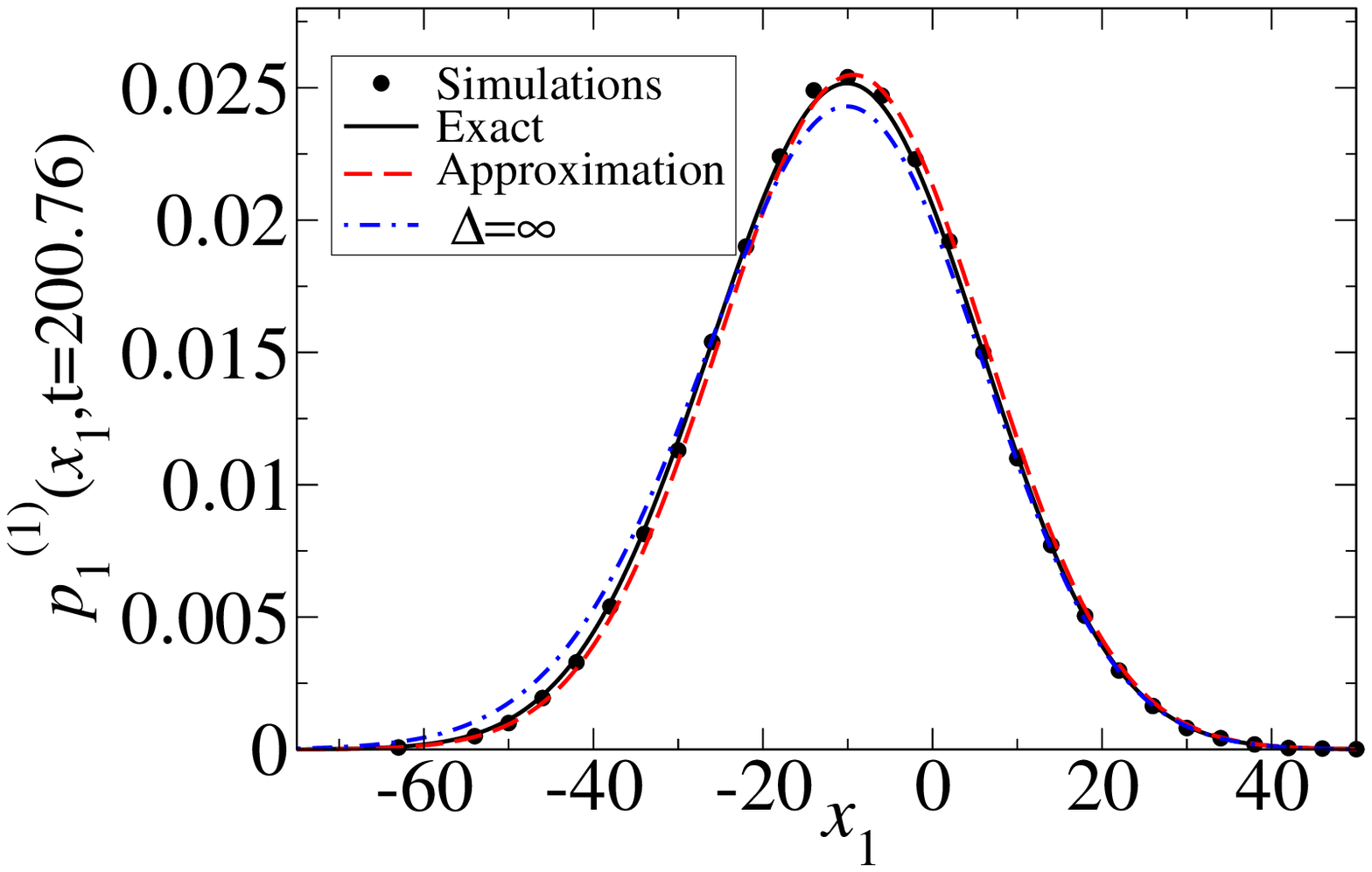}
\end{center}
\caption{\label{figp1xt}
Simulated pdf $p_1^{(1)}(x_1,t)$ for $N=2$ (circles). We have taken $D=1$, $t=200.76$, and $\Delta=50$, which gives $\delta=1.7644$. The solid line represents the exact solution \eqref{exactp1x}. The dashed lined is the approximate theoretical expression \eqref{p1x}.   Finally, the dashed-dotted line corresponds to the case in which the impenetrable particles can move without any further limitation (i.e., $\Delta=\infty$). }
\end{figure}

In Fig.~\ref{figp1xt} we compare the simulation results for the reduced pdf $p_1^{(1)}(x_1,t)$ with the approximate theoretical expression \eqref{p1x} and Aslangul's result for
$\Delta=\infty$. The exact analytical solution given by Eq.~\eqref{exactp1x} is also included in the figure (see Sec.~\ref{ExactSol}). The agreement between the approximation and the simulations/exact result is good as long as the diffusion approximation holds, which requires that $\Delta$ remain large in comparison with the typical jump length used in the simulations.

In our simulations we consider both continuum and on-lattice models
(this is relevant to explore their differences \cite{Aslangul1999}; see Appendix C).
In both cases, we employ two protocols of the MC algorithm: the classical one with discrete time steps \cite{Binder2015}, or the continuous-time variant known as the kinetic MC algorithm \cite{Bortz1975,Gillespie1976,Gillespie2001,Fichthorn1991}.
In both the discrete-time and continuous-time algorithms, particles perform many jumps during a single time step (one MC cycle). We typically used $\sim 10^7-10^8$ jumps per cycle and averaged up to $10^5$ MC cycles.
The diffusion constant is given by $D=\langle \lambda^2 \rangle W$, where
$\langle \lambda^2 \rangle$ is the second moment of the single-jump length
$\lambda$, and $W$ is the one-sided hopping rate. In our simulations, the time unit is $\tau=1/W$, and the length unit is chosen in such a way that one obtains the desired value of $D$.

\subsection{First-order moment}
\label{fom}

Using the result
\begin{equation}
\int_{-\infty}^\infty du \, u\, e^{-u^2} \text{Erf}(u+C)=\frac{1}{\sqrt{2}} e^{-C^2/2},
\label{I7}
\end{equation}
 one finds upon use of Eq.~\eqref{p1N2u} the result
\begin{equation}
\label{uN2a}
\langle u_2 \rangle=-\langle u_1 \rangle=\frac{1-e^{-\delta^2/2}}{\text{Erf}(\delta/\sqrt{2})}\,\sqrt{\frac{1}{2\pi}}.
\end{equation}
In terms of the unscaled variable, one thus has
\begin{equation}
\label{xN2a}
\langle x_2 \rangle=-\langle x_1 \rangle
=\frac{1-e^{-\Delta^2/8Dt}}{\text{Erf}(\Delta/\sqrt{8Dt})}\,\sqrt{\frac{2}{\pi}Dt}.
\end{equation}
For large $\delta$, i.e., for $\Delta\to\infty$ or $t\to 0$ (early time regime in the sense described in Ref.~\onlinecite{footnote}), one recovers the result given by Eq.~(6) in Ref.~\onlinecite{Aslangul1998}, namely, $\langle x_2 \rangle=-\langle x_1 \rangle =\sqrt{\frac{2}{\pi}Dt}$. For the $\Delta=\infty$ case addressed in that reference, this expression is valid for arbitrarily large times, and the particle drift grows without bound in this limit. Returning to the case of finite $\Delta$, if we incorporate the next correction term in the expansion of the rhs of Eq.~\eqref{xN2a}, we are then left with the expression
\begin{equation}
\label{doubleapprox}
 \langle x_2 \rangle \approx\sqrt{\frac{2}{\pi}Dt}\,\left(1-e^{-\Delta^2/8Dt}\right),
\end{equation}
that is, our approximation shows that the influence of a finite $\Delta$-value becomes noticeable when it becomes comparable to $\sqrt{Dt}$. Note, however, that the prefactor of the subleading term is not correct, as it differs from the one obtained from the exact solution [see Eq.~\eqref{correctfom}].

In the opposite limit of small $\delta$ or long times $t$, one can use the asymptotic expansion
\begin{equation}
\frac{1-e^{-z^2}}{\text{Erf}(z)}=\frac{\pi z}{2}-\frac{\sqrt{\pi} z^3}{12}+{\cal O}(z^5), \qquad z\to 0
\end{equation}
in Eq.~\eqref{xN2a} to obtain
\begin{equation}
\label{long-time-approx}
\frac{\langle x_2 \rangle}{\Delta}= \frac{1}{4}-\frac{\Delta^2}{192 Dt}+{\cal O}\left(\frac{\Delta^2}{Dt}\right)^2, \qquad t\to \infty.
\end{equation}
Therefore, for finite $\Delta$, Eq.~\eqref{xN2a} yields a finite  first-order moment as $t\to \infty$, that is, $\langle x_2 \rangle=-\langle x_1 \rangle \to \Delta/4$ (or, equivalently, $\langle u_2 \rangle=-\langle u_1 \rangle =\delta/4$ as $\delta\to 0$). While the limiting value $\langle x_2 (t\to\infty)\rangle=\Delta/4$ agrees with that obtained from the exact solution, the subleading term proportional to $(\Delta^2 Dt)^{-1}$ does not correspond to the exponential long-time relaxation predicted by the exact solution [see Eq.~\eqref{exactFM}]. Likewise, higher order terms in $(\Delta^2 Dt)^{-1}$ also turn out to be incorrect.

\begin{figure}[t]
\includegraphics[width=0.46\textwidth,angle=0]{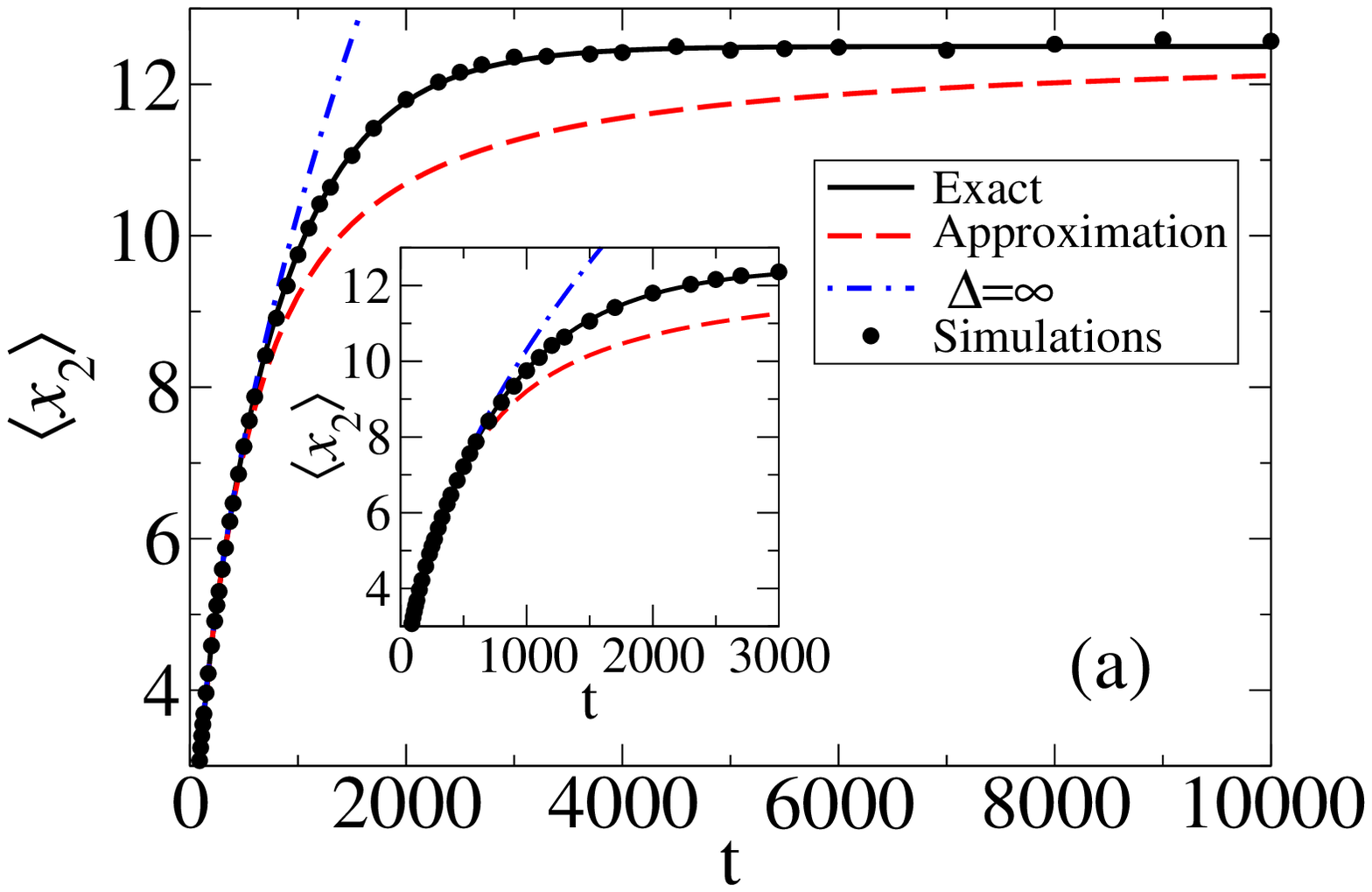}
\includegraphics[width=0.46\textwidth,angle=0]{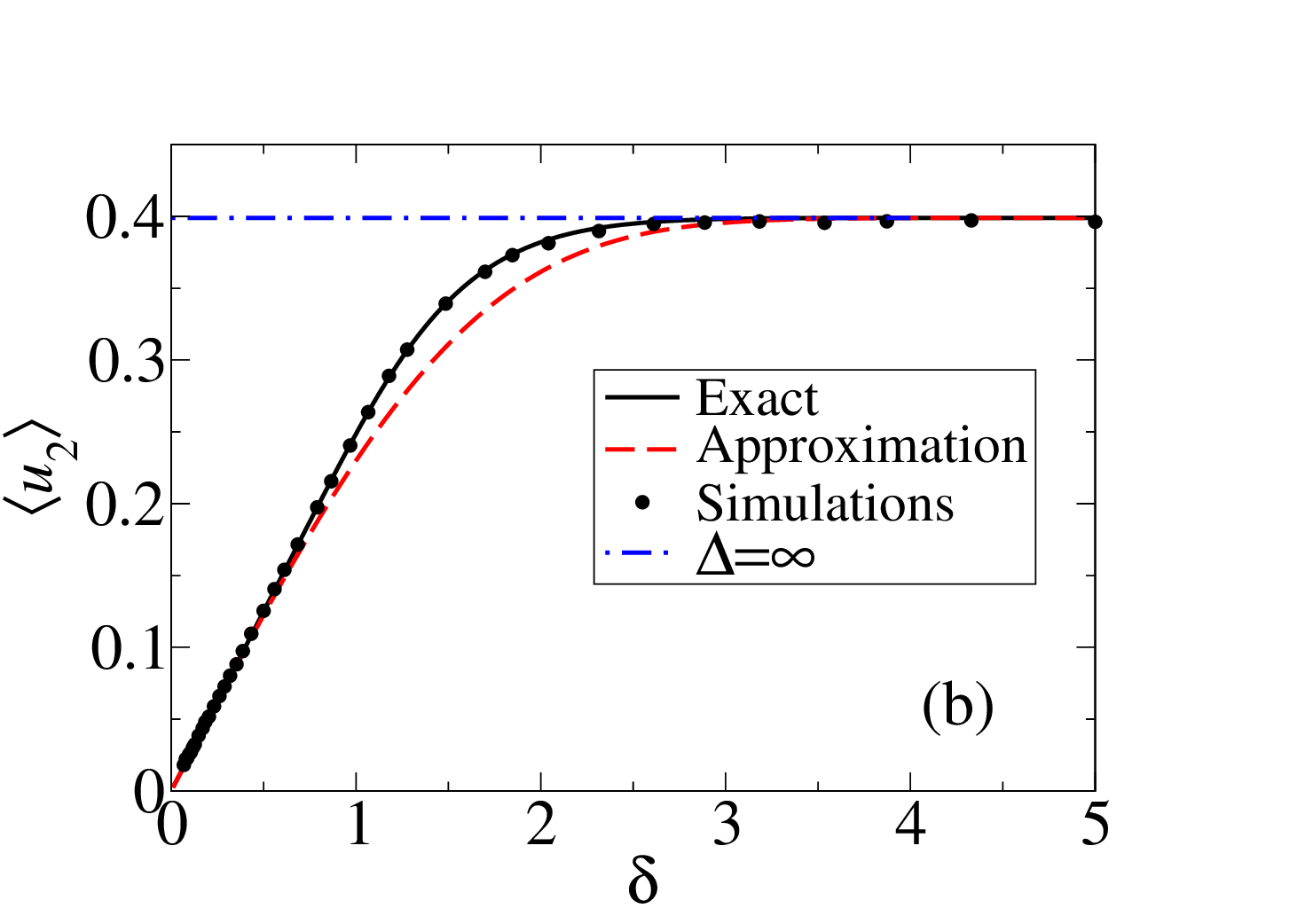}
\caption{\label{x-u-plots} Exact and approximate time dependence of
$\langle x_2 \rangle$ (a) and $\delta$-dependence of the scaled moment
$\langle u_2 \rangle$ (b). For panel (a), we have used
$\Delta=50$ and $D=1/6$. Circles correspond to simulation data that confirm the exact result. The inset on the left panel shows how the exact solution given by the solid line already starts to depart from the approximate one (dashed line) at relatively short times. The dot-dashed lines in both panels correspond to the exact solution for $\Delta=\infty$, i.e., $\langle x_2 \rangle=(2Dt/\pi)^{1/2}$ or, equivalently, $\langle u_2 \rangle=1/(2\pi)^{1/2}$. Note that, for fixed values of $\Delta$ and $D$, increasing time $t$ amounts to decreasing the value of $\delta$, i.e., moving from the right to the left along the horizontal axis.}
\end{figure}

Fig.~\ref{x-u-plots}(a) displays the time dependence of $\langle x_2 \rangle$. Clearly, the exact solution \eqref{exactFM} converges to the long-time limit $\Delta/4$ much faster than the approximate solution. The exact result is confirmed by MC simulations. For completeness, Fig.~\ref{x-u-plots}(b) also shows the universal curve of the scaled moment $\langle u_2 \rangle$ as a function of $\delta$. In contrast to the curve for $\langle x_2 (t)\rangle$, the latter curve does not depend on the specific values of $\Delta$ and $D$.

While Fig.~\ref{figp1xt} shows good agreement between the exact and approximate position pdfs, Fig.~\ref{x-u-plots} reveals, at first sight,  surprising discrepancies between the exact and approximate values of the first moment for certain $\delta$. Specifically, Fig.~\ref{figp1xt} shows these pdfs for $t=200.76$, $D=1$, and $\Delta=50$, that is, $\delta \simeq 1.764$. In this case, $\langle x^2 \rangle \simeq 9.67$ ($\langle u^2 \rangle \simeq 0.341$) for the approximate pdf and $\langle x^2 \rangle \simeq 10.4$ ($\langle u^2 \rangle \simeq 0.368$) for the exact pdf. The error in the corresponding first moment is then about 7\%. This discrepancy can be seen clearly in Fig.~\ref{x-u-plots}(b). But this is not strange: to illustrate that similar-looking pdfs can have different first moments, consider a Gaussian pdf centered at $x_0$:
$\exp[-(x-x_0)^2/(4Dt)]/\sqrt{4\pi D t}$,
whose first moment is, obviously, $\langle x \rangle = x_0$. For the values $t=200.76$ and $D=1$ of Fig.~\ref{figp1xt}, the Gaussian pdf takes on a very similar appearance to the pdfs of Fig.~\ref{figp1xt}  if $x_0 = -10$ is taken. But if we take $x_0 = -11$, the Gaussian pdf is very close to that of $x_0 = -10$ with differences similar to those observed in the pdfs of Fig.~\ref{figp1xt}, and yet $\langle x \rangle$ (or $\langle u \rangle$) differ by 10\%. In summary, this simple example reminds us that pdfs can look very similar and, nevertheless, lead to clearly distinct moments. Little differences between exact and approximate pdfs add up in the corresponding integrals, resulting in significant discrepancies in the moments.

Finally, we note that it is possible to extend the above approximate analysis for the first-order moment to arbitrary odd-order moments (see Appendix B).
The case where both particles evolve on a one-dimensional lattice rather than on the real line also lends itself to analytical treatment (see Appendix C).

\subsection{Second-order moment}
\label{som}

The evaluation of the scaled second-order moment
\begin{equation}
\label{u2N2}
\langle u^2\rangle\equiv \langle u_2^2\rangle=\langle u_1^2\rangle=\int_{-\infty}^\infty du\, u^2 p_1^{(1)}(u,\delta)
\end{equation}
is complicated by the fact that an analytic expression for the integral
\begin{equation}
\int_{-\infty}^\infty du\, u^2 e^{-u^2} \text{Erf}(u+\delta)
\end{equation}
must be found. We have not succeeded in doing so and thus we are led to perform series expansions for small and large values of $\delta$.

\subsubsection{Small-$\delta$ expansion}

To evaluate the rhs of Eq.~\eqref{u2N2} for small values of $\delta$ (large values of $t$), one could directly expand $\text{Erf}(u+\delta)$ in powers of $\delta$ and perform the corresponding integrals. However, we will follow a less direct route which will later be conveniently extended to the case $N>2$. From the definition of the error function, we have
\begin{equation}
p_1^{(1)}(u_1,\delta)=
\frac{2 \,e^{-u_1^2}}{\pi\,\text{Erf}(\delta/\sqrt{2})}
 \int_{u_1}^{u_1+\delta} e^{-u_2^2}\; du_2.
\end{equation}
First, we expand $e^{-u_2^2}$ around $u_1$ as follows (we will keep three terms in the expansions; finding higher-order terms is immediate):
\begin{equation}
e^{-u_2^2}/ e^{-{u_1}^2}=1-2u_1(u_2-u_1)-(1-2u_1^2) (u_2-u_1)^2+{\cal O}\left((u_2-u_1)^3\right).
 \end{equation}
Consequently,
\begin{equation}
e^{u_1^2} \int_{u_1}^{u_1+\delta} du_2  e^{-u_2^2} = \delta -u_1 \delta^2 -\frac{1}{3}(1-2u_1^2)\delta^3+O(\delta^4).
\end{equation}
Therefore, the reduced one-particle pdf can be written as follows:
\begin{equation}
\label{ExpandedReducedPdf}
p_1^{(1)}(u_1,\delta)=
\frac{2 e^{-2u_1^2} }{\pi\,\text{Erf}(\delta/\sqrt{2})}
\left[\delta -u_1 \delta^2-\frac{1}{3}(1-2u_1^2)\delta^3+O(\delta^4) \right].
\end{equation}
Note that the expression also follows directly by using the pertinent expansion in Eq.~\eqref{p1N2u}.

The second-order moment is obtained by inserting Eq.~\eqref{ExpandedReducedPdf} into Eq.~\eqref{u2N2}, carrying out the integrations, and finally expanding the result in powers of $\delta$. This gives
\begin{equation}
\label{u2N2power}
\langle u^2_1\rangle=\langle u^2_2\rangle=\frac{1}{4}+\frac{\delta^2}{12}-\frac{\delta^4}{90} +{\cal O}(\delta^{6}).
\end{equation}
Note that $\langle u^2_{1,2}\rangle=1/4=1/(2N)$ as $\delta\to 0$, that is, as $t\to\infty$ for a prescribed $\Delta$. Thus, $\langle x^2_{1,2}\rangle=Dt$, that is, the effective diffusion coefficient of either particle tends to $D/2$  (recall that $\langle x^2\rangle=2Dt$ for a single particle). This result is consistent with the Rouse model for polymers, where the diffusion coefficient of a chain with $N$ beads is $D/N$ given the diffusion coefficient $D$ of a single bead \cite{Rubinstein2003}. While the correction provided by the $\delta^2$-term in Eq.~\eqref{u2N2power} is still correct, the $\delta^4$-term and higher order ones disagree with the exact solution: the latter exhibits a much faster, multiexponential decay as $\delta\to 0$. Indeed, from Eq.~\eqref{x22t}, we have:
\begin{equation}
\langle u_2^2 \rangle=\frac{1}{4}+\frac{\delta^2}{12}
+\frac{\delta^2}{\pi^2}\sum_{m=1}^\infty \frac{(-1)^m}{m^2} \, e^{-m^2\pi^2/(2\delta^2)}.
\end{equation}

The moments in terms of the original coordinates are obtained by undoing the scaling in Eq.~\eqref{u2N2power}:
\begin{equation}
\label{u2N2power2}
\langle x^2_1\rangle=\langle x^2_2\rangle=Dt+\frac{\Delta^2}{12}-\frac{\Delta^4}{360 Dt}+{\cal O}\left(\frac{\Delta^6}{D^2 t^2}\right).
\end{equation}
The first two terms in the rhs agree with those provided by the exact solution \eqref{x22t}; however, the higher order terms provided by \eqref{u2N2power2}  [which are of the form $C_n \Delta^{2n}/(Dt)^n$, $n$ being a positive integer], fail to reproduce the multiexponential behavior of the exact solution.

\subsubsection{Large-$\delta$ expansion}

The small-$\delta$ expansion of $\langle u^2\rangle$  can be complemented by an analysis of the opposite limit (large $\delta$).
In the following, we use the asymptotic expansion \eqref{exp-erf} of the error function to obtain approximations for the integral leading to $\langle u^2 \rangle$. Setting $z=u+\delta$ in Eq.~\eqref{exp-erf}, we find
\begin{equation}
\label{asexp}
\text{Erf}(u+\delta)=1-\frac{e^{-(u+\delta)^2}}{{\sqrt{\pi}\delta}}\left[ 1+\frac{u}{\delta}+\frac{\frac{1}{2}-u}{\delta^2}+{\cal O}(\delta^{-3})\right],\quad \delta\to\infty.
\end{equation}
The above expansion can be used as a starting point to obtain approximations for
\begin{equation}
\label{intexpu2}
\langle u^2\rangle=\frac{1}{\sqrt{\pi}\,\text{Erf}(\delta/\sqrt{2})} \int_{-\infty}^\infty du \, u^2 e^{-u^2} \text{Erf}(u+\delta).
\end{equation}
For example, retaining terms up to the order $\delta^{-2}$ in brackets in the rhs of Eq.~\eqref{asexp}, we obtain
\begin{equation}
\label{asappru2}
\langle u^2\rangle \approx \frac{1}{2\text{Erf}(\delta/\sqrt{2})} \left[1-\frac{e^{-\delta^2/2}(1+14\delta^2+7\delta^4)}{8\sqrt{2\pi}\delta^3}\right]
\end{equation}
(expressions for higher even-order moments can also be obtained using the expansion \eqref{asexp} in the corresponding integrals). From Eqs.~\eqref{asappru2},  \eqref{uN2a} and \eqref{Deff} we find that the (dimensionless) effective diffusion coefficient is
\begin{equation}
\label{DeffN2}
\frac{D_\text{eff}}{D}= \frac{1}{2}\left(1-\frac{1}{\pi}\right)-\frac{7}{16\sqrt{2\pi}} \delta  \,e^{-\delta^2/2}+\ldots.
\end{equation}

In terms of unscaled coordinates,
\begin{equation}
\label{2ndshortime}
\langle x^2\rangle=2Dt- \frac{7}{8\sqrt{2\pi}} \Delta \sqrt{Dt} \,e^{-\Delta^2/(8Dt)}+\ldots.
\end{equation}
Together with our result \eqref{doubleapprox} for the first-order moment, Eq.~\eqref{2ndshortime}
implies
\begin{equation}
\label{approxvar}
\text{Var}(x) =2\left(1-\frac{1}{\pi}\right) Dt-\frac{7}{8\sqrt{2\pi}} \Delta \sqrt{Dt} \,e^{-\Delta^2/(8Dt)}+\ldots.
\end{equation}
For times short enough to ensure that the $\Delta$-constraint does not play a noticeable role, one recovers Aslangul's result $\text{Var}(x)\sim 2(1-\pi^{-1})Dt$ in the $\Delta=\infty$ case; thus, the diffusion is normal, but the particle diffusivity decreases with respect to the case of free particles by the single file constraint. Physically, this makes sense, but we recall that our solution here is only approximate (based on the factorization ansatz); therefore, the validity of this expression for the long-time limit must be assessed by simulations and by the exact solution, which is available in the present $N=2$ case.
In fact, the first subleading term, i.e., the term proportional to $\Delta \sqrt{Dt} \,e^{-\Delta^2/(8Dt)}$ in the rhs of Eq.~\eqref{approxvar}, is wrong [the correct term is provided by Eq.~\eqref{correctvar}].

\subsubsection{Comparison with exact solution and simulations}

\begin{figure}[ht]
\begin{center}
 \includegraphics[width=0.46\textwidth,angle=0]{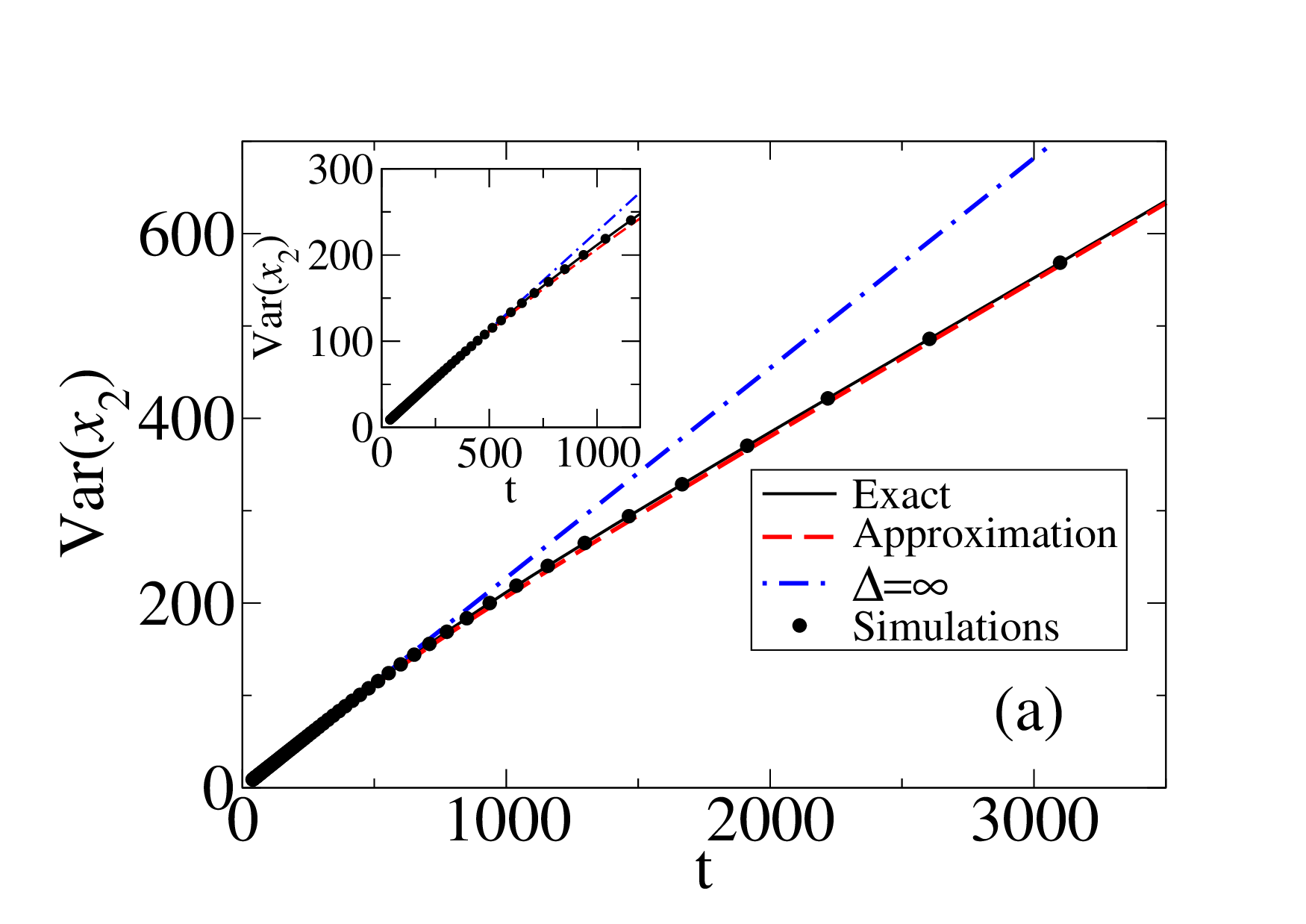}
 \includegraphics[width=0.46\textwidth,angle=0]{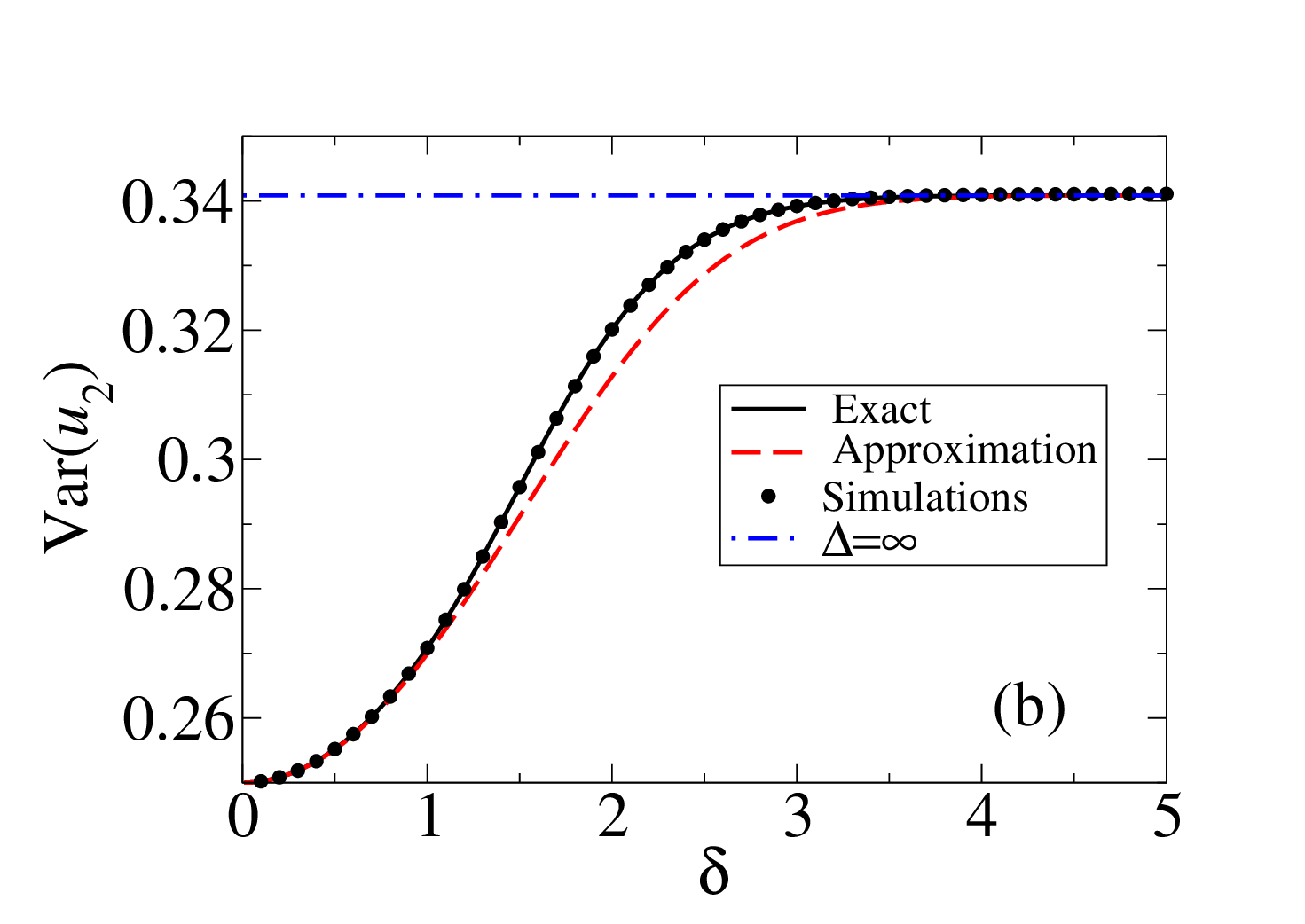}
\end{center}
\caption{\label{x-u-variance}  Time evolution of the exact solution (solid line) and approximate solution (dashed line) for $\text{Var}(x_2)$ [panel (a)]. Displayed are also simulation results (circles). The inset displays the short time behavior. The corresponding solutions for $D_\text{eff}/D=\text{Var}(u_2)$ are also displayed as a function of $\delta$ [panel (b)]. The latter have been computed from data with $\Delta=50$ and $D=1/6$, but in the representation the analytical results collapse into the same curve regardless of the specific values of $D$ and $\Delta$. Dot-dashed lines in both panels correspond to the exact solution for $\Delta=\infty$, namely,
$\text{Var}(x_2)=2D(1-\pi^{-1})t$ and $\text{Var}(u_2)=(1-\pi^{-1})/2$.}
\end{figure}

Fig.~\ref{x-u-variance}(a) shows the time evolution of the variance $\text{Var}(x_2)$. The crossover from the early time value $2D(1-\pi^{-1})t$ to the late time value $Dt$ is clearly observed. Note that these two regimes of normal diffusion at short and long-times have different diffusivities; therefore, they must be connected to one another by a transient anomalous diffusion regime at intermediate times characterized by a nonlinear time dependence of the MSD. The approximate solution (dashed curve) deviates from the exact solution at intermediate times. MC simulations (circles) are in good agreement with the theory. The difference between the approximate solution and the exact one becomes much more evident if the scaled variance $\text{Var}(u)\equiv \text{Var}(x)/(4Dt)$ is considered as a function of $\delta$ [see Fig.~\ref{x-u-variance}(b)].

\subsection{Correlator}

The two-particle correlation function $C(t)\equiv \langle x_1 x_2 \rangle-\langle x_1 \rangle \langle x_2 \rangle$ can be written in terms of rescaled coordinates as follows:
\begin{equation}
C(t)=4 D t \, c(t)
\end{equation}
with
\begin{equation}
  c(t)=c(\delta)=\langle u_1 u_2 \rangle-\langle u_1 \rangle \langle u_2 \rangle.
\end{equation}
After elementary calculations, we find
\begin{equation}
\label{twopointu}
\langle u_1 u_2 \rangle=\int_{-\infty}^\infty du_1 du_2 u_1 u_2 \, p(u_1,u_2)=
\frac{\delta e^{-\delta^2/2}}{\sqrt{8\pi}\text{Erf}(\delta/\sqrt{2})}.
\end{equation}
Thus, using Eq.~\eqref{twopointu}
and Eq.~\eqref{uN2a}, we eventually arrive at the result
\begin{equation}
c(\delta)=\frac{(1-e^{-\delta^2/2})^2}{2\pi \text{Erf}^2(\delta/\sqrt{2})}+
\frac{\delta e^{-\delta^2/2}}{\sqrt{8\pi}\text{Erf}(\delta/\sqrt{2})}.
\end{equation}
Even for large $\delta$, one has non-zero correlations owing to particle repulsion as a result of collisions and recollisions at short times. At smaller $\delta$ (or sufficiently long times), both particles start to feel the additional effect of the $\Delta$-constraint, resulting in an increase in the correlation.

The limiting values are $c(\delta\to 0)=1/4$  and $c(\delta \to \infty)=1/(2\pi)$. The latter is consistent with the expression $C(t)=2 D t/\pi$ obtained in Ref.~\onlinecite{Aslangul1998} for $\Delta=\infty$, which is valid for an arbitrary $t$. As expected, the influence of the finite interaction range is negligible at short times, and the result for $\Delta=\infty$ is recovered.

\begin{figure}[ht]
\begin{center}
 \includegraphics[width=0.46\textwidth,angle=0]{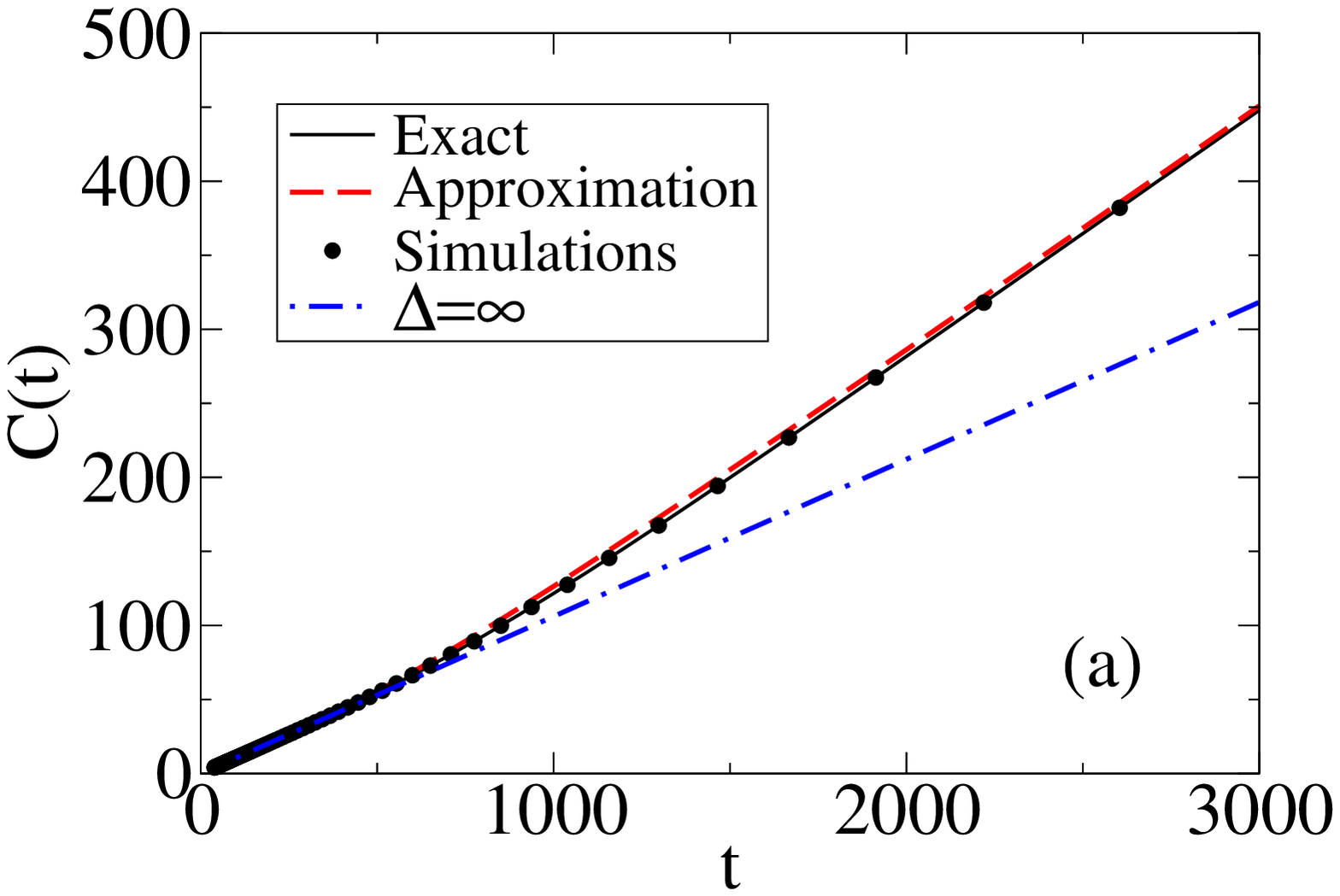}
 \includegraphics[width=0.46\textwidth,angle=0]{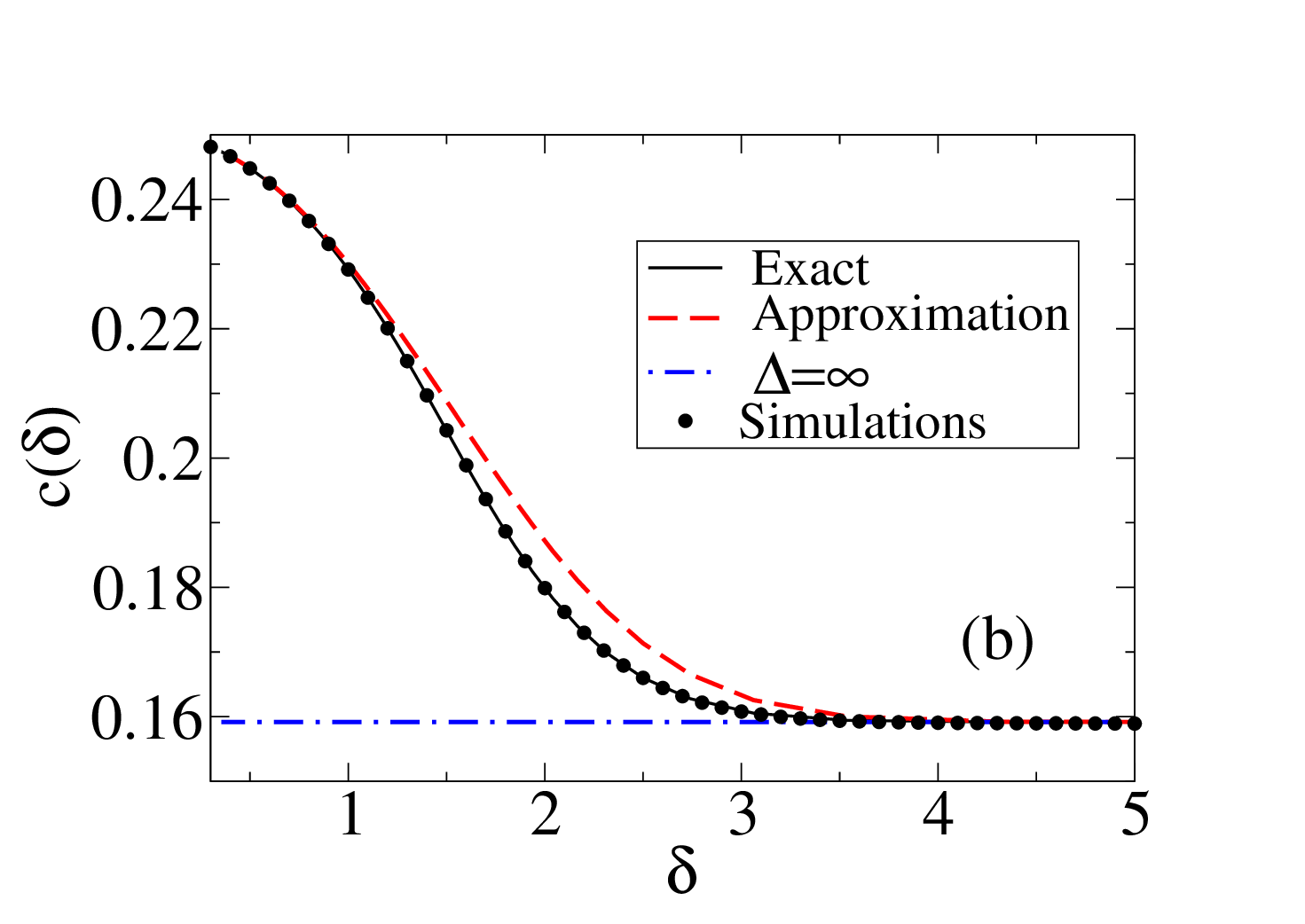}
\end{center}
\caption{\label{x-u-correlator} Unscaled correlator $C(t)$ as a function of time (a), and scaled correlator $c(\delta)$ as a function of
$\delta$ (b). Solid lines correspond to the exact solution, dashed lines to the approximate one. Dot-dashed lines correspond to the case $\Delta=\infty$. In this case $C(t)=(2/\pi)Dt$, and consequently $c(\delta)=1/(2\pi)$.}
\end{figure}

Fig.~\ref{x-u-correlator}(a) shows the crossover of the correlator from the early time value $(2/\pi)Dt$ to the late time value $Dt$. The deviations between the exact and approximate solutions are once again clearly shown in the universal curve for the scaled correlator $c(\delta)\equiv\langle u_1 u_2 \rangle-\langle u_1 \rangle \langle u_2 \rangle=C(t)/(4Dt)$ with $t=t(\delta)=\Delta^2/(4D\delta)$ [see panel (b)].

\section{$\bm{N=3}$ case}
\label{sec5}

In the three-particle case, it does not seem possible to obtain an exact solution for the positional pdf and associated moments. Therefore, we resort to the factorization ansatz introduced earlier to obtain approximate results, and we will assess the accuracy of this method using MC simulations. We again focus on the case of a fully packed initial condition (the three particles start from the origin).

\subsection{General expressions for one-particle pdfs}

The one-particle pdfs for the left- and rightmost particles (respectively labeled $1$ and $3$) are symmetric with respect to the origin and different from the pdf for the central particle (walker $2$). In terms of scaled variables, the one-particle pdf for the leftmost particle takes the form:
\begin{equation}
\label{p1N3}
p_1^{(1)}(u_1,\delta)=A_1 e^{-u_1^2} \int_{u_1}^{u_1+\delta} du_2 e^{-u_2^2} \int_{u_2}^{u_2+\delta} du_3 e^{-u_3^2},
\end{equation}
where $A_1$ is a normalization constant (more generally, $A_n$ denotes the normalization constant for the $n$-th one-particle pdf; see below). For $\delta\to\infty$ (short times or large $\Delta$) one recovers Aslangul's result for $\Delta=\infty$: \cite{Aslangul1998}
\begin{equation}
\label{p1N3dinf}
p_1^{(1)}(u_1,\delta\to\infty)=\frac{\pi A_1 }{8} e^{-u_1^2} \left[1-\text{Erf}(u_1)\right]^2.
\end{equation}
For completeness, even though no new physics is involved, we also provide the pdf for the rightmost particle in the general case:
\begin{equation}
p_3^{(1)}(u_3,\delta)=A_3 e^{-u_3^2} \int_{u_3}^{u_3-\delta} du_2 e^{-u_2^2} \int_{u_2}^{u_2-\delta} du_1 e^{-u_1^2}.
\end{equation}
As for the central particle, one has
\begin{equation}
\label{centralpdf}
p_2^{(1)}(u_2,\delta)=A_2 e^{-u_2^2} \int_{u_2}^{u_2-\delta} du_1 e^{-u_1^2} \int_{u_2}^{u_2+\delta} du_3 e^{-u_3^2},
\end{equation}
whence the $\Delta=\infty$ result already obtained by Aslangul follows \cite{Aslangul1998}:
\begin{equation}
\label{p2N3dinf}
p_2^{(1)}(u_2,\delta\to\infty)=\frac{\pi A_2}{4} e^{-u_2^2} \left[1+\text{Erf}(u_2)\right]\,\left[1-\text{Erf}(u_2)\right].
\end{equation}

\subsection{$\boldsymbol{\delta}$-expansion}

By analogy with the procedure employed in the $N=2$ case (see \ref{fom} and \ref{som}), we now seek to obtain approximations for the reduced pdfs and the corresponding moments by means of a $\delta$-expansion. To this end, we use the expansion $f(z)=f(u)+f'(u) (z-u)+\ldots$ in integrals of the form $\int_{u}^{u\pm \delta} dz f(z)$ , and then carry out the integration term by term. From Eq.~\eqref{p1N3}, we find
\begin{equation}
p_1^{(1)}(u_1,\delta)=A_1\sqrt{\frac{\pi }{3}}\,e^{-3u_1^2}\,
\left[
1-3 u_1\delta -\frac{8}{9}(1-6u_1^2)\delta^2 + \ldots
\right]
\end{equation}
for the leftmost particle. This yields the expansion
\begin{equation}
\label{u1deltaexp}
\langle u_1 \rangle= -\frac{\delta}{2 } +\frac{\delta^3 }{ 12} -\frac{\delta^5 }{120 } +\ldots
\end{equation}
for the first-order moment, implying $\langle u_1\rangle=-\delta/2$ when $\delta\to 0$, for example in the long-time limit $t\to\infty$. Consequently, $\langle x_1\rangle=-\Delta/2=-\langle x_3\rangle$  as $t\to\infty$. According to our analytical findings, while the first term in the rhs of Eq.~\eqref{u12deltaexp} provides the correct behavior in the $\delta\to 0$ limit, it is unlikely that higher order terms are still correct, since this was already not the case for $N=2$ (note, however, that an exact solution to assess the validity of these terms is not available in the present case).

Proceeding in a similar way for the second-order moment, we obtain the result
\begin{equation}
\label{u12deltaexp}
\langle u_1^2 \rangle=  \frac{1}{6 } +\frac{8\delta^2}{ 27} -\frac{85\delta^4 }{972 } +\ldots
\end{equation}
Note that $\langle u_1^2\rangle=1/6=1/(2N)$ in the limit of small $\delta$. For the centered MSD, this implies $\text{Var}(x_1) \to 2Dt/3$ as $t\to\infty$, that is, the effective diffusion coefficient of particle 1 tends to $D/3$ as $t\to\infty$ (which, by symmetry, is also the diffusion coefficient of particle 3). As was the case for $N=2$, the first term in the rhs of Eq.~\eqref{u12deltaexp} provides the correct behavior, and expectedly also the $\delta^2$-term. In contrast, we expect the $\delta^4$-term and higher order ones to be wrong.

The central particle 2 has, of course, different dynamics, as it is confined by its two neighbors 1 and 3. The corresponding expansion yield
\begin{align}
\label{p2N3b}
p_2^{(1)}(u_2,\delta)
&
=A_2 \frac{\pi }{4}  \,e^{-u_2^2}\, \left[\text{Erf}(u_2)-\text{Erf}(u_2-\delta)\right]
\left[\text{Erf}(u_2+\delta)-\text{Erf}(u_2)\right]
\\
&
=A_2\sqrt{\frac{\pi }{3}}\,e^{-3u_2^2}\,
\left[1-\frac{1}{18}(1-6u_2^2)\delta^2+\frac{5-66u_2^2 +72u_2^4 }{1620}\delta^4 +\ldots\right]
\end{align}
for the pdf. The first-order moment vanishes trivially, $\langle u_2 \rangle=0$, whereas the second-order moment can be expanded as follows:
\begin{equation}
\label{u22deltaexp}
\langle u_2^2 \rangle=  \frac{1}{6} +\frac{ \delta^2  }{ 54} +\frac{\delta^4 }{4860 } +\ldots
\end{equation}
 Thus, $\langle u_2^2\rangle\to1/6\to1/(2N)$ as $\delta\to 0$. This implies $\text{Var}(x_2) \approx \text{Var}(x_1)\approx 2Dt/3$ at long times. Thus, the effective diffusion coefficient of the central particle is asymptotically the same as that of each neighbor $D_{2,\text{eff}}=D_{1,\text{eff}}=D_{3,\text{eff}}\to D/3$ as $t\to\infty$. In other words, at long times the three-particle system behaves as a single entity with an effective diffusion coefficient $D/N=D/3$. Moreover, as we will soon see, this $D/N$-behavior at long times holds for arbitrary $N$.

Finally, we note that the $\delta^2$-term in the rhs of Eq.~\eqref{u22deltaexp} is still expected to be correct, but this is no longer the case for the $\delta^4$-term and higher-order ones.

\subsection{Comparison with MC simulations}

\begin{figure}[ht]
\begin{center} \includegraphics[width=0.46\textwidth,angle=0]{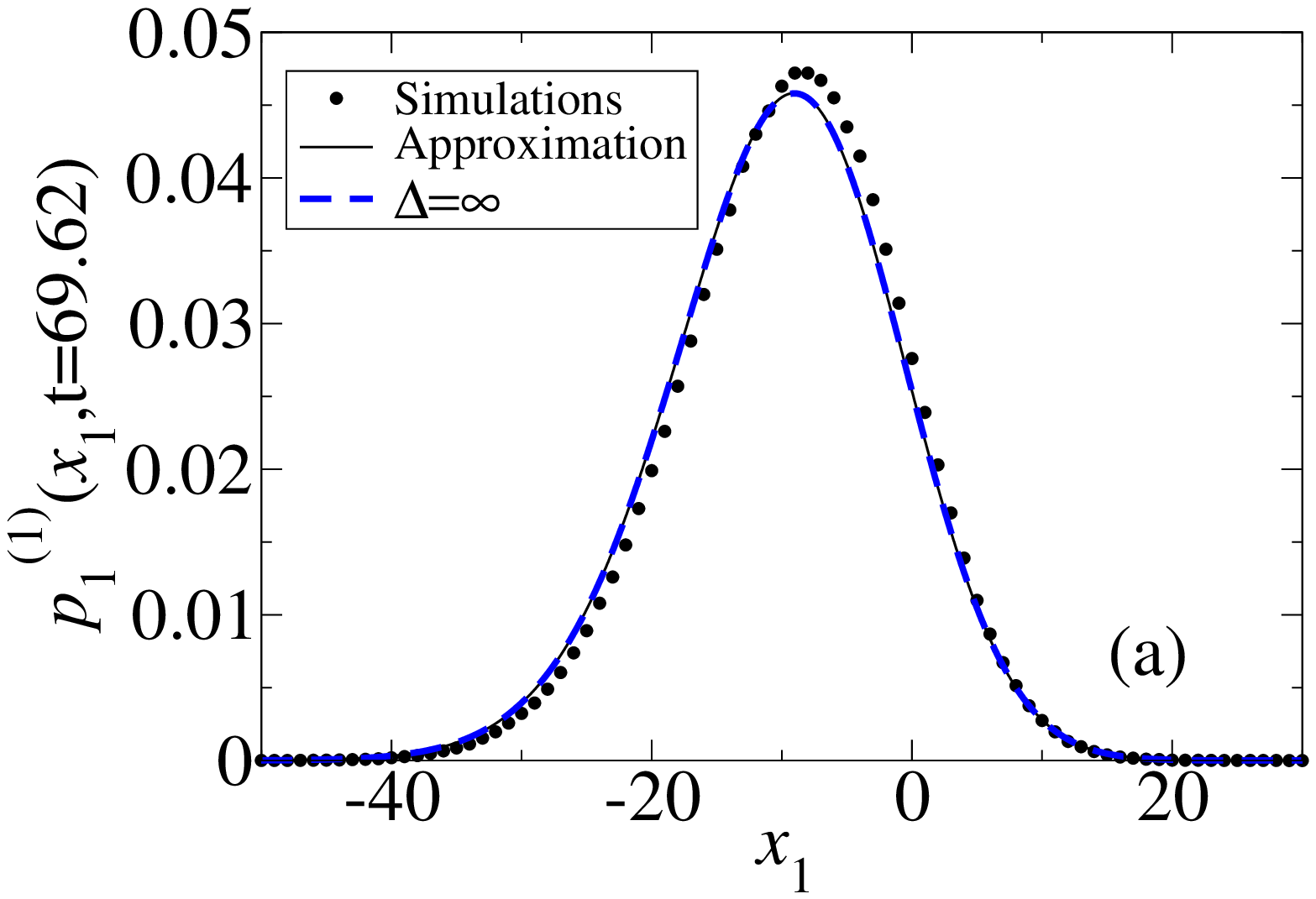}
\includegraphics[width=0.46\textwidth,angle=0]{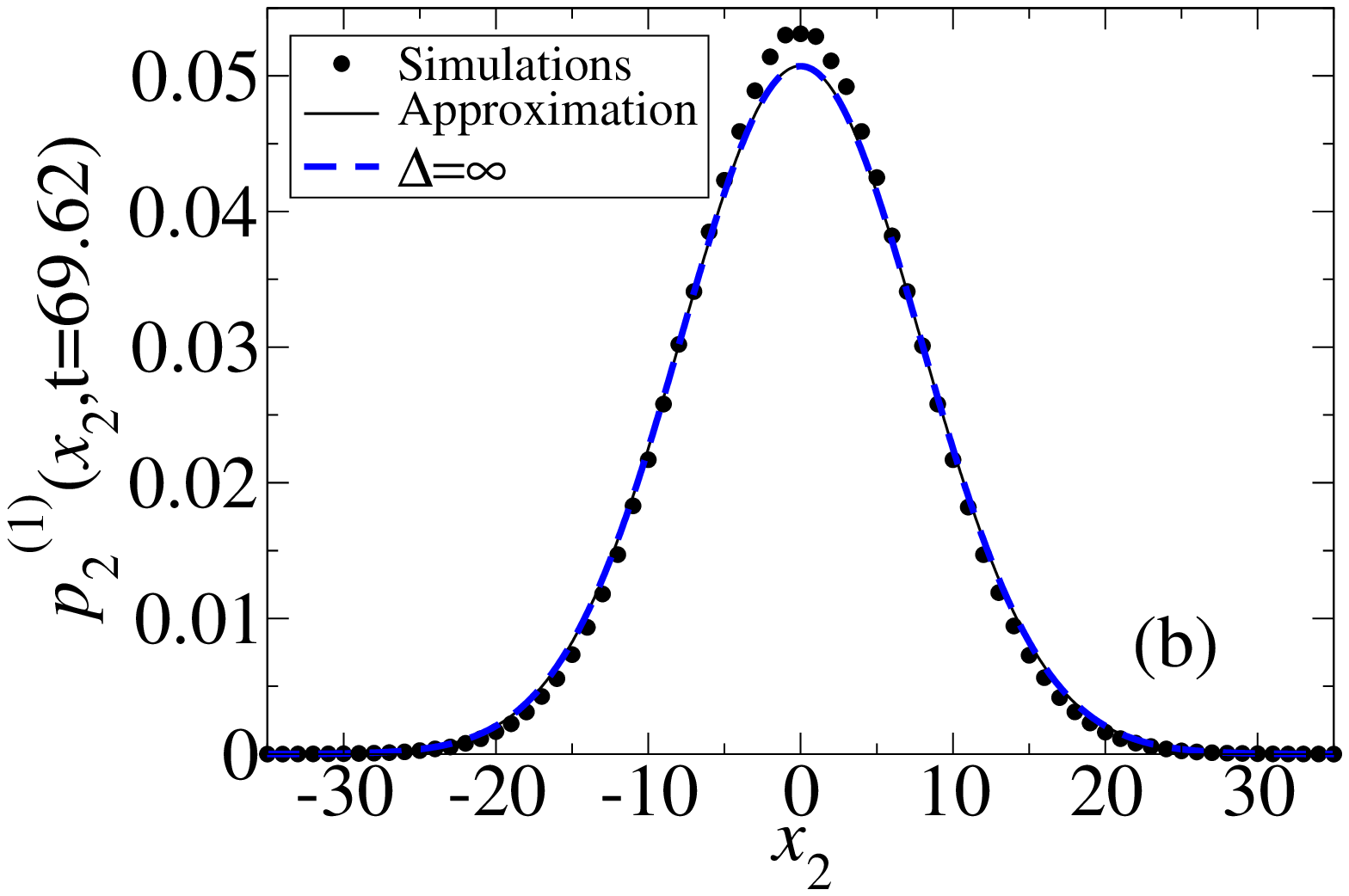}
\end{center}
\caption{\label{FigpmN3t10}
Simulation results for $p_m^{(1)}(x,t,\Delta)$ with $D=1$, $\Delta=50$, $t=69.62$ (i.e, a short time, corresponding to
$\delta=3$) and $N=3$. Panel (a) corresponds to the left particle $m=1$, whereas panel (b) corresponds to the central particle ($m=2$). Solid lines correspond to the approximate solution, whereas dashed lines correspond to Aslangul's solution for $\Delta=\infty$.}
\end{figure}

\begin{figure}[ht]
\begin{center}
\includegraphics[width=0.46\textwidth,angle=0]{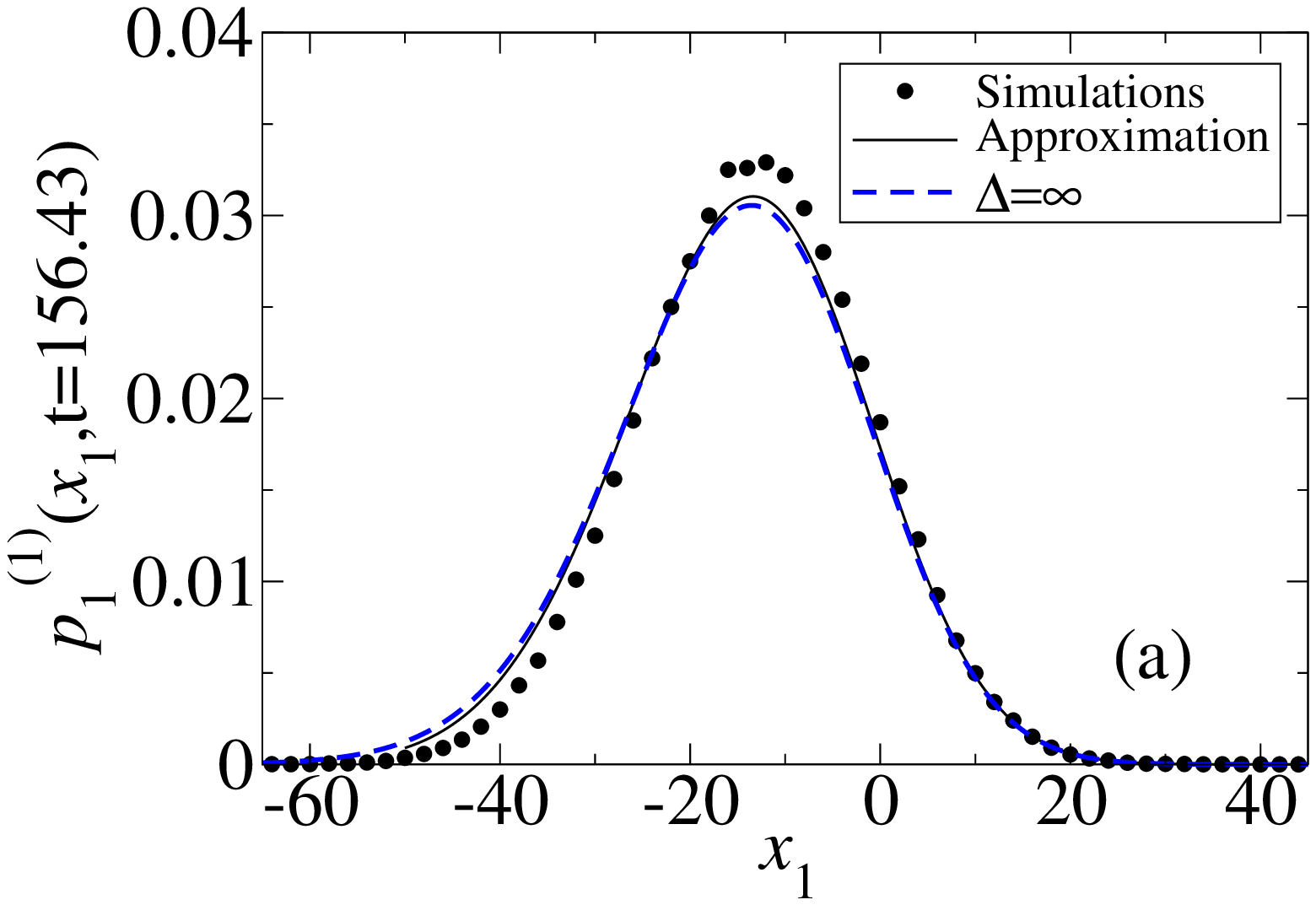}
\includegraphics[width=0.46\textwidth,angle=0]{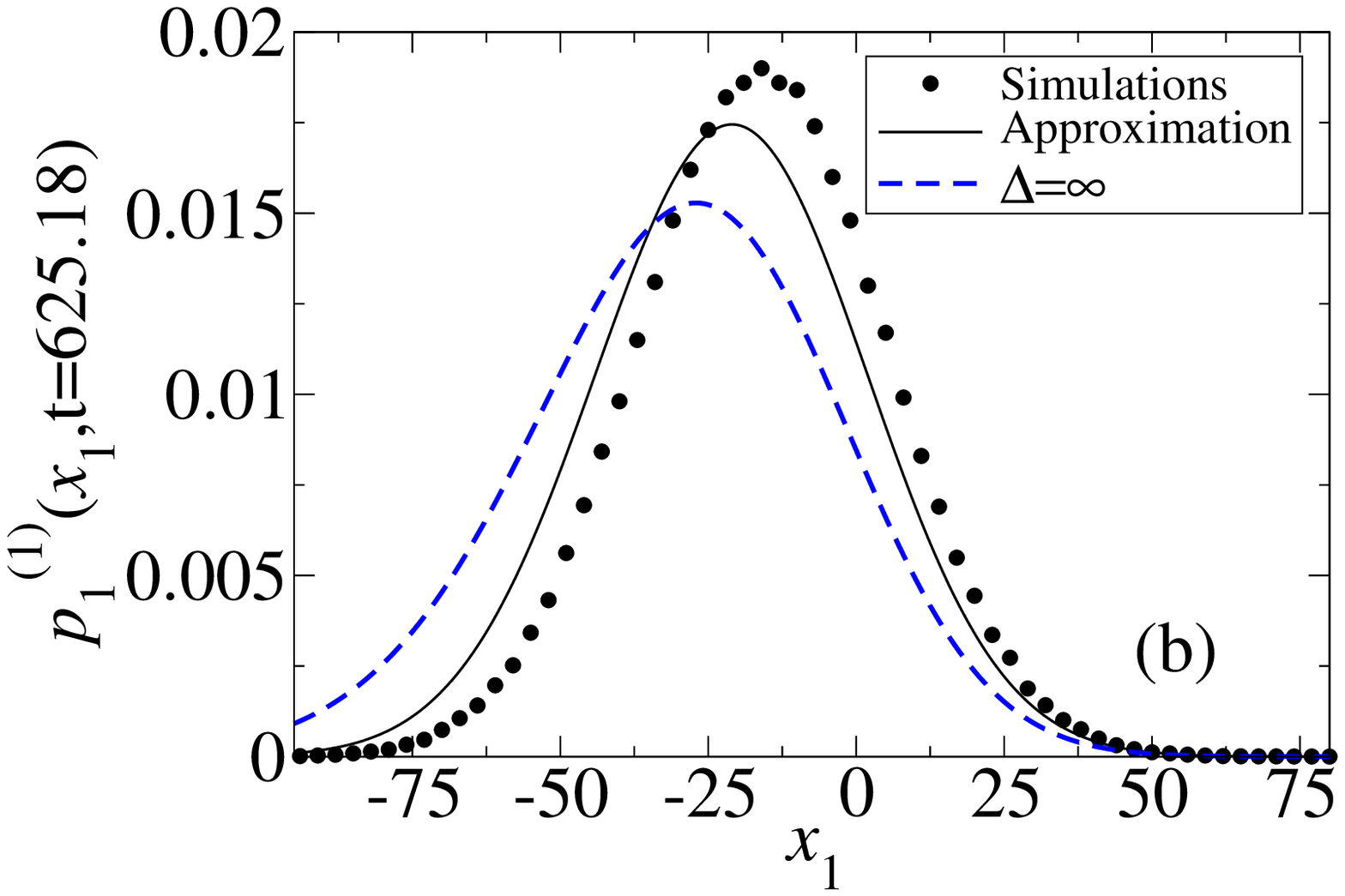}
\includegraphics[width=0.46\textwidth,angle=0]{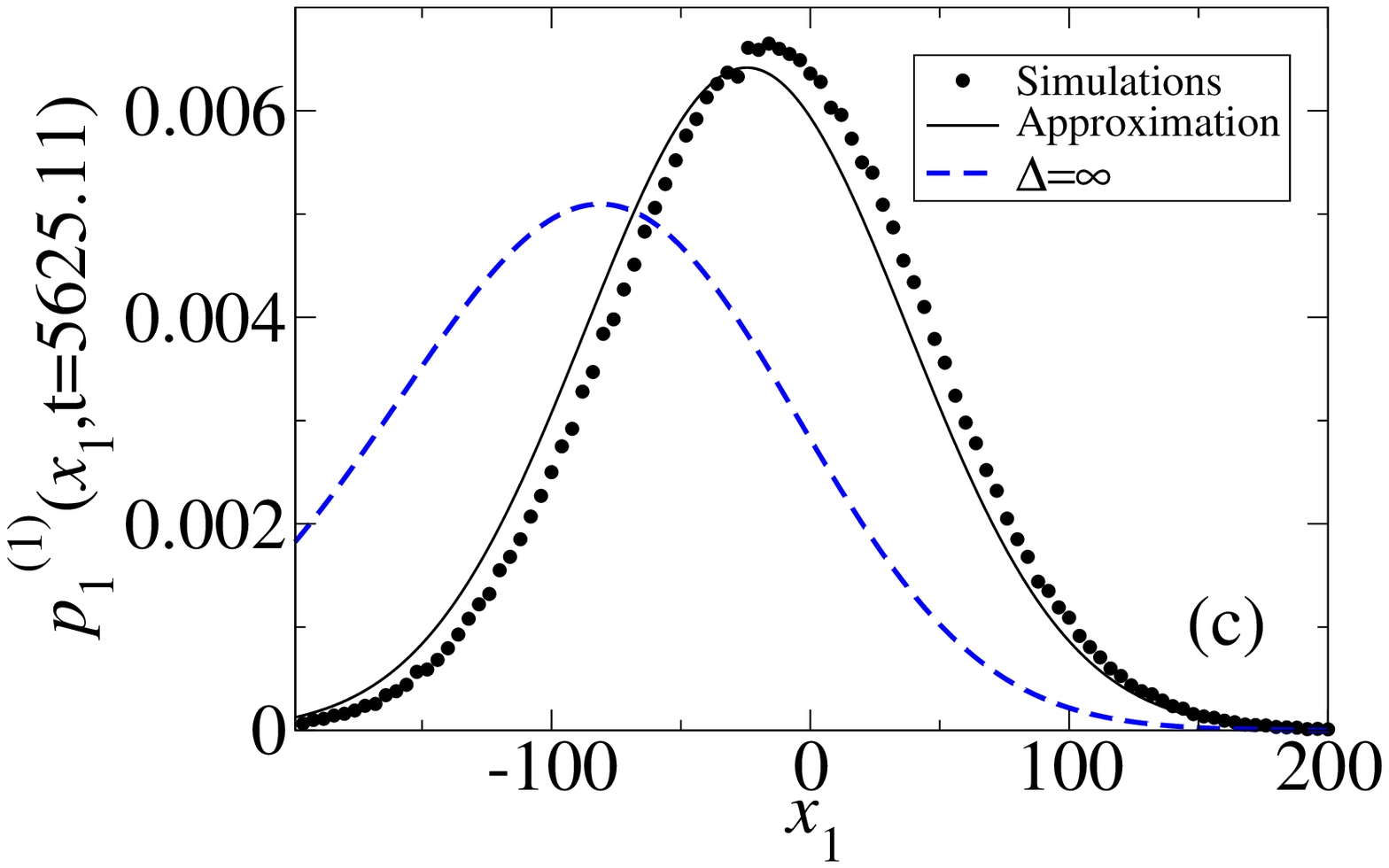}
\end{center}
\caption{\label{Figp12N3t100y1000}
Simulation vs. approximate analytical results for $p_1^{(1)}(x,t,\Delta)$ with $D=1$ and $\Delta=50$ for $t=156.43$ ($\delta=2$) (a), $t=625.18$ ($\delta=1$) (b) and $t=5625.11$ ($\delta=1/3$) (c) for three particles, $N=3$. The dashed curves correspond to the analytical result for $\Delta=\infty$.
}
\end{figure}

To assess the validity of the Gaussian (diffusive) approximation
based on the factorization ansatz [cf. Eqs.~\eqref{p1N3} and \eqref{centralpdf}], we have performed numerical simulations for the particles' motion with $\tau=1$ and $D=1$ in all cases.
For short times (up to $\sim 10$) the Gaussian approximation is poor. However, the agreement improves at somewhat larger times, for example, for times such that $\delta\approx 3$ (see Fig.~\ref{FigpmN3t10}). In Fig.~\ref{FigpmN3t10} we compare the simulation results for the reduced pdfs $p_m^{(1)}$ (where $m=1$ for the left particle and $m=2$ for the central particle) with the approximate reduced pdfs obtained by reexpressing Eqs.~\eqref{p1N3} and \eqref{centralpdf} in terms of the original (i.e., unscaled) space and time variables. For both the left and the central particle, the agreement between the approximate theory and the numerical pdfs is quite acceptable, but the effect of the $\Delta$ constraint in this time regime is not very relevant yet (the approximate analytical solutions overlap with the curves for $\Delta=\infty$, which are also displayed in Fig.~\ref{FigpmN3t10}).
The agreement between the simulations, approximate solution and the
$\Delta=\infty$ curve progressively worsens at longer times, as shown in Fig.~\ref{Figp12N3t100y1000} for the left particle in the time regime where
$\delta$ varies from 2 to 1/3. In this figure, it can also be observed that the agreement between the simulations and the approximate solution first deteriorates with increasing time [see panels (b) and (c)], yielding an important discrepancy at times for which $\delta \approx 1$. However, at longer times the agreement improves again (see panel (c) for the case $\delta=1/3$).

\begin{figure}[ht]
\begin{center}
\includegraphics[width=0.46\textwidth,angle=0]{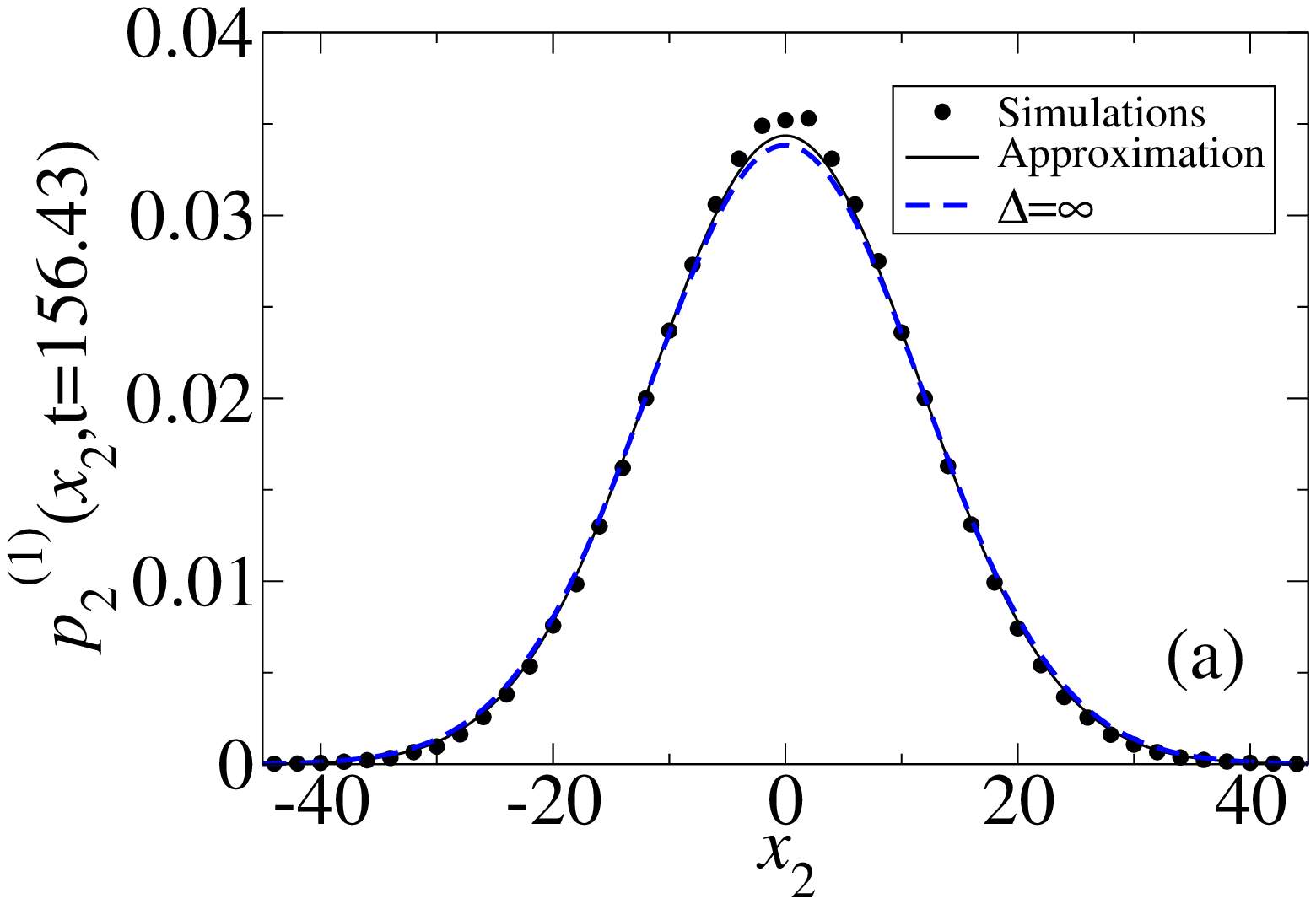}
\includegraphics[width=0.46\textwidth,angle=0]{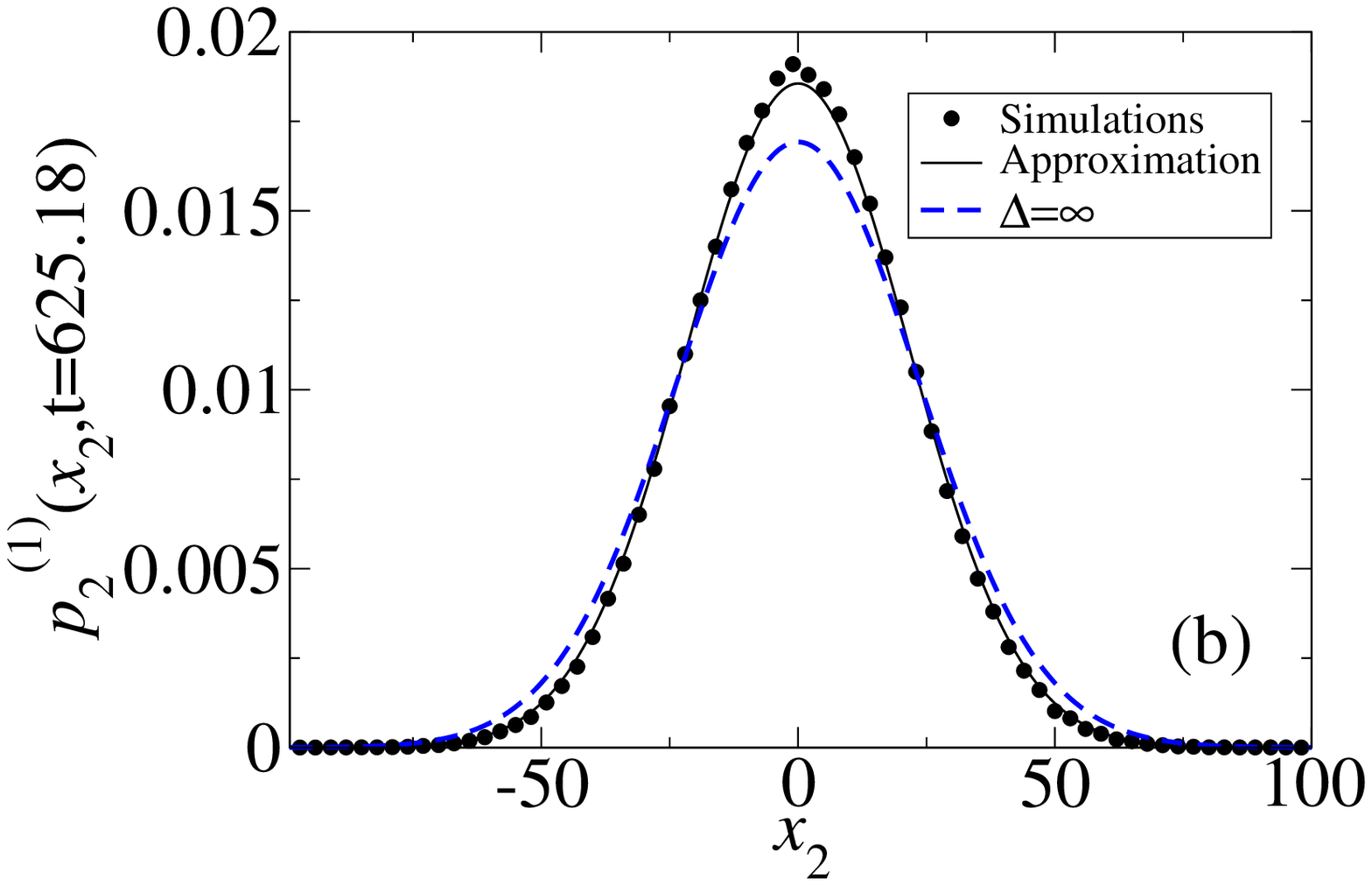}
\includegraphics[width=0.46\textwidth,angle=0]{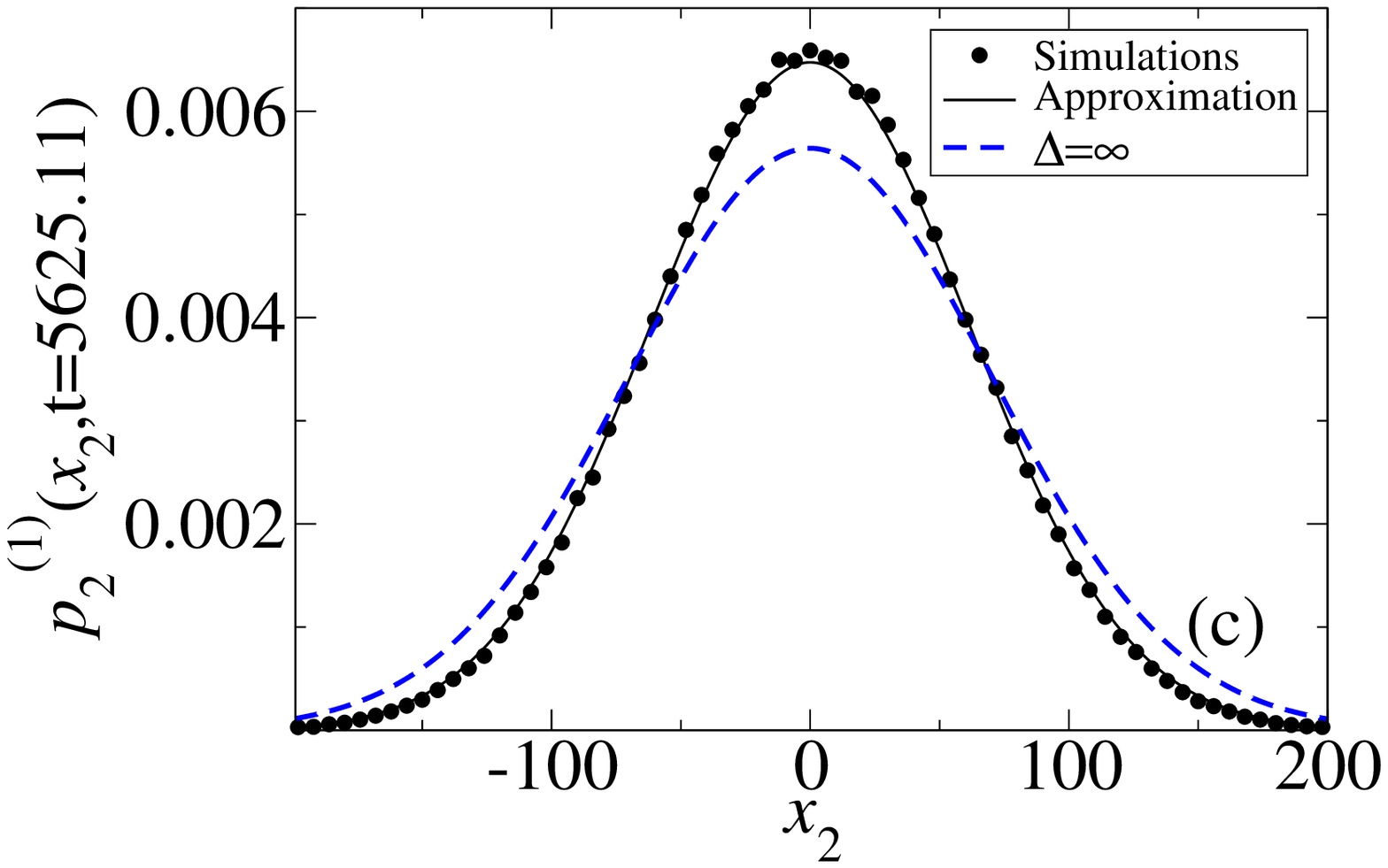}
\end{center}
\caption{\label{Figp2N3t100}
Simulation vs approximate analytical results for $p_2^{(1)}(x,t,\Delta)$ with $D=1$ and $\Delta=50$ for $t=156.43$ ($\delta=2$) (a), $t=625.18$ ($\delta=1$) (b) and $t=5625.11$ ($\delta=1/3$) (c) for three particles, $N=3$. The dashed curves correspond to the analytical result for $\Delta=\infty$.
}
\end{figure}

Interestingly, the Gaussian approximation appears to perform better for the central particle than for the edge particle. Figure~\ref{Figp2N3t100} shows a comparison of the pdfs obtained from the numerical simulations with the curves computed from Eq.~\eqref{centralpdf} at the same times as shown in Fig.~\ref{Figp12N3t100y1000}. The discrepancy between the simulation data and the approximate curves remains small as $\delta$ decreases from 2 at $t=156$ to $1/3$ at $t=5625$.

\section{$\bm{N}$-particle case}
\label{sec6}

In this section, we present results for the general case of arbitrary $N$. To this end, we exploit the iterative relations between the integrals that appear in the general expressions for one-particle pdfs.

\subsection{Iterative formulas}
\label{itform}

The one-particle pdf for the $m$-th particle is
 \begin{align}
 \label{pm1Anidada}
p_m^{(1)}(u_m;\delta)&=A_m
 \int_{-\infty}^{\infty} du_1\int_{-\infty}^{\infty} du_{m-1} \ldots \int_{-\infty}^{\infty} du_{m+1}\int_{-\infty}^{\infty} du_{N}  \prod_{i=1}^N e^{-u_i^2} \prod_{i=1}^{N-1} R(u_{i+1}-u_i,\delta)
\end{align}
i.e.,
\begin{equation}
\label{pm1b}
p_m^{(1)}(u_m,\delta)=A_m e^{u_m^2} I_L(m-1,\delta;u_m) I_R(N-m,\delta;u_m),
\end{equation}
where
 \begin{align}
I_R(N-m,\delta;u_m)&=
\int_{-\infty}^{\infty} du_{m+1} e^{-u_{m+1}^2} R(u_{m+1}-u_{m},\delta) \ldots\int_{-\infty}^{\infty} du_{N} e^{-u_N^2} R(u_{N}-u_{N-1},\delta) \nonumber \\
&=
\int_{u_{m}}^{u_{m}+\delta} du_{m+1} e^{-u_{m+1}^2}   \ldots
\int_{u_{N-1}}^{u_{N-1}+\delta} du_{N} e^{-u_N^2}
\label{IR2}
 \end{align}
and
 \begin{align}
I_L(m-1,\delta;u_m)&=
\int_{-\infty}^{\infty} du_1 e^{-u_1^2} R(u_{2}-u_1,\delta) \ldots\int_{-\infty}^{\infty} du_{m-1} e^{-u_{m-1}^2} R(u_{m}-u_{m-1},\delta) \nonumber \\
&=
\int_{u_{m}-\delta}^{u_{m}} du_{m-1} e^{-u_{m-1}^2}
  \ldots
\int_{u_{2}-\delta}^{u_2} du_1 e^{-u_1^2} \nonumber \\
&=
(-1)^{m-1}\int_{u_{m}}^{u_{m}-\delta} du_{m-1} e^{-u_{m-1}^2}
  \ldots
\int_{u_{2}}^{u_2-\delta} du_1 e^{-u_1^2}.
 \end{align}

For the particular case $\Delta=\infty$, Eq.~\eqref{pm1b} leads to Aslangul's result (cf. Eq.~(3) in Ref.~\onlinecite{Aslangul1998}). To this end, one uses the explicit expressions
\begin{align}
\int_{u }^{\infty} dz  e^{-z^2}  \text{Erfc}^n(z) =  \frac{\sqrt{\pi}}{2(n+1)}\text{Erfc}^{n+1}(u), \qquad n=0,1,\ldots
 \end{align}
 and
\begin{align}
\int_{-\infty }^{u} dz  e^{-z^2}  \left(1+\text{Erf}(z)\right)^n  =  \frac{\sqrt{\pi}}{2(n+1)} \left(1+\text{Erf}(z)\right)^{n+1}, \qquad n=0,1,\ldots
 \end{align}
to simplify the integrals $I_R(N-m,\infty;u_m)$ and $I_L(m-1,\infty;u_m)$.

In order to deal with the case of a finite $\delta$, we observe that
\begin{equation}
I_L(n,\delta;u)=(-1)^n\, I_R(n,-\delta;u).
\end{equation}
Therefore, Eq.~\eqref{pm1b} is equivalent to
\begin{equation}
\label{pm1c}
p_m^{(1)}(u_m,\delta)=A_m e^{u_m^2}\,(-1)^{m-1}\, I_R(m-1,-\delta;u_m)\, I_R(N-m,\delta;u_m).
\end{equation}
On the other hand, one has
\begin{equation}
I_R(n,\delta;u)=\int_{u}^{u+\delta} du_1 e^{-u_{1}^2} I_R(n-1,\delta;u_1),
\end{equation}
with $I_R(0,\delta;u)=1$.
We now express $I_R(n,\delta;u)$ in powers of $\delta$ by expanding $e^{-u_{1}^2} I_R(n-1,\delta;u_1) $ about $u$ and performing the integration term by term. The result is
\begin{equation}
\label{IRiter}
I_R(n,\delta;u)=\sum_{k=0}^\infty \frac{\delta^{k+1}}{(k+1)!}\;\frac{d^k}{du^k} \left\{e^{-u^2}I_R(n-1,\delta;u) \right\}.
\end{equation}
Thus, we arrive at an \emph{iterative} formula that is very convenient for the evaluation of the series expansion of $I_R(n,\delta;u)$ in powers of $\delta$. By means of Eq.~\eqref{IRiter}, it is not difficult to compute the $\delta$-expansions of the first two moments (for the central and leftmost particles) for large values of $N$ (e.g., up to $N=200$). In this way we find
\begin{align}
\label{u1N}
\langle u_1\rangle&=-\frac{N-1}{4}\,\delta-\frac{N(1-N^2)}{288} \delta^3  +{\cal O}(\delta^5),\\
\label{uleftmost}
\langle u_1^2\rangle&=\frac{1}{2N}+ \frac{2+3N-14N^2+9N^3}{144N} \delta^2+{\cal O}(\delta^4),\\
\label{ucentral}
\langle u_\text{central}^2\rangle &\equiv \langle u_{\floor*{\frac{N}{2}}}^2\rangle=\frac{1}{2N}+ \frac{N^2-1}{144N} \delta^2+{\cal O}(\delta^4).
\end{align}
In Eq.~\eqref{ucentral}, $\left\lfloor \cdots \right\rfloor$ denotes the floor function.

The constant term $(2N)^{-1}$ obtained from Eqs.~\eqref{uleftmost} and \eqref{ucentral} confirms the $D/N$ asymptotic behavior already observed for $N=2$ and $N=3$, i.e., each particle moves with effective diffusivity $D/N$. The $1/N$ dependence of $D_{\floor*{\frac{N}{2}},\text{eff}}$ is in qualitative agreement with Aslangul's findings for the $\Delta=\infty$ case; however, in the case of the edge particles, the behavior is different. For these particles, Aslangul obtained a much slower decay of the effective diffusivity with $N$, $D_{1,\text{eff}}=D_{N,\text{eff}}\propto D/\ln(N)$ in the large-$N$ limit. In contrast,  we see that a finite $\Delta$ induces a much faster decrease of the diffusivity of the edge particles with increasing particle number and homogenizes the asymptotic behavior of the collection of particles. This common diffusivity $D/N$ at long times exhibited by our model is a well-known result of an ideal model of polymers (e.g., the Rouse model), where $D$ is the diffusivity of a single (free) monomer.

We close the present subsection by noting that, in view of the above results based on \eqref{IRiter}, one might think that it is possible to obtain general asymptotically exact expressions for the $N$-dependence of both reduced pdfs and associated moments. However, this hope is greatly reduced by the practical difficulties encountered during the evaluation of nested integrals. This difficulty also arises in the numerical evaluation of the nested integrals that provide the one-particle pdf according to Eq.~\eqref{pm1Anidada} [or Eq.~\eqref{pm1c}]. That said, for small $N$ our Gaussian ansatz provides good approximations to the pdfs obtained in simulations (see Sec. \ref{compMCtheory}).

\subsection{Comparison between approximate theory and MC simulations}
\label{compMCtheory}

\begin{figure}[t]
\includegraphics[width=0.46\textwidth,angle=0]{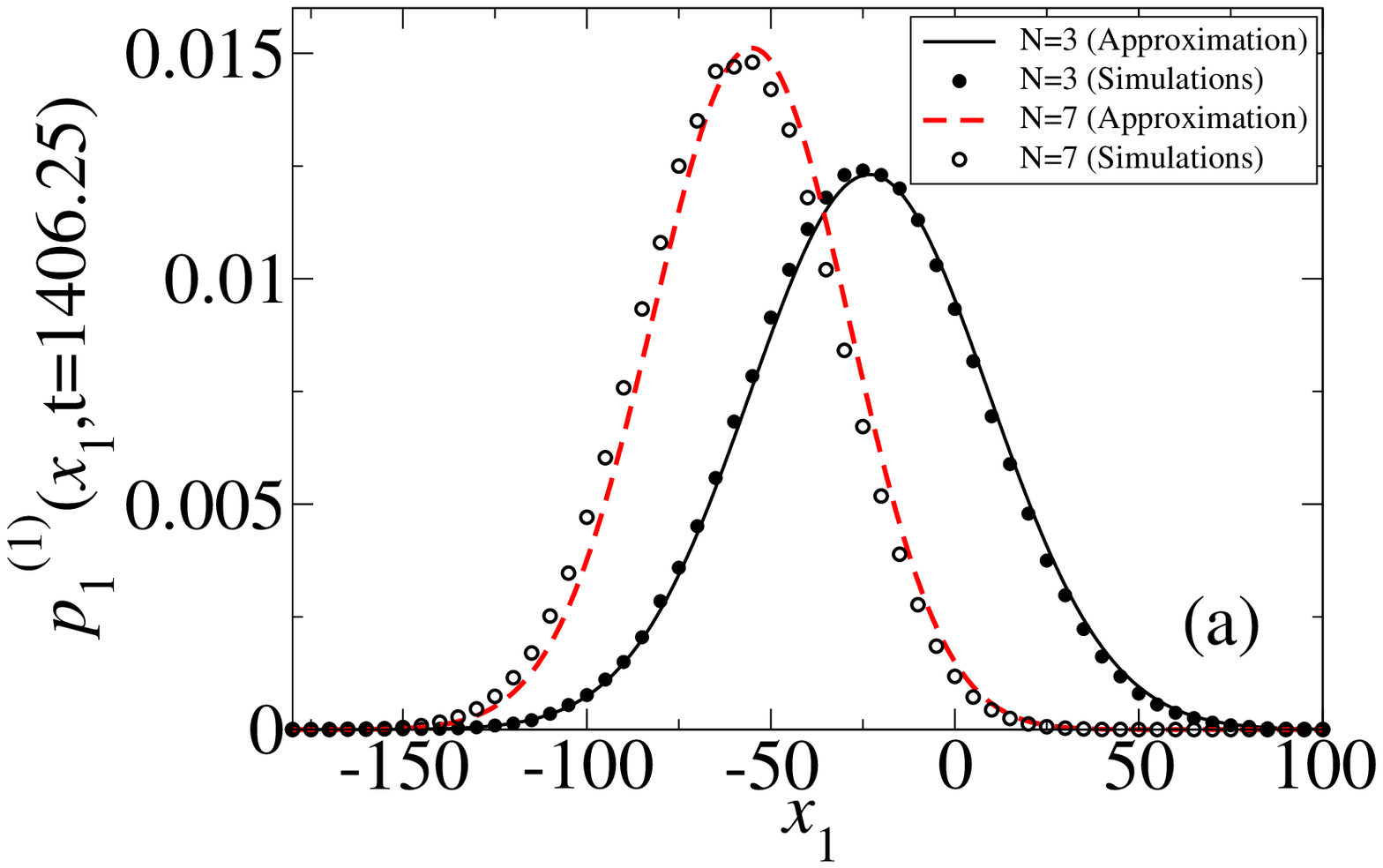}
\includegraphics[width=0.46\textwidth,angle=0]{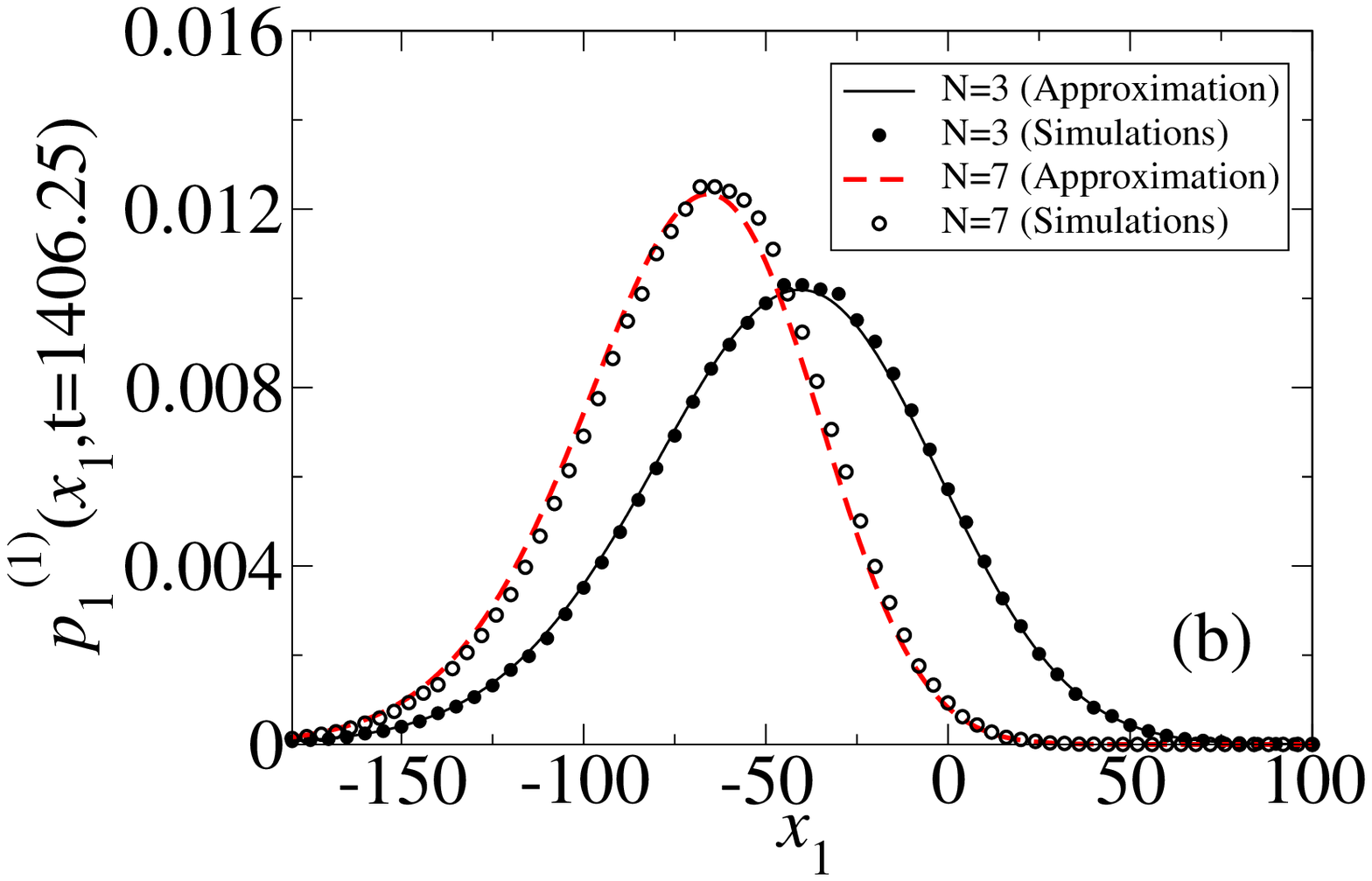}
\caption{\label{leftparticleN3N7} Behavior of the leftmost particle. Open and filled circles correspond to simulation results, solid and dashed lines to the approximate analytical solutions. Panel (a) corresponds to $\Delta=50$, and panel (b) corresponds to $\Delta=7500$. The time was set to $t=1406.25$, and the diffusivity $D$ was set to unity, leading to $\delta=2/3$ [panel (a)] and to $\delta=100$ [panel (b)].
}
\end{figure}

Figure \ref{leftparticleN3N7} shows a comparison of the behavior of the leftmost particle for $N=3$ and for $N=7$. Both the simulation results and approximate solutions based on the factorization ansatz are shown (see \ref{itform}). It can be seen that the approximation agrees very well with the simulations for $N=3$, and that the agreement is still good for $N=7$. For a given value of $\Delta$, one clearly sees that the leftmost particle drifts further away from the origin with increasing $N$ because of the increased push of a larger number of particles. For a given value of $N$, the peak of the pdf moves to the left with increasing $\Delta$, since a higher value of this quantity implies that the leftmost particle has a less severe restriction to move in the negative direction of the real axis (increasing $\Delta$ has, however, a stronger effect for $N=3$ than for $N=7$).

\begin{figure}[t]
\includegraphics[width=0.46\textwidth,angle=0]{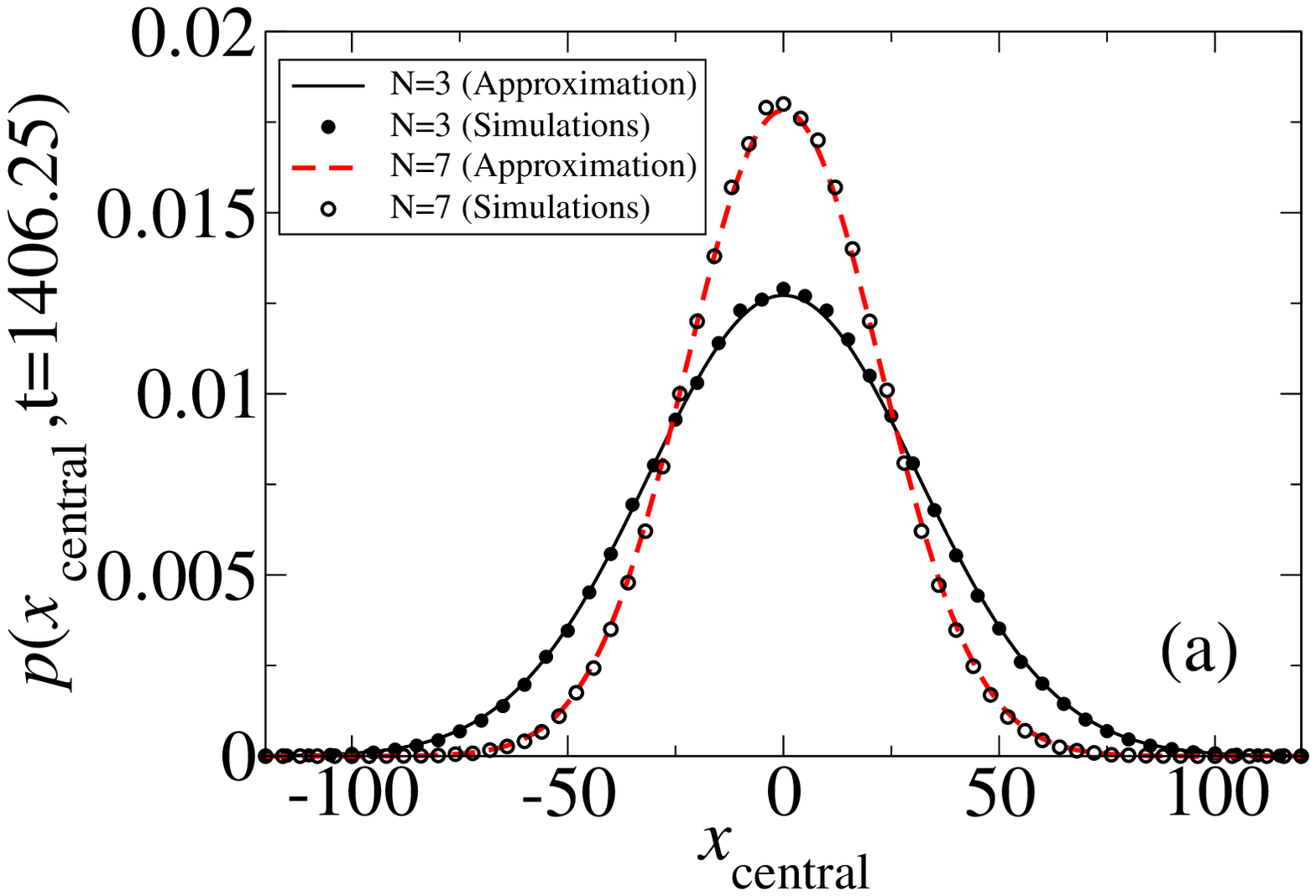}
\includegraphics[width=0.46\textwidth,angle=0]{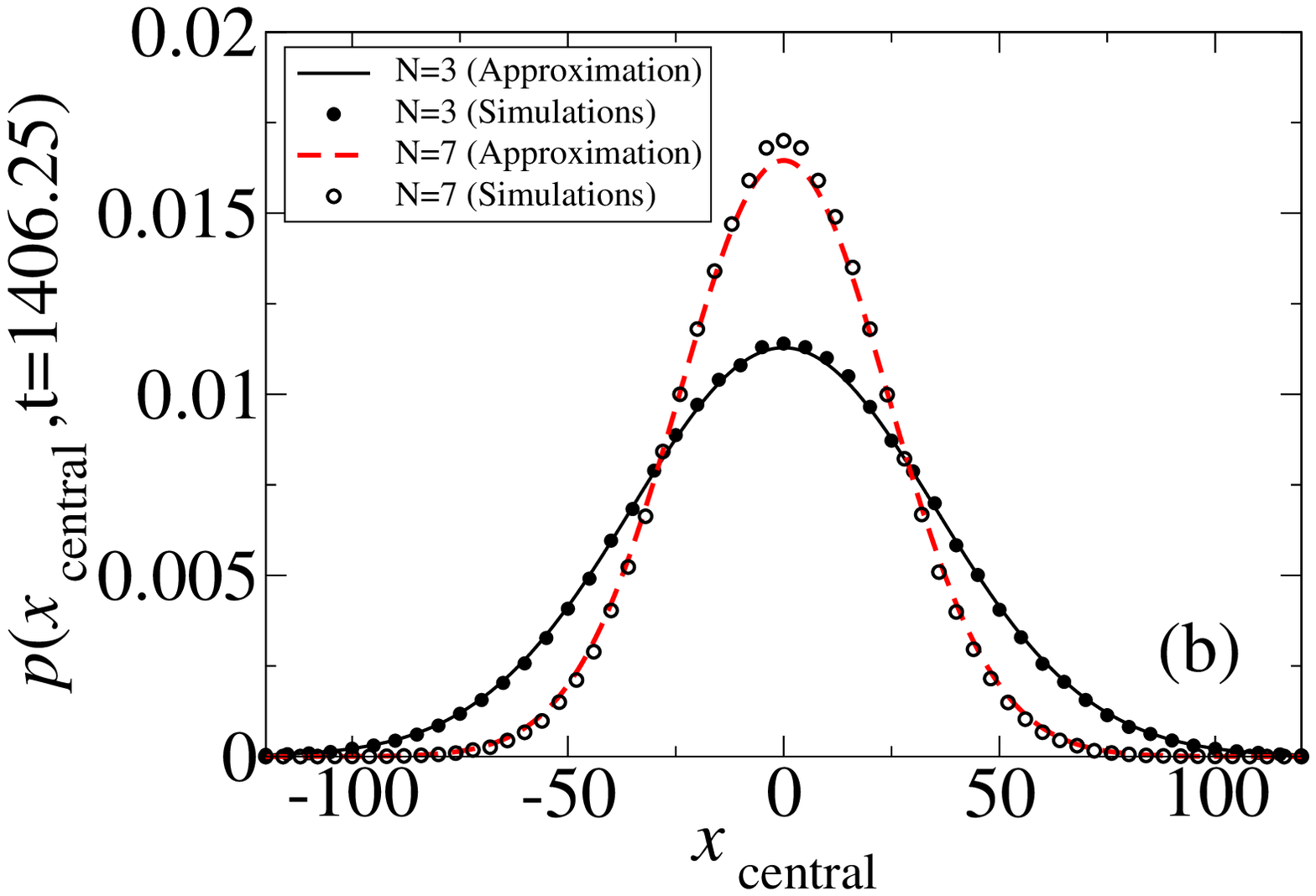}
\caption{\label{centralparticleN3N7} Behavior of the central particle. Open and filled circles correspond to simulation results, solid and dashed lines to the approximate analytical solutions. As in Fig.~\ref{leftparticleN3N7}, here $t=1406.25$, $D=1$, $\Delta=50$ [panel (a)], and
$\Delta=7500$ [panel (b)].}
\end{figure}

Similar plots for the pdf of the central particle $p(x_\text{central},t)\equiv p_{\lfloor N/2 \rfloor}^{(1)}(x_{\lfloor N/2 \rfloor},t)$ are shown in Fig.~\ref{centralparticleN3N7} (the parameters for panels (a) and (b) in this figure are the same as those used in Fig.~\ref{leftparticleN3N7} for the leftmost particle). As expected, the localization of the central particle becomes stronger with increasing $N$, as a larger number of particles enhances the confinement effects. For a given $N$, the pdf is higher and narrower (i.e., more localized) for a smaller value of $\Delta$ because of the stronger confinement induced by mutual interactions. As we mentioned at the end of Sec.~\ref{itform}, the numerical evaluation of the nested integrals that define the one-particle pdf according to Eq.~\eqref{pm1Anidada} [or Eq.~\eqref{pm1c}] becomes increasingly complex as $N$ increases. In our case, the computation time required to evaluate the one-particle pdf for $N=7$ has been very long, making it prohibitive to go to much larger values of $N$.

\subsection{Relaxation to the equilibrium state}

\subsubsection{Prediction based on factorization ansatz}
\label{appanres}

For small $\delta=\Delta/\sqrt{4Dt}$ (e.g., in the long-time limit), our factorization ansatz provides analytical expressions for $\langle u_1 \rangle$ and, by symmetry, also for $\langle u_N \rangle=-\langle u_1 \rangle$. Using Eq.~\eqref{u1N}, we obtain
\begin{equation}
\frac{ \langle x_1 \rangle}{\Delta} = -\frac{N-1}{4} +\frac{N(N^2-1)}{288}\, \delta^2 + {\cal O}(\delta^4).
\end{equation}
In terms of the time $t$, we thus have
\begin{equation}
\label{x1exp}
\frac{ \langle x_1 \rangle}{\Delta} = -\frac{N-1}{4} +\frac{N(N^2-1)}{288}\, \frac{\Delta^2}{4Dt}+ {\cal O}\left(\frac{\Delta^2}{4Dt}\right)^2.
\end{equation}
The average length (end-to-end distance) of our system is $\langle L\rangle =\langle  x_N-x_1\rangle =2|\langle x_1\rangle |=-2\langle x_1\rangle$. Eq.~\eqref{x1exp} then yields
\begin{equation}
\frac{ \langle L  \rangle}{\Delta} =  \frac{N-1}{2} -\frac{N(N^2-1)}{144}\, \frac{\Delta^2}{4Dt} + {\cal O}\left(\frac{\Delta^2}{4Dt}\right)^2.
\end{equation}
In terms of the average equilibrium system length, $\langle L\rangle_{\text{eq}}=(N-1)\,\Delta/2$ (see Sec. VII), this result can be rewritten as follows:
\begin{equation}
\label{normLapp}
\frac{ \langle L  \rangle}{\langle L\rangle_\text{eq}}=
  1 -\frac{N(N+1)}{72}\, \frac{\Delta^2}{4Dt}+ {\cal O}\left(\frac{\Delta^2}{4Dt}\right)^2.
\end{equation}
Thus, the Gaussian approximation predicts that the long-time approach to the equilibrium length is dictated by an inverse power law:
\begin{equation}
\frac{\langle L \rangle-\langle L\rangle_\text{eq}}{\Delta} \propto  \frac{\Delta^2}{4Dt}, \quad t\gg \Delta^2/D.
\end{equation}
However, note that this prediction based on the factorization ansatz contradicts the exact result \eqref{exactFM}, which yields exponential relaxation in the long-time limit for the specific case $N=2$.

For large $N$, Eq.~\eqref{normLapp} becomes
\begin{equation}
\frac{ \langle L  \rangle}{\langle L\rangle_\text{eq}}=
 1 -\frac{N^2}{72}\, \frac{\Delta^2}{4Dt} +{\cal O}\left(\frac{\Delta^2}{4Dt}\right)^2, \quad N\gg 1.
\end{equation}
On the basis of the above results, one would expect that the time (relaxation time $\tau_r$) required to reach a given fraction of the equilibrium length (99\%, say) scales as $N^2\Delta^2/D$ for large $N$, that is, as $\tau_r\sim\langle L\rangle_{\text{eq}}^2/D$.
Despite the peculiar form of the particle-particle interactions in our system (string-bead chain), it is interesting to note that this result coincides with that yielded by the Rouse model for an ideal polymer chain \cite{Grosberg2002}.

As far as higher-order moments are concerned, we note that the computation of $\langle L^2  \rangle$ requires the evaluation of the corresponding correlator $\langle x_N x_1 \rangle$, since
$\langle L^2  \rangle= \langle (x_N-x_1)^2  \rangle= \langle  x_N^2 \rangle+\langle x_1^2  \rangle-2\langle x_N x_1 \rangle=2 \langle x_1^2  \rangle-2\langle x_N x_1 \rangle$. While $\langle x_1^2\rangle $ can be evaluated approximately in the framework of the factorization ansatz, it does not seem possible to obtain explicit results for the integrals appearing in the definition of the correlator in the case of arbitrary $N>2$ (in contrast, we recall that the case $N=2$ is amenable to analytical treatment, see \ref{ssscorr}).

\subsubsection{Comparison with MC simulations}

\begin{figure}[ht]
\begin{center}
        \includegraphics[width=0.46\textwidth,angle=0]{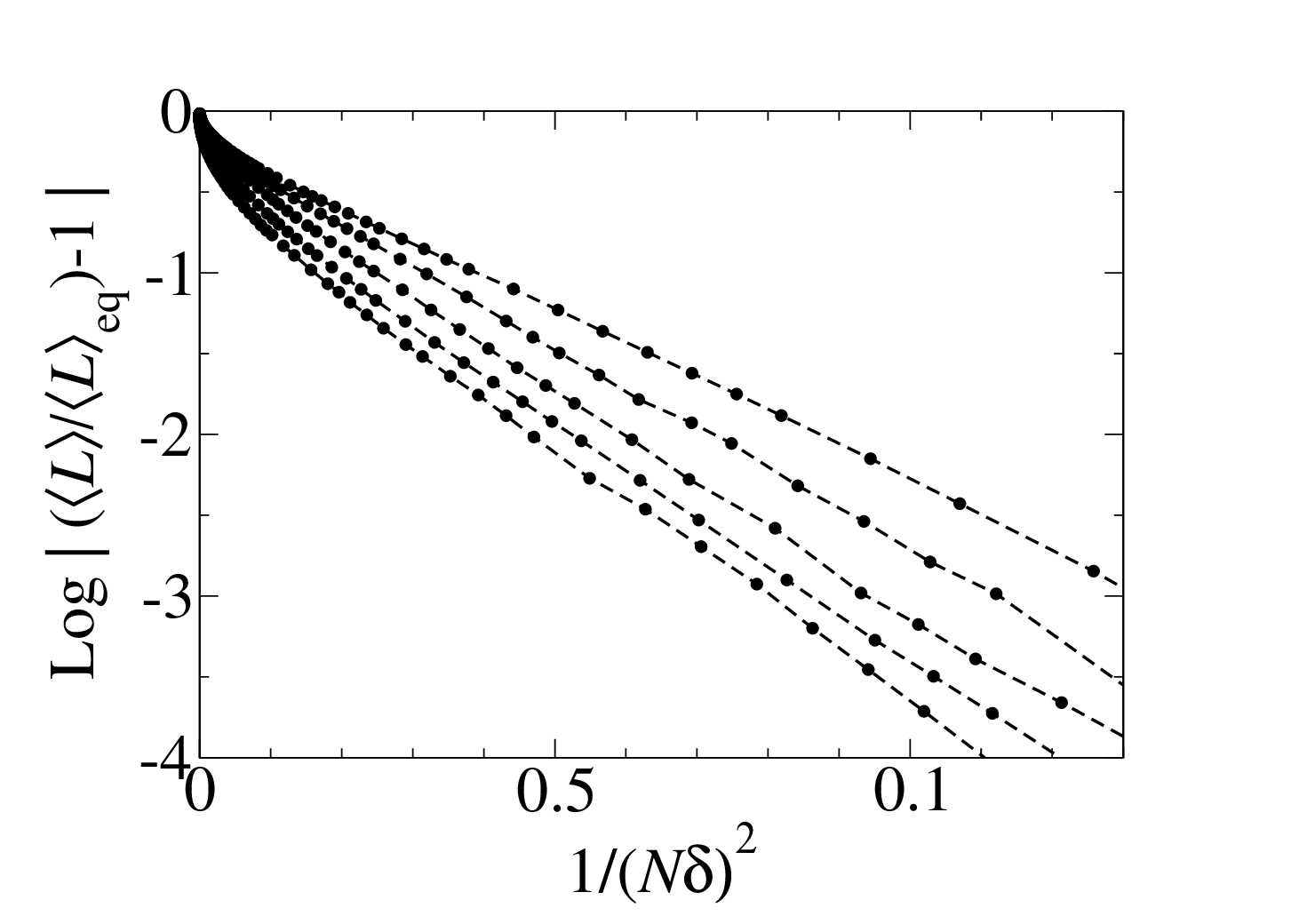}
\end{center}
\caption{\label{FigLoLeqm1Nvs1oNdelta2}
Simulation results for the relaxation of the (properly scaled) average end-to-end distance $\langle L \rangle$ as a function of $1/(N\delta)^2$. The (dashed) curves correspond, from top to bottom, to the values $N=2,3,5,11,101$. For the first four curves we have taken $\Delta=200$, whereas for $N=101$ we have chosen
$\Delta=50$. In all cases, $D=1$. The lines are an aid to the eye.
 }
\end{figure}

In order to assess the analytical predictions of \ref{appanres}, numerical simulations were performed for different values of $N$. As it turns out, the simulation results \emph{contradict} the predictions of the Gaussian approximation and thus question its validity in the long-time regime. Rather than an inverse power law decay, one finds an \emph{exponential} long-time ($\delta\to 0)$ decay in the long limit $\delta\to 0$:
\begin{equation}
\log\left[\frac{ |\langle L  \rangle-\langle L\rangle_\text{eq}|}{\langle L\rangle_\text{eq}}\right]\sim -\frac{c(N)}{N^2 \delta^{2}} \sim-c(N)\,\frac{t}{\tau_r} .
\end{equation}
This divergence in the limit of small $\delta$ explains the failure of the expansion of $\langle L\rangle$ in terms of this parameter. According to our numerical estimates
(cf. Fig.~\ref{FigLoLeqm1Nvs1oNdelta2}), the function $c(N)$ depends only weakly on $N$ and seems to converge to a common value $c(\infty)$ when $N$ increases. The exponential relaxation observed at long times is strongly at odds with the much slower $1/t$ asymptotic decay predicted in \ref{appanres},  but in line with the exact analytic result obtained for $N=2$ and in the Rouse model. Thus, we see that the Gaussian approximation not only fails to predict the correct relaxation behavior for $N=2$, but also for larger $N$. In any case, our string-bead model demonstrates that, while assuming an elastic interaction between the beads is useful for a detailed analysis of relaxation modes, it is not essential for observing exponential relaxation. In fact, the abrupt interactions considered here are quite different from those prescribed by harmonic potentials.

\begin{figure}[ht]
\begin{center}
\includegraphics[width=0.46\textwidth,angle=0]{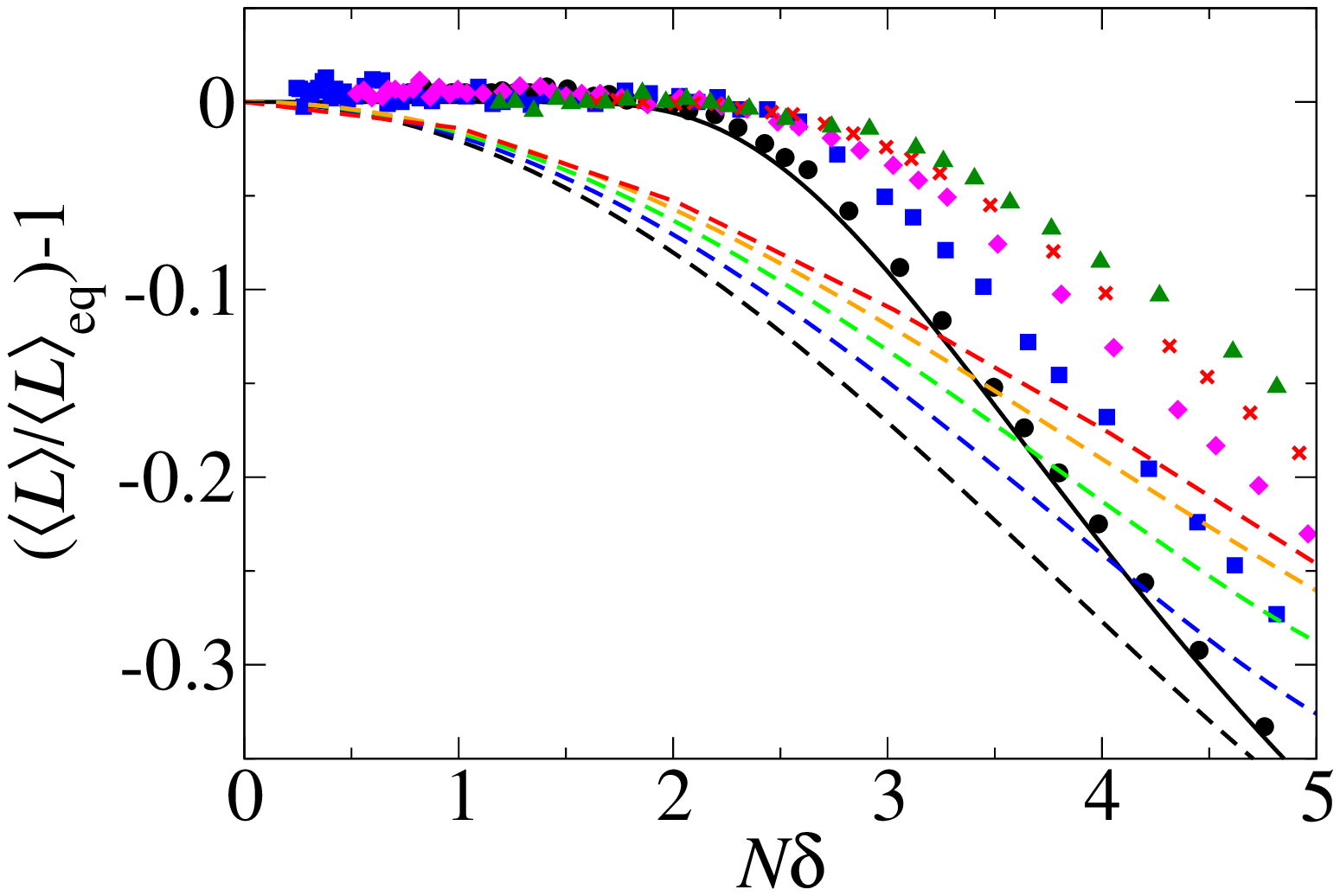}
\end{center}
\caption{\label{FigLoLeqm1NvsNdelta}
Relaxation of $\langle L \rangle$ (properly scaled) as a function of $N\delta$. Notice that the equilibrium state is reached for $N\delta\approx 2$. Symbols correspond to data from numerical simulations (with $N=101,11,5,3,2$ from top to bottom). The dashed lines depict the corresponding approximate solutions obtained from $\delta$-expansions for the first order moment up to terms of order $\delta^{12}$ (again with $N=101,11,5,3,2$ from top to bottom). The solid line corresponds to the solution provided by Eq.~\eqref{rhoseries} for the $N=2$ case. In all cases, $D=1$. The values of $\Delta$ for each $N$ are the same as in Fig.~\ref{FigLoLeqm1Nvs1oNdelta2}.}
\end{figure}

Finally, note that in Fig.~\ref{FigLoLeqm1NvsNdelta} the relaxation time $\tau_r$ needed to practically reach the equilibrium state scales in such a way that $N\delta_\text{eq}\approx 2$, where $\delta_\text{eq}=\Delta/\sqrt{4D\tau_r}$; in other words, $\tau_r\approx \langle L\rangle_\text{eq}^2/(4D)$, in agreement with the Rouse model \cite{Grosberg2002} (as pointed out in Sec.~\ref{appanres}).

\begin{figure}[t]
\begin{center}
\includegraphics[width=0.46\textwidth,angle=0]{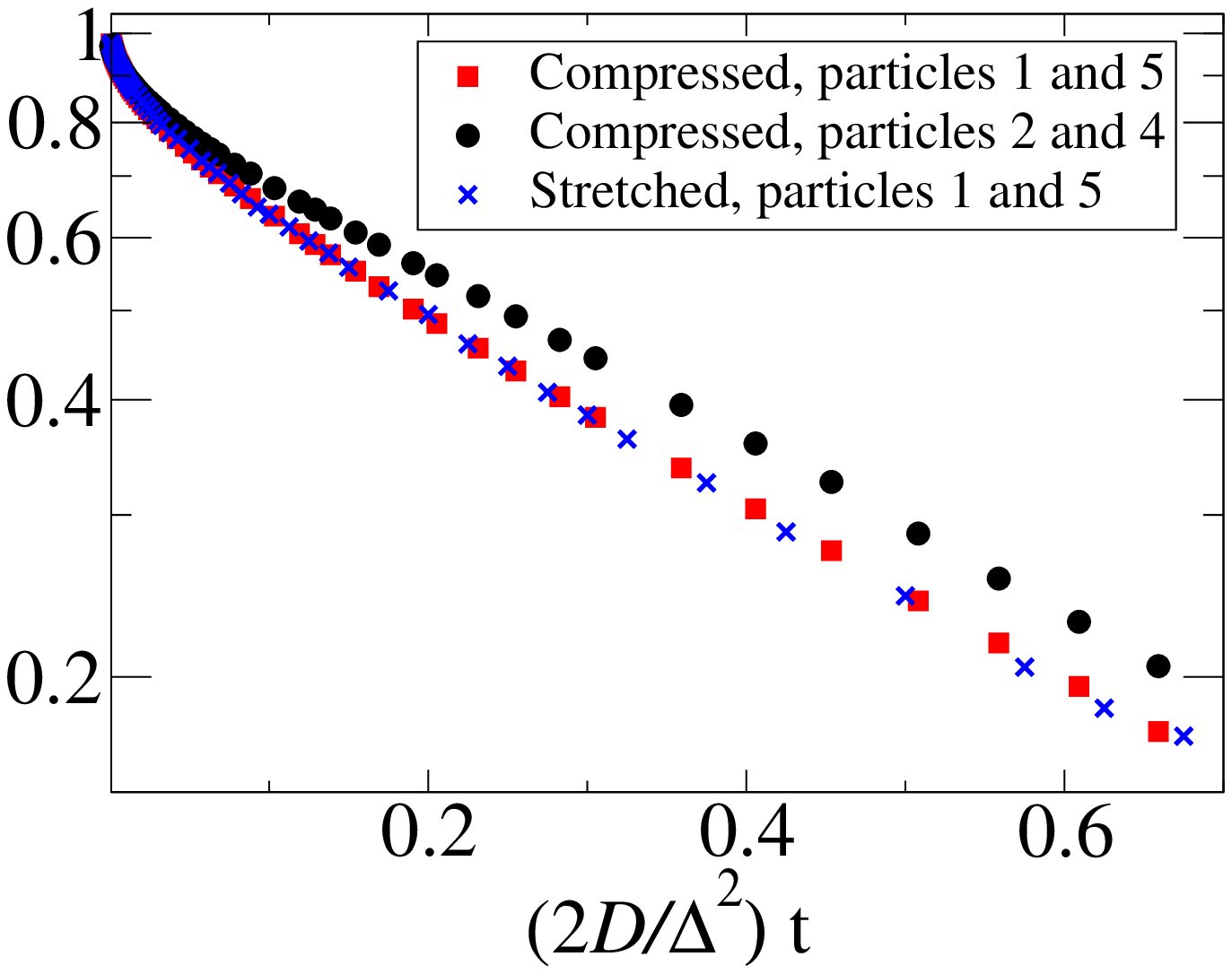}
\end{center}
\caption{\label{FigRelaxN5}
Relaxation of $|\langle L(t)\rangle-\langle L \rangle_\text{eq}|/\langle L\rangle_\text{eq}=|1-(\langle x_5\rangle-\langle x_1\rangle)/(2\Delta)|$ for a system with $N=5$  (in this case $\langle L\rangle_\text{eq}=2\Delta$) for stretched initial condition ($L(0)=4\Delta=2\langle L\rangle_\text{eq}$; crosses) and  compressed  initial condition ($L(0)=0$; squares)]. Relaxation of the distance between the \emph{neighbors} of the edge particles $[\Delta-(\langle x_4\rangle-\langle x_2\rangle)]/\Delta$  for  the compressed initial condition (circles). For the simulations, the values $D=1$ and
$\Delta=200$ have been used.
}
\end{figure}

\subsubsection{Role of the initial condition}

In Fig.~\ref{FigRelaxN5} we show the simulation results describing the relaxation towards the equilibrium state for $N=5$.  Specifically, the horizontal axis represents the values of $t$ in units of $\Delta^2/(2D)$, that is, the dimensionless time $t_\Delta=(2D/\Delta^2)\,t$. The vertical axis represents the different functions of the mean end-to-end distance $\langle L(t) \rangle$ (that is, the mean separation between the edge particles 1 and 5), so that all curves take the value one for $t_\Delta=0$ (see figure caption). We consider the case of a fully compressed initial condition treated so far (all particles start at $x=0$), but also another initial arrangement where the particles are separated from each other by the maximum distance $\Delta$ (fully stretched initial condition). Fig.~\ref{FigRelaxN5} shows that the relaxation for the fully stretched and fully compressed initial conditions \emph{is the same}. This behavior is also found in the Rouse model \cite{Grosberg2002}.

We also find that in the early and intermediate time regimes, the fit to a simple monoexponential decay is not satisfactory. As clearly seen in the semilog plot shown in Fig.~\ref{FigRelaxN5}, the numerical data deviate significantly from a straight line at short times, implying that the decay is more complicated than a simple exponential relaxation. This is in line with the analytical findings for $N=2$ in this regime,  which predict a multimodal decay [Eq.~\eqref{rhoseries}]. This is also similar to what occurs in polymer models, such as in the Rouse model, where the evolution of the polymer towards the equilibrium state is given in terms of a superposition of exponentially decaying modes, each of which has a different characteristic decay time \cite{Grosberg2002}.

From Fig.~\ref{FigRelaxN5} it is clear that the relaxation of the edge particles $1$ and $5$ is different from that of the particles $2$ and $4$. This is sensible because particles $2$ and $4$ have two nearest neighbors, whereas particles $1$ and $5$ only have a single neighbor to interact with during the relaxation process.

\section{Characterization of the equilibrium state}
\label{sec8}

So far, we have mainly discussed relaxation properties without giving many details about the final state beyond macroscopic properties such as the first moments of the end-to-end distance $L$ (recall that $\langle L\rangle =-2\langle x_1 \rangle=2\langle x_N \rangle$ for odd $N$) or the long-time asymptotic diffusion coefficient of the system as a whole. We have noted in Sec.~\ref{itform} that the diffusion coefficient of the tethered chain depends on $N$ as in the Rouse model for polymers. We also found in Sec.~\ref{appanres} that the long-time relaxation of the tethered chain scales with the diffusivity $D$ and with $N$ (or, equivalently, with the system length), as is also the case with the Rouse model.

In this section, we further explore the possible similarities between the tethered chain and an ideal polymer. To this end, we compute the number of states (microstates) compatible with a given size of the chain at equilibrium and infer the exact distribution of this quantity, as well as the entropy and Helmholtz free energy of our system. As it turns out, the tethered chain follows a linear relationship between an applied force and the change in the end-to-end distance (Hooke's law), as is the case in ideal polymer models such as the spring-bead chain (Rouse model), where the interaction between neighboring beads is mediated by harmonic springs.

This finding is surprising at first sight given the striking differences between the tethered (string-bead) chain and the spring-bead model. However, it can be rationalized as follows. In the absence of thermal noise, the spring-bead model has a well-defined mechanical equilibrium length (the length equal to the sum of the distances between successive beads when the springs connecting them are neither stretched nor compressed). However, there is no mechanical equilibrium length in the tethered chain. In fact, if the thermal noise were zero, one would find that the system length could take any value between zero and $N\Delta$, and the separation distances between nearest-neighbor beads would then take any value between 0 and $\Delta$. In other words, in the tethered chain, the beads do not rattle (owing to thermal noise) about a mechanical equilibrium position (determined by elastic forces between beads). Thus, the energy of this ``string-bead system'' does not depend on the specific configuration. In this respect, this one-dimensional model resembles ideal three-dimensional models (such as the freely rotating chain model) of polymers \cite{Rubinstein2003,Doi1986}.

\subsection{Microstates and associated equilibrium probabilities}
\label{micro}

In this section, we count the number of states (microstates) of a tethered chain with a given $N$ and $\Delta$ that are compatible with a chain length $L$ at equilibrium [In this section, for brevity, we use the notation $\langle L\rangle$ instead of the mean (or ensemble average) $\langle L\rangle_\text{eq}$].
Each microstate is characterized by the set $\left\{ r_1,\cdots,r_{N-1}\right\}$ of $N-1$ interparticle distances ($r_i=x_{i+1}-x_i$). The corresponding macroscopic variable is the length $L=\sum_{i=1}^{N-1} r_i$ of the chain.  The set of relevant macrovariables then is $\left\{ L, N,\Delta\right\}$.

For the reasoning below, we consider a microscopic lattice with unit mesh size $a=1$. In addition, we assume that each lattice site can be occupied by a single particle at most, and that each particle symmetrically attempts to perform jumps to nearest-neighbor sites in continuous time subject to this restriction and to the $\Delta$ constraint. When implementing the latter constraint we adopt the microscopic rule that the separation between any two neighboring particles cannot become equal to $\Delta$, that is, its maximum value is $\Delta-1$. Similarly, the minimum value of the separation between two neighboring particles is 1. Jump attempts implying the occupation of the same site or violation of the $\Delta$-constraint result in the resetting of the chosen particle to its starting position.

If one does not fix the value of $L$, the number of possible microstates (that is, the number of different relative positions $\left\{ r_1,\cdots,r_{N-1}\right\}$ of the $N$ particles in the lattice)  is $\Omega\equiv (\Delta-1)^{N-1}$.  Of course,  $\Omega=\sum_{L} \Omega(L,N,\Delta)$ where $\Omega(L, N,\Delta)$ denotes the number of
microstates with a given value of $L$ (and $N$ and $\Delta$).

The task of computing $\Omega(L,N,\Delta)$ can be addressed using standard combinatorial arguments. Note that $\Omega(L,N+1,\Delta+1)$ can be seen as the number of different ways, $\left\{r_1,\cdots,r_{N}\right\}$, to obtain the sum $L=\sum_{i}^N r_i$  (with $N\le L\le N\Delta$) when rolling $N$ dice if the outcome $r_i$ of each die is restricted to the set of values $1,2,\ldots \Delta$.  The solution to this classic combinatorial problem is \cite{Uspensky1937}:
\begin{equation}
\Omega(L, N+1,\Delta+1)=  \sum _{k=0}^{\left\lfloor \frac{L-N}{\Delta}\right\rfloor } (-1)^k \binom{N}{k} \binom{L-k
   \Delta-1}{N-1}.
\end{equation}

Because the probability of each microstate $\left\{ r_1,\cdots,r_{N}\right\}$ is the same, the probability of having a chain of length $L$ is   $p(L,N+1,\Delta+1)=\Omega(L,N,\Delta)/\Omega$, that is,
\begin{equation}
\label{pLNdis}
p(L,N+1,\Delta+1)=\Delta^{-N} \sum _{k=0}^{\left\lfloor \frac{L-N}{\Delta}\right\rfloor } (-1)^k \binom{N}{k} \binom{L-k
   \Delta-1}{N-1}.
\end{equation}
For a chain with macrovariables $\left\{ L, N+1,\Delta+1\right\}$ the mean value of $L$ of the system length is $ \langle L\rangle = N(\Delta+1)/2 $ and its variance is $\sigma^2=N(\Delta^2-1)/12$. Note that, because $L$ comes from the sum of $N$ independent variables, $r_i$, the mean and variance of $L$ are the mean, $(\Delta+1)/2$, and variance, $(\Delta^2-1)/12$, of $r_i$ multiplied by a factor $N$.

Equation \eqref{pLNdis} provides the exact probability of finding a tethered chain of length $L$. However,  it is not very useful when $N$ is very large.  Fortunately, in the large-$N$ limit this probability can be well approximated in the form of a Gaussian distribution centered on $ \langle L\rangle$ and with variance $\sigma^2$.
 For large $N$ and $\Delta$ one can approximate the average and variance of the length by  $N\Delta/2$ and   $N\Delta^2/12$, respectively, and we obtain
\begin{equation}
\label{pLGauss1}
p(L,N,\Delta) \approx
   \frac{1}{ \sqrt{\pi N \Delta^2/6}} \exp\left[ -\frac{(L-N\Delta/2)^2}{N\Delta^2/6} \right].
\end{equation}

\subsection{Statistical thermodynamics}

We now assume that the diffusive (Brownian) motion of the particles (beads) of the tethered (or string-bead) chain is due to the thermal forces coming from the medium (bath) in which the chain is located. According to Einstein's fluctuation-dissipation relation, $D=k_B T/\xi$, where $k_B$ is the Boltzmann constant, $T$ is the temperature of the bath, and $\xi$ is the friction coefficient.

The entropy of the tethered chain  $S(L,N,\Delta)=k_B \ln \Omega(L,N,\Delta)$ is obtained from the relation $\Omega(L,N,\Delta)=p(L,N,\Delta)\Delta^{N-1}$ and Eq.~\eqref{pLGauss1}:
\begin{equation}
\label{SLND}
S(L,N,\Delta)=  - k_B\frac{6(L-\langle L \rangle)^2}{N\Delta^2} - k_B\ln\left(\frac{\pi N\Delta^2}{6}\right) +  k_B(N-1)\ln\Delta.
\end{equation}
The Helmholtz free energy of the chain is $F=U-TS$  where $U$ denotes the energy of the chain. As in models of ideal polymers  \cite{Rubinstein2003}, $U$ does not depend on $L$ because the particles (beads) in the allowed configurations do not have interaction energy.  This means that, according to Eq.~\eqref{SLND}, the free energy of the tethered chain increases quadratically with $L$.  This implies, as in the ideal chain polymer models, that the elasticity of the tethered chain satisfies Hooke's law: the force  $f$ required to stretch the chain a length $x=L-\langle L\rangle$ from the averaged (equilibrium) value $\langle L\rangle$ is proportional to $x$:
\begin{equation}
 f = k_B T \,\frac{\partial F}{\partial L}= \frac{6 k_B T}{N\Delta^2} \left(L-  \langle L \rangle \right).
\end{equation}
The associated (entropic) spring constant is  $C_L={6 k_B T}/(N\Delta^2)$.  Taking into account that $\ell=\langle r_i\rangle =\Delta/2$ is the mean particle separation at equilibrium, the elastic constant can be rewritten as  $C_L={3 k_B T}/(2N\ell^2)$, that is just the expression of the spring constant for ideal polymers \cite{Rubinstein2003}. Note that the force $f$ and the associated  spring constant $C_L$ are completely of entropic origin because there is no elastic interaction and hence no springs modeling mechanical forces between the beads in our string-bead chain.

\section{Summary and outlook}
\label{sec9}

We briefly recap the main results of this study. Inspired by Aslangul's work \cite{Aslangul1998}, we have considered a one-dimensional system of $N$ Brownian walkers undergoing SFD, but subject to the additional constraint that the separation distance of any two walkers cannot exceed the threshold value $\Delta$ (we have coined the term ``tethered walkers'' to reflect this constraint). The dynamics displays remarkable differences with respect to the $\Delta=\infty$ case treated by Aslangul. For finite $\Delta$, a factorization ansatz reasonably describes the behavior of reduced pdfs over a wide range of parameters, but fails to capture other features such as long-time relaxation properties. As we have seen, an exact solution can only be found for the $N=2$ case (an exact solution for the on-lattice case is also available in the Laplace space). The first positional moment relaxes exponentially to the equilibrium state at long times, as opposed to the inverse power law behavior predicted by our ansatz. The MSD of either particle behaves linearly at short and long times, but with different (yet $\Delta$-independent) diffusion coefficients (at intermediate times, transient anomalous diffusion is observed). Another effect of introducing a maximum interparticle separation $\Delta$ is the expected enhancement of particle-particle correlations, which has been precisely quantified from the exact solution.

For arbitrary $N > 2$, an iterative procedure has been developed to compute the one-particle pdf of any particle and the positional moments (in the latter case, as a power series in terms of $\delta = \Delta/\sqrt{4Dt}$).
The resulting single-particle pdfs accurately capture the corresponding simulation results, both qualitatively and, in many cases (especially for short and long times), quantitatively.

In the long-time regime, every particle in the system moves with an effective diffusivity $\propto 1/N$, as opposed to the large-$N$ behavior $\propto 1/\ln(N)$ observed in the $\Delta=\infty$ case for the edge particles \cite{Aslangul1998}. According to our analytical findings for $N=2$, the positional first-order moment for each particle decays exponentially at long times; this result is also obtained in numerical simulations for $N>2$, the relaxation time being $\tau_r=N^2\Delta^2/4D=\langle L\rangle_\text{eq}^2/D$. Finally, we discussed the statistical properties of the equilibrium state reached when $t\to \infty$ (in this state, the average system length does not depend on time). Using combinatorial arguments, we evaluated the number of microstates compatible with a given system length, the probability of this length, the associated mean and variance, the entropy of the chain, and the force required to change it by a given amount. As in the Rouse spring-bead model, this (entropic) force is Hookean despite the non-analiticity of the interaction potential imposed by the $\Delta$-constraint.

This problem can be extended in several ways. For example, one could investigate the role of different initial conditions in a more comprehensive manner. One could also consider the effect of an external force \cite{Barkai2009,Barkai2010,Lapolla2018,Lapolla2019}, which amounts to replacing the unbiased Gaussian solution for each individual particle in the factorization ansatz with the corresponding biased propagator. Finally, our approach can also be adapted to the situation where the maximum separation $\Delta_L$ between a given particle and its left neighbor is different from the maximum allowed distance $\Delta_R$ to the right neighbor. Moreover, one could even consider the case where either $\Delta_L$ or $\Delta_R$ becomes infinite, whereas the other unilateral reach remains finite.

\section*{Acknowledgements}
S. B. Y. and E. A. acknowledge financial support from Grant
No. PID2020-112936GB-I00 funded by \\
MCIN/AEI/10.13039/501100011033. A. B. thanks the Department of Physics at the University of Extremadura for its hospitality.
\section*{Author declarations}
\subsection*{Conflict of interest declaration}
The authors have no conflicts to disclose.
\subsection*{Author contributions}
S. B. Yuste: Conceptualization (equal), Formal analysis (equal), Investigation (equal), Methodology (equal), Writing-original draft (equal), Writing-review \& editing (equal).  A. Baumgaertner: Methodology (equal), Formal analysis (equal), Investigation (equal), Writing-review \& editing (equal). E. Abad: Conceptualization (equal), Formal analysis (equal), Investigation (equal), Methodology (equal), Writing-original draft (equal), Writing-review \& editing (equal).
\section*{Data availability statement}
The data that support the findings of this study are available from the corresponding author upon reasonable request.

\appendix

\section{$\bm{N=2}$ case with a finite initial particle separation}

In what follows, our aim is to compute the two-particle distribution $p(x_1,x_2,t;\Delta)$ for the case in which the initial interparticle separation $d$ is strictly positive, thereby generalizing the $d=0$ case described by Eq.~\eqref{exactpdf} (without loss of generality, we take $x_1(0)=0$ and $x_2(0)=d$). In the $\Delta=\infty$ case, the solution for the corresponding problem with a reflecting horizontal boundary is given by:
\begin{equation}
p_\infty^\text{hor}(x_1,x_2,t)=\frac{e^{-(x_1-\frac{d}{\sqrt{2}})^2/4Dt}}{4\pi Dt}\left(e^{-(x_2-\frac{d}{\sqrt{2}})^2/4Dt}+e^{-(x_2+\frac{d}{\sqrt{2}})^2/4Dt} \right) \,\Theta(x_2).
\end{equation}
This solution satisfies the initial condition $p(x_1,x_2, 0)=\delta(x_1-d/\sqrt{2})\,\delta(x_2-d/\sqrt{2})$ as well as the zero normal flux condition $\left. \partial_{x_2} \, p(x_1,x_2,t)\right|_{x_2=0}$. The tilted solution
\begin{equation}
p_\infty (x_1,x_2,t)=\frac{e^{-(x_1+x_2-d)^2/8Dt}}{4\pi Dt}\left(e^{-(x_2-x_1-d)^2/8Dt}+e^{-(x_2-x_1+d)^2/8Dt} \right) \,\Theta(x_2-x_1)
\end{equation}
satisfies the imposed initial condition $p_\infty(x_1,x_2, 0)=\delta(x_1)\,\delta(x_2-d)$, as well as the reflecting boundary condition
$\left.{\bm n}_\perp \cdot \nabla p_\infty(x_1,x_2,t)\right|_{x_2=x_1}=0$.
In the case of a finite reach $\Delta$, the solution is more involved, but can again be obtained by a procedure similar to that for the case $x_1(0)=x_2(0)=0$. To this end, one must consider the solution of the two-dimensional diffusion problem for two reflecting horizontal boundaries at $x_2=0$ and $x_2=\Delta/\sqrt{2}$ and the particle starting at $(x_1(0),x_2(0))=(d/\sqrt{2},d/\sqrt{2})$. This particular initial condition is still covered by the general solution given in Ref.~\onlinecite{CarslawJaeger1959} (Eq.~(19) on p. 374):
\begin{equation}
\label{rotsol0}
p^\text{hor}(x_1,x_2,t;\Delta)= \sum_{n=-\infty}^\infty \!\!\left(e^{-(\sqrt{2}n\Delta+ x_2-\frac{d}{\sqrt{2}})^2/4Dt}+ e^{-(\sqrt{2}n\Delta+ x_2+\frac{d}{\sqrt{2}})^2/4Dt} \right)
\frac{e^{-(x_1-\frac{d}{\sqrt{2}})^2/4Dt}}{4\pi Dt}R(x_2,{\scriptstyle \frac{\Delta}{\sqrt{2}}}).
\end{equation}
 The solution for the initial condition $(0,d)$ is then determined by tilting the above expression anticlockwise, which gives
\begin{align}
p(x_1,x_2,t;\Delta)=&
\sum_{n=-\infty}^\infty \left(e^{-(2n\Delta+ x_2-x_1-d)^2/8Dt}+ e^{-(2n\Delta+ x_2-x_1+d)^2/8Dt} \right)\times\nonumber \\
&
\label{rotsol1}
\times\frac{e^{-(x_1+x_2-d)^2/8Dt}}{4\pi Dt} R(x_2-x_1,\Delta).
\end{align}
The above expression can be rewritten in terms of elliptic functions as follows:
\begin{align}
p(x_1,x_2,t)=&\frac{ e^{-(x_2+x_1-d)^2/8Dt}}{\Delta (8 \pi D t)^{1/2}} \left[
\vartheta_3\left(\frac{\pi (x_2-x_1-d)}{2\Delta}, e^{-2\pi^2 Dt/\Delta^2}\right)+\right.\nonumber\\
&
\label{rotsol2}
\left.+\vartheta_3\left(\frac{\pi (x_2-x_1+d)}{2\Delta},
e^{-2\pi^2 Dt/\Delta^2}\right)\right] R(x_2-x_1,\Delta).
\end{align}

\section{arbitrary odd-order moments for $\bm{N=2}$}
In this appendix, we obtain analytic expressions for the odd-order moments of the approximate single-particle pdf proposed (factorization ansatz) in Sec.~\ref{sec:pdfN2approx}. To obtain the $n$-th order moment when $N=2$, the following expression must be evaluated:
\begin{equation}
\label{oddmoment}
\langle u_2^n \rangle=\frac{1}{\sqrt{\pi}\,\text{Erf}(\delta/\sqrt{2})}\left[ \int_{-\infty}^\infty du_2 \, u_2^n \,
\text{Erf}\left(u_2\right) \,e^{-u_2^2}-\int_{-\infty}^\infty du_2 \, u_2^n \,\text{Erf}\left(u_2-\delta \right)\, e^{-u_2^2}\right].
\end{equation}
The analytic form of the first integral in the rhs of \eqref{oddmoment} is known. One has
\begin{equation}
 \int_{-\infty}^\infty du_2 \, u_2^n \,
\text{Erf}\left(u_2\right) \,e^{-u_2^2}=\frac{2}{\sqrt{\pi}} \, {}_2F_1 (1/2, n/2+1, 3/2,-1) \Gamma(n/2+1).
\end{equation}
On the other hand, the second integral in the rhs of Eq.~\eqref{oddmoment} can be expressed as the sum of Laplace
transforms $\widetilde{g}(s)$ of the function $g(z)=z^m \, \text{Erf}(a \sqrt{z}+b)$ (where $a$ and $b$ are real constants and $m$ is a positive integer). To this end, one performs the change of variable $z=u_2^2$ and subsequently setting the Laplace variable $s$ equal to one. Given that  \cite{Prudnikov1992}
\begin{equation}
\widetilde{g}(s)=(-1)^m  \frac{d^m}{ds^m}\left[\frac{1}{s}\text{Erf}(b)+\frac{a}{s\sqrt{s+a^2}}e^{-b^2 s/(s+a^2)}
\text{Erfc}\left(\frac{ab}{\sqrt{s+a^2}}\right)\right],
\end{equation}
we find, upon performing the pertinent simplifications,
\begin{align}
\langle u_2^n \rangle=-\langle u_1^n \rangle=& \frac{1}{\sqrt{\pi}\,\text{Erf}(\delta/\sqrt{2})}\left[\frac{2}{\sqrt{\pi}} \,\, {}_2F_1 (1/2, n/2+1, 3/2,-1) \Gamma(n/2+1)\right.\nonumber \\
&\left.\left.-(-1)^{(n-1)/2} \frac{d^{(n-1)/2}}{ds^{(n-1)/2}} \frac{1}{s \sqrt{1+s}}e^{-\delta^2 s/(1+s)}
\right|_{s=1}\right], \quad n \mbox{ odd}.
\end{align}

\section{On-lattice solution for $\bm{N=2}$}

It is interesting to characterize the differences between the dynamics of the continuum two-walker system and its corresponding lattice counterpart. An on-lattice solution for the $\Delta=\infty$ case was obtained by Aslangul \cite{Aslangul1999}. He showed that lattice effects may persist for up to unexpectedly long times. For a finite $\Delta$, we expect that the associated cage effect acts as an additional perturbation that makes it even more difficult for the particles to reach a regime in which their motion is well described by continuum diffusion.

To study the impact of lattice effects on the behavior of positional moments, we consider that both particles start from contiguous sites on an infinite one-dimensional lattice with mesh size $a$. Let us label lattice sites with consecutive integers. We assume that one of the particles starts at the origin (site $0$), and the other starts at the right neighbor site (site $1$). Each particle performs a nearest-neighbor random walk in continuous time with a symmetric hopping rate $2W$ (the one-sided jump rate is thus $W$, and consequently the single-particle diffusion coefficient becomes $D=Wa^2$). Following Aslangul \cite{Aslangul1999}, we neglect events in which both particles jump during the same interval $dt$; thus, the dynamics are asynchronous. The single file constraint is implemented by precluding double site occupation; specifically, when a particle attempts to hop on a site that is already occupied by the other, the jump attempt is rejected, and the particle remains at the same position. On the other hand, we implement the $\Delta$-constraint on the discrete support as follows.  The interparticle separation before an attempted particle jump is $(n_2-n_1)a$, where $n_1$ and $n_2$ denote the lattice sites at which walker $1$ and walker $2$ are located. If the interparticle separation becomes larger or equal to $\Delta$ as a result of an attempted jump by either particle, then no hopping occurs (this is consistent with the interaction rule used in Sec. \ref{micro}).

As in the continuum case (see \ref{subs-exact-moments}), it is possible to decompose the joint motion of the pair of particles in c.o.m. and relative coordinates. The evolution of the relative coordinate $L$ may be viewed as that of a random walker of diffusivity $2D=2Wa^2$ performing a nearest-neighbor walk on a one-dimensional lattice of spacing $a$ and two reflecting end sites $0$ and $N_\Delta=\Delta/a$ (we assume that $\Delta$ is a multiple of the lattice spacing $a$). Let us denote by $P_j^L(t)$ the probability that such a walker is found at site $j$ (with $0<j<N$) at time $t$, and by $\widetilde{P}_j^L(s)
\equiv \int_0^\infty dt \, e^{-st} P_j^L(t)$ its Laplace transform with respect to time. Then, we obtain the result \cite{Mazo1987}:
\begin{equation}
\label{laptransprob}
\widetilde{P}_j^L(s)=\frac{\cosh[\beta (N_\Delta-j-\frac{1}{2})]\cosh(\frac{\beta}{2})}
{2W\sinh(\beta)\sinh[\beta(N_\Delta-1)]},
\end{equation}
where $\beta=\cosh^{-1}(1+(4W)^{-1}s)$.
From Eq.~\eqref{laptransprob}, the calculation of Laplace-transformed (integer) moments of the interparticle separation $L$ is immediate:
\begin{equation}
\langle \widetilde{L}^n(s) \rangle=a^n\sum_{j=1}^{N_\Delta-1} j^n \widetilde{P}_j^L(s).
\end{equation}
It does not seem possible to derive an exact analytical expression for these moments in the time domain because neither the above sum nor the individual $\widetilde{P}_j^L(s)$'s are easy to invert. For $n=1$, the sum evaluates to the following expression:
\begin{equation}
\label{Laplaceseplength}
\langle \widetilde{L}(s) \rangle=\frac{\sqrt{2 a^2 W}\sech[\frac{N_\Delta-1}{2}\cosh^{-1}
(1+\frac{s}{4W})]\sinh[\frac{N_\Delta}{2}\cosh^{-1}(1+\frac{s}{4W})]}{s^{3/2}}.
\end{equation}
 Similarly, for $n=2$, a lengthy expression is obtained in terms of hyperbolic trigonometric functions. In general, one must resort to numerical inversion to obtain results in the time domain. However, asymptotic analytic results for the short-time behavior can be obtained via Tauberian theorems. For $n=1$, say, one has the general result
\begin{equation}
\label{Lexp}
\langle L(t) \rangle=2\langle x_2 \rangle-a=a\left\{1+2Wt-\sum_{j=2}^\infty (-1)^j
c_j (Wt)^j\right\},
\end{equation}
where the coefficients $c_j>0$ do not depend on $N_\Delta$ up to $j\ge N_\Delta-1$. Thus, a finite value of $\Delta$ impacts the term of order $(Wt)^{N_\Delta-1}$ and the higher-order terms, while the lower-order terms remain the same as for $\Delta=\infty$ (of course, the effect of a finite $\Delta$ is to decrease $\langle L(t) \rangle$ at a given time $t$). The first drift term [the second term in the rhs of Eq.~\eqref{Lexp}] is linear in time, in agreement with Aslangul's result for $\Delta=\infty$ (see Ref.~\onlinecite{Aslangul1999}). As expected, at early times the particle drift is slowed down with respect to the off-lattice case, since the dominant term of the lattice solution is proportional to $\sqrt{t}$, and therefore faster than the linear time dependence obtained for on-lattice diffusion.

We now examine the behavior of the c.o.m. motion. In this case, one may regard its motion as that of a (fictitious) random walker with the same jump rate $2W$ as that for the interparticle distance, but effectively moving on an infinite one-dimensional lattice with spacing $a/2$ and starting at site $1$. The diffusivity of this walker is $2W(a/2)^2=D/2$. In this case, the expression of the Laplace transformed probability $\widetilde{P}_m^X(s)\equiv \int_0^\infty dt \, e^{-st} P_m^X(t)$ for the walker to be at site $m$ is known. One has \cite{Mazo1987}
\begin{equation}
\widetilde{P}_m^X(s)=\frac{e^{-\beta |m-1|}}{4W\sinh\beta}.
\end{equation}
In general, the evaluation of the associated positional moments of lattice walks is usually easier if one first carries out the corresponding sums in the Laplace space, and then inverts the resulting expressions either exactly or in asymptotic regimes amenable to analysis via Tauberian theorems. However, the inverse Laplace transform of $\widetilde{P}_m^X(s)$ is well-known in the present case of a simple nearest-neighbor walk:
\begin{equation}
P_m^X(t)=e^{-4Wt} I_{|m-1|}(4Wt),
\end{equation}
where $I_n(\cdot)$ denotes the $n$-th order modified Bessel function of the first kind. The $P_m^X$'s are indeed normalized, since $\sum_{m=-\infty}^\infty I_{|m-1|}(x)=\sum_{m=-\infty}^\infty I_{|m|}(x)=\exp(x)$. Now, the discrete moments of the c.o.m. can be computed from their Laplace transformed definition
\begin{equation}
\langle \widetilde{X}^n(s) \rangle=\left(\frac{a}{2}\right)^n\sum_{j=-\infty}^{\infty} j^n \widetilde{P}_j^X(s),
\end{equation}
or directly in the time domain:
\begin{equation}
\langle X^n(t) \rangle=\left(\frac{a}{2}\right)^n\sum_{j=-\infty}^{\infty} j^n P_j^X(t).
\end{equation}
The formulas given in subsection \ref{subs-exact-moments} continue to hold in the discrete case, except for the fact that one must now take into account that the first moment of the c.o.m. no longer vanishes,
$\langle X \rangle= a/2$. Thus, one has
\begin{align}
\langle x_2 \rangle &=\langle X \rangle+\frac{\langle L \rangle}{2}=\frac{a}{2}+\frac{\langle L \rangle}{2}, \label{lattfm}\\
\langle x_2^2 \rangle &= \langle X^2 \rangle+\langle X \rangle \langle L \rangle+\frac{\langle L^2 \rangle}{4}=\langle X^2 \rangle+
\frac{a}{2} \langle L \rangle +\frac{\langle L^2 \rangle}{4}, \\
\text{Var}(x_2)&=\text{Var}(X) +\frac{\langle L^2 \rangle-\langle L \rangle^2}{4}=\langle X^2 \rangle-\frac{a^2}{4}+\frac{\langle L^2 \rangle-\langle L \rangle^2}{4},\\
C(t)&=\langle X^2 \rangle -\frac{\langle L^2 \rangle-\langle L \rangle^2}{4}.
\end{align}
The second moment
\begin{equation}
\langle X^2 \rangle=\frac{a^2}{4}\sum_{j=-\infty}^{\infty} j^2 P_j^X(t)
\end{equation}
of the c.o.m. coordinate can be explicitly computed in the time domain. Using the identity
\begin{equation}
\sum_{m=-\infty}^\infty m^2 I_{|m-1|}(x)= e^x (x+1),
\end{equation}
one easily finds $\langle X^2 \rangle =\frac{a^2}{4}+ Wa^2 t$, that is
\begin{equation}
\langle X^2 \rangle-\langle X \rangle^2=W a^2 t
\end{equation}
Finally, to compare with the continuum case, one must set $W=D/a^2$ in all of the above formulas (this expression leads to the c.o.m. diffusivity $D/2$, as it should). To perform the comparison in terms of scaled variables, one sets (as before) $\langle u_2^n \rangle= \langle x_2^n \rangle/(4Dt)^{n/2}$.

\begin{figure}[ht]
\begin{center}
 \includegraphics[width=0.46\textwidth,angle=0]{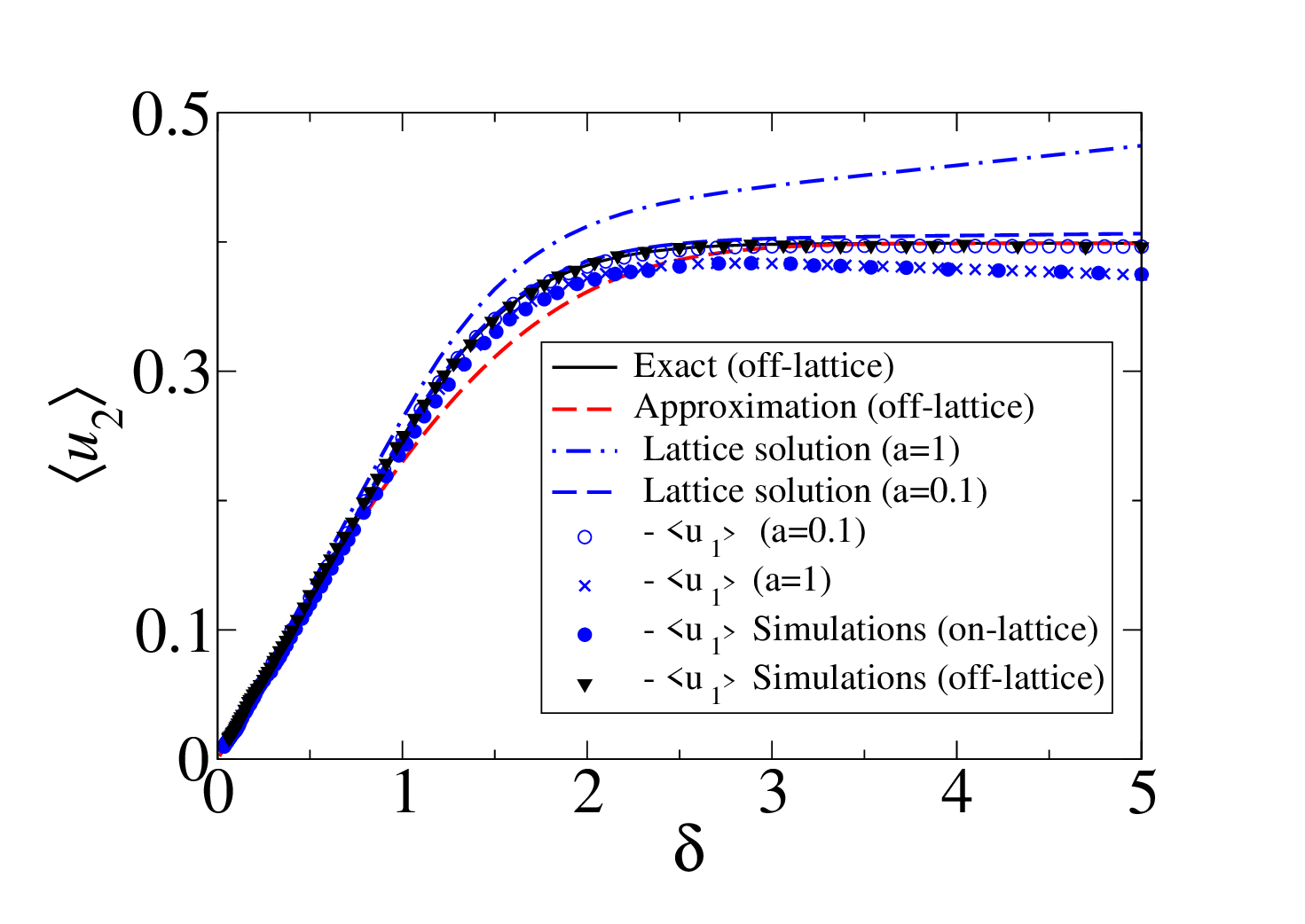}
\end{center}
\caption{\label{u-continuum-lattice} Lattice solution for the scaled first-order moment $\langle u_2 \rangle$ of the right particle obtained from Eq.~\eqref{lattfm} and the numerical inversion of Eq.~\eqref{Laplaceseplength}. We have used $\Delta=50$. A comparison with the off-lattice solution is also included.}
\end{figure}

A comparison between the off-lattice solution (continuum) and the on-lattice solution is shown in Fig.~\ref{u-continuum-lattice} at the level of the scaled first-order moment $\langle u_2(\delta)\rangle$. In the case of the lattice solution, one of course no longer has $\langle u_2 \rangle=-\langle u_1 \rangle$, but rather $\langle u_2 \rangle=-\langle u_1 \rangle+(\delta/\Delta)a$, and one must therefore consider these two solutions separately. However, both are seen to converge to the off-lattice solution when one takes the diffusive limit by simultaneously letting the mesh size $a$ (in this case identical to the initial particle separation) go to zero and the unilateral hopping rate $W$ increase so as to maintain the value of the diffusivity $D=a^2 W$ constant (in the case of Fig.~\ref{u-continuum-lattice}, we have taken $D=1$). Recall that a fixed $\Delta$ implies decreasing values of $\delta$ with increasing time (large values of $\delta$ therefore correspond to early times). The differences between the on- and off-lattice solution are most pronounced for large values of $\delta$, corresponding to the early time regime. The results are confirmed by MC simulations (see Fig.~\ref{u-continuum-lattice}).


\begin{thebibliography}{99}

\bibitem{Ryabov2016} A. Ryabov, \emph{Stochastic Dynamics and Energetics of Biomolecules} (Springer, Switzerland, 2016).

\bibitem{Benichou2018} O. B\'enichou, P. Illien, G. Oshanin, A. Sarracino, and R. Voituriez, ``Tracer diffusion in crowded narrow channels,'' J. Phys.: Condens. Matter {\bf 30}, 443001 (2018).

\bibitem{Gupta1995} V. Gupta, S. S. Nivarthi, A. V. McCormick, and H. T. Davis, ``Evidence for single file diffusion of ethane in the molecular sieve $AlPO_4-5$,'' Chem. Phys. Lett. {\bf 247}, 596 (1995).

\bibitem{Meersmann2000} T. Meersmann, J. W. Logan, R. Simonutti, S. Caldarelli, A. Comotti, P. Sozzani, L. G. Kaiser, and A. Pines, ``Exploring single-file diffusion in one-dimensional nanochannels by laser-polarized Xe-129 NMR spectroscopy,'' J. Phys. Chem. A {\bf 104}, 11665 (2000).

\bibitem{Jepsen1965} D. W. Jepsen, ``Dynamics of a Simple Many-Body System of Hard Rods,'' J. Math. Phys. {\bf 6}, 405 (1965).

\bibitem{Levitt1973} D. G. Levitt, ``One-dimensional Time-Dependent Distributions,'' J. Stat. Phys.  {\bf 7}, 329 (1973).

\bibitem{Chowdhury2000} D. Chowdhury, L. Santen, and A. Schadschneider, ``Statistical physics of vehicular traffic and some related systems,'' Phys. Rep. {\bf 329}, 199 (2000).

\bibitem{Graneli2006} A. Graneli, C. Yeykal, R. B. Robertson, and E. C. Greene, ``Long-distance lateral diffusion of human Rad51 on double-stranded DNA,'' Proc. Natl. Acad. Sci. U.S.A. {\bf 103}, 1221 (2006).

\bibitem{Li2009} G.-W. Li, O. G. Berg, and J. Elf, ``Effects of macromolecular crowding and DNA looping on gene regulation kinetics,'' Nat. Phys. {\bf 5}, 294 (2009).

\bibitem{John2009} A. John, A. Schadschneider, D. Chowdhury, and K. Nishinari, ``Trafficlike collective movement of ants on trails: absence of a jammed phase,'' Phys. Rev. Lett. {\bf 102}, 108001 (2009).

\bibitem{Bressloff2013}  P. C. Bressloff and J. M. Newby, ``Stochastic models of intracellular transport,'' Rev. Mod. Phys. {\bf 85}, 135 (2013).

\bibitem{Muthukumar2011} M. Muthukumar, \emph{Polymer Translocation} (CRC Press, Boca Raton, Florida, 2011).

\bibitem{Perkins1994} T. T. Perkins, D. E. Smith, and S. Chu, ``Direct observation of tube-like motion of a single polymer chain,'' Science {\bf 264}, 819 (1994).

\bibitem{Hahn1995}  K. Hahn, J. K\"arger, and V. Kukla, ``Single-file diffusion observation,'' Phys.Rev. Lett. {\bf 76}, 2762 (1995).

\bibitem{Kukla1996} V. Kukla, J. Kornatowski, D. Demuth, I. Girnus, H. Pfeifer, L. V. C. Rees, S. Schunk, K. K. Unger, and J. K\"arger, ``NMR studies of single-file diffusion in unidimensional channel zeolites,''
Science {\bf 272}, 702 (1996).

\bibitem{Keffer1999} D. Keffer, ``The temperature dependence of single-file separation mechanisms in onedimensional nanoporous materials,'' Chem. Eng. J. {\bf 74}, 33 (1999).

\bibitem{Kaerger2012} J. K\"arger, D. M. Ruthven, and D. N. Theodorou,
\emph{Diffusion in Nanoporous Materials (Vol. 1)}, (Wiley, Weinheim, 2012).

\bibitem{Cheng2007} C.-Y. Cheng and C. R. Bowers, ``Observation of Single-File diffusion in dipeptide nanotubes by continuous-flow hyperpolarized xenon-129 NMR spectroscopy,'' ChemPhysChem {\bf 8}, 2077 (2007).

\bibitem{Wei2000} Q.-H. Wei, C. Bechinger, and P. Leiderer, ``Single-file diffusion of colloids in one-dimensional channels,'' Science {\bf 287}, 625 (2000).

\bibitem{Zia2010} R. N. Zia and J. F. Brady, ``Single-particle motion in colloids: force-induced diffusion,'' J. Fluid Mech. {\bf 658}, 188 (2010).

\bibitem{Kollmann2003} M. Kollmann, ``Single-file Diffusion of Atomic and Colloidal Systems,'' Phys. Rev. Lett. {\bf 90}, 180602 (2003).

 \bibitem{Rodenbeck1998} C. R\"odenbeck, J. K\"arger, and C. Hahn, ``Calculating exact propagators in syngle file systems via the reflection principle,'' Phys. Rev. E {\bf 57}, 4382 (1998).

\bibitem{Aslangul1998} C. Aslangul, ``Classical diffusion of $N$ interacting particles in one dimension: General results and asymptotic laws,''  Europhys. Lett. {\bf 44}, 284 (1998).

\bibitem{Aslangul1999} C. Aslangul, ``Diffusion of two repulsive particles in a one-dimensional lattice,''
J. Phys. A Math. Gen. {\bf 32}, 3993 (1999).

\bibitem{Lizana2008} L. Lizana and T. Ambj\"ornsson, ``Single file diffusion in a box,'' Phys. Rev. Lett. {\bf 100}, 200601 (2008).

\bibitem{Lizana2009} L. Lizana and T. Ambj\"ornsson, ``Diffusion of finite-sized hard-core interacting particles in a one-dimensional box: Tagged particle dynamics,'' Phys. Rev. E {\bf 80}, 051103
(2009).

\bibitem{Illien2013} P. Illien, O. B\'enichou, C. Mej\'{i}a-Monasterio, G. Oshanin, and R. Voituriez, ``Active Transport in Dense Diffusive Single-File Systems,'' Phys. Rev. Lett. {\bf 111}, 038102 (2013).

\bibitem{Poncet2021} A. Poncet, A. Grabsch, P. Illien, and O. B\'enichou, ``Generalized Correlation Profiles in Single-File Systems,'' Phys. Rev. Lett. {\bf 127}, 220601 (2021).

\bibitem{Krapivsky2014} P. L. Krapivsky, K. Mallick, and T. Sadhu, ``Large Deviations in Single-File Diffusion,'' Phys. Rev. Lett. {\bf 113}, 078101 (2014).

\bibitem{Krapivsky2015} P. L. Krapivsky, K. Mallick, and T. Sadhu, ``Tagged Particle in Single-File Diffusion,'' J. Stat. Phys. {\bf 160}, 885 (2015).

 \bibitem{Lapolla2018} A. Lapolla and A. Godec, ``Unfolding tagged particle histories in single-file diffusion: exact single- and two-tag local times beyond large deviation theory,'' New J. Phys. {\bf 20}, 113021 (2018).

\bibitem{Mallmin2021} E. Mallmin, J. du Buisson, and H. Touchette, ``Large deviations of currents in diffusions with reflective boundaries,'' J. Phys. A: Math. Theor. {\bf 54}, 295001 (2021).

\bibitem{Grabsch2022} A. Grabsch, A. Poncet, P. Rizkallah, P. Illien,  and O. B\'enichou, ``Exact closure and solution for spatial correlations in single-file diffusion,'' Sci. Adv. {\bf 8}, eabm5043 (2022).

\bibitem{Grabsch2024} A. Grabsch and O. B\'enichou, ``Tracer Diffusion beyond Gaussian Behavior: Explicit Results for General Single-File Systems,'' Phys. Rev. Lett. {\bf 132}, 217101 (2024).

\bibitem{Sanders2012} L. P. Sanders and T. Ambj\"ornsson,
``First passage times for a tracer particle in single file diffusion and fractional Brownian motion,''
J. Chem. Phys. {\bf 136}, 175103 (2012).

\bibitem{Lapolla2022} A. Lapolla, ``The First Exit Time Statistics and the Entropic Forces in Single File Diffusion,'' arXiv:2205.02339v1.

\bibitem{Harris1965} T. E. Harris, ``Diffusion with ``collisions'' between particles,'' J. Appl. Prob. {\bf 2}, 323 (1965).

\bibitem{Lizana2010} L. Lizana, T. Ambj\"ornsson, A. Taloni, E. Barkai, and M. A. Lomholt, ``Foundation of fractional Langevin equation: Harmonization of a many-body problem,'' Phys. Rev. E {\bf 81}, 051118 (2010).

\bibitem{Metzler2000} R. Metzler and J. Klafter, ``The random walk's guide to anomalous diffusion: a fractional dynamics approach,'' Phys Rep. {\bf 339}, 1 (2000).

\bibitem{Metzler2004} R. Metzler and J. Klafter, ``The restaurant at the end of the random walk: recent developments in the description of anomalous transport by fractional dynamics,'' J. Phys. A Math. Gen. {\bf 37}, R161 (2004).

\bibitem{Metzler2014a} R. Metzler, J.-H. Jeon, A. G. Cherstvy, and E. Barkai, ``Anomalous diffusion models and their properties: non-stationarity, non-ergodicity, and ageing at the centenary of single particle tracking,'' Phys. Chem. Chem. Phys. {\bf 16}, 24128 (2014).

\bibitem{Metzler2014b} R. Metzler, L. Sanders, M. A. Lomholt, L. Lizana, K. Fogelmark, and T. Ambj\"ornsson, ``Ageing single file motion,'' Eur. Phys. J. Spec. Top. {\bf 223}, 3287 (2014).

\bibitem{Levitt1986} D. G. Levitt, ``Interpretation of Biological Ion channel flux data: Reaction-Rate versus Continuum Theory,'' Ann. Rev. Biophys. Biophys. Chem. {\bf 15}, 29 (1986).

\bibitem{Bauer2006} R. W. Bauer and W. Nadler, ``Molecular transport through channels and pores: Effects of in-channel interactions and blocking,'' Proc. Natl. Acad. Sci. U.S.A. {\bf 103}, 11446 (2006).

\bibitem{Abad2009} E. Abad, J. Reingruber, and M. S. P. Sansom, ``On a novel rate theory for transport in narrow ion channels and its application to the study of flux optimization via geometric effects,'' J. Chem. Phys. {\bf 130}, 085101 (2009).

\bibitem{Benichou2015} O. B\'enichou and J. Desbois, ``Occupation times for single-file diffusion,''
J. Stat. Mech. P03001 (2015).

\bibitem{Skorokhod1961} A. V. Skorokhod, ``Stochastic equations for diffusion processes in a bounded region,'' Theory Probab. Appl. {\bf 6}, 264 (1961).

\bibitem{Grebenkov2007a} D. S. Grebenkov, ``NMR survey of reflected Brownian motion,'' Rev. Mod. Phys. {\bf 79}, 1077 (2007).

\bibitem{Grebenkov2007b} D. S. Grebenkov, ``Residence times and other functionals of reflected Brownian motion,'' Phys. Rev. E {\bf 76}, 041139 (2007).

\bibitem{Grebenkov2019} D. S. Grebenkov, ``Probability distribution of the boundary local time of reflected Brownian motion in Euclidean domains,'' Phys. Rev. E {\bf 100}, 062110 (2019).

\bibitem{Dai2016} L. Dai, C. B. Renner, and P. S. Doyle, ``The polymer physics of single DNA confined in nanochannels,'' Adv. Colloid Interface Sci. {\bf 232}, 80 (2016).

\bibitem{Hille2001}
B. Hille, \emph{Ionic Channels of Excitable Membranes}, 3rd ed. (Sinauer, Sunderland, 2001).

\bibitem{Lutz2004} C. Lutz, M. Kollmann, and C. Bechinger, ``Single-File Diffusion of Colloids in One-Dimensional Channels,'' Phys. Rev. Lett. {\bf 93},
026001 (2004).

\bibitem{Dzubiella2003} J. Dzubiella, H. L\"owen, and C. N. Likos,
``Depletion Forces in Nonequilibrium,'' Phys. Rev. Lett. {\bf 91},
248301 (2003).

\bibitem{Doi1986} M. Doi and S. F. Edwards, \emph{The Theory of Polymer Dynamics} (Oxford University Press, New York, 1986).

\bibitem{Grosberg2002} A. J. Grosberg and A. R. Khokhlov, \emph{
Statistical physics of macromolecules} (AIP Press, Woodbury, New York, 1994).

\bibitem{Delfau2012} J.-B. Delfau, C. Coste, and M. Saint Jean, ``Single-file diffusion of particles in a box: Transient behaviors,'' Phys. Rev. E {\bf 85}, 061111 (2012).

\bibitem{Potts2011} J. R. Potts, S. Harris, and L. Giuggioli, ``An anti-symmetric exclusion process for two particles on an infinite 1D lattice,''
J. Phys. A: Math. Theor. {\bf 44}, 485003 (2011).

\bibitem{CarslawJaeger1959}
H. S. Carslaw and J. C. Jaeger, \emph{Conduction of Heat in Solids}, 2nd ed. (Clarendon Press, Oxford, 1959).

\bibitem{Prudnikov1998}
A. P. Prudnikov, Yu. A. Brychkov, and O. I. Marichev, \emph{Integrals and series, Vol. 1:Elementary functions} (Gordon and Breach, Amsterdam, 1998), p. 689.

\bibitem{Abramowitz1972} M. Abramowitz and I. Stegun, \emph{Handbook of Mathematical Functions}, 10th ed. (Dover, Washington, 1972).

\bibitem{Binder2015} D. P. Landau and K. Binder, \emph{A guide to Monte Carlo simulations in statistical physics}, 4th ed. (Cambridge University Press, Cambridge, 2015).

 \bibitem{Bortz1975} A. B. Bortz, M. H. Kalos, and J. L. Lebowitz, ``A new algorithm for Monte Carlo simulations of Ising spin systems,'' J. Comput. Phys. {\bf 17}, 10 (1975).

\bibitem{Gillespie1976} D. J. Gillespie, ``A General Method for Numerically Simulating the Stochastic Time Evolution of Coupled Chemical Reactions,'' J. Comput. Phys. {\bf 22}, 403 (1976).

\bibitem{Gillespie2001} D. J. Gillespie, ``Approximate accelerated stochastic simulation of chemically reacting systems,'' J. Chem. Phys. {\bf 115}, 1716 (2001).

\bibitem{Fichthorn1991} K. A. Fichthorn and W. H. Weinberg,
``Theoretical foundations of dynamical Monte Carlo simulations,''
J. Chem. Phys. {\bf 95}, 1090 (1991).

\bibitem{footnote} Note that the limit $t\to 0$ must be understood in the usual sense in the context of diffusion, i.e.
on the one hand the times $t$ are still long enough for the walker to have performed a large number of jumps $N_\text{jump} \gg 1$, but on the other hand short enough to ensure that $Dt$ remains small for a given value of $D$.

\bibitem{Rubinstein2003} M. Rubinstein and R. H. Colby, \emph{Polymer Physics} (Oxford University Press, New York, 2003).

\bibitem{Uspensky1937}
J.V. Uspensky, \emph{Introduction to Mathematical Probability}
(McGraw-Hill, New York, 1937), pp. 23-24 (Problem 13).

\bibitem{Barkai2009} E. Barkai and R. Silbey, ``Theory of Single File Diffusion in a Force Field,'' Phys. Rev. Lett. {\bf 102}, 050602 (2009).

\bibitem{Barkai2010} E. Barkai and R. Silbey, ``Diffusion of tagged particle in an exclusion process,''
Phys. Rev. E {\bf 81}, 041129 (2010).

\bibitem{Lapolla2019} A. Lapolla and A. Godec, ``Manifestations of Projection-Induced
Memory: General Theory and the Tilted Single File,'' Front. Phys. {\bf 7}, 182 (2019).

\bibitem{Prudnikov1992}
A. P. Prudnikov, Yu. A. Brychkov, and O. I. Marichev, \emph{Integrals and series, Vol. 4:
Direct Laplace transforms} (Gordon and Breach, Amsterdam, 1992), p. 173.

\bibitem{Mazo1987}
R. M. Mazo, ``On the Green's Function for a One-Dimensional Random Walk,'' Cell Biophys. {\bf 11},
19 (1987).

\end{thebibliography}
\end{document}